\documentclass[prd,showpacs,nofootinbib,floatfix,superscriptaddress,11pt]{revtex4-1}
\usepackage{amsmath}
\usepackage{amsfonts}
\usepackage{tikz}
\usetikzlibrary{snakes}
\usepackage{amsmath}
\usepackage{amssymb}
\usepackage{CJKutf8}
\usepackage{verbatim,graphicx}
\usepackage{mathtools}
\usepackage{color}
\usepackage[all]{xy}
\usepackage{simplewick}
\usepackage{hyperref}
\usepackage{setspace}
\usepackage{array}
\usepackage{tikz}
\usepackage{tkz-euclide}
\usepackage{amsthm}
\usepackage{enumitem}
\usepackage{bbm}
\usepackage{bbold}

\newcounter{jvcc}
\newcounter{dario}

\newcounter{yjcc}

\newcommand{\llangle}{\left\langle}
\newcommand{\rrangle}{\right\rangle}
\newcommand{\Tr}{{\rm Tr}}

\newcommand{\eref}[1]{(\ref{#1})}
\newcommand{\nn}{\nonumber}

\newcommand{\be}{\begin{eqnarray}}
\newcommand{\ee}{\end{eqnarray}}
\newcommand{\vev}[1]{\langle #1\rangle}

\newcommand{\bmat}{\left ( \begin{array}{cc} }
	\newcommand{\emat}{\end{array} \right ) }

\usetikzlibrary{shapes.misc}

\tikzset{cross/.style={cross out, draw=black, fill=none, minimum size=2*(#1-\pgflinewidth), inner sep=0pt, outer sep=0pt}, cross/.default={2pt}}

\graphicspath{{figures/}}

\def\Tr{\textrm{Tr}}

\usetikzlibrary{shapes.misc}

\newcommand{\beq}{\begin{equation}}
\newcommand{\beqs}{\begin{equation*}}
\newcommand{\eeq}{\end{equation}}
\newcommand{\eeqs}{\end{equation*}}

\allowdisplaybreaks

\begin{document}

%%%%%%%%%%%%%%%%%%%%%%%%%%%%%%%%%%%%%%%%%%%%%%%%%%%%%%%%%%%%%%%%
\title{Replica
  Symmetry Breaking in Random Non-Hermitian Systems}
\author{Antonio M. Garc\'\i a-Garc\'\i a}
\email{amgg@sjtu.edu.cn}
\affiliation{Shanghai Center for Complex Physics,
	School of Physics and Astronomy, Shanghai Jiao Tong
	University, Shanghai 200240, China}
\author{Yiyang Jia\begin{CJK*}{UTF8}{gbsn}
(贾抑扬)
\end{CJK*}}
\email{yiyang.jia@weizmann.ac.il}
\affiliation{Department of Particle Physics and Astrophysics, Weizmann Institute of Science, Rehovot 7610001, Israel}
\author{Dario Rosa}
\email{dario\_rosa@ibs.re.kr}
\affiliation{Center for Theoretical Physics of Complex Systems,
  Institute for Basic Science (IBS), Daejeon - 34126, Korea}
\affiliation{Basic Science Program, Korea University of Science and Technology (UST), Daejeon - 34113, Korea}
\author{Jacobus J. M. Verbaarschot}
\email{jacobus.verbaarschot@stonybrook.edu}
\affiliation{Center for Nuclear Theory, Department of Physics and Astronomy, Stony Brook University, Stony Brook, New York 11794, USA}
\begin{abstract}
  Recent studies have revealed intriguing similarities between the contribution of wormholes to the gravitational path integral and the phenomenon of replica symmetry breaking observed in spin glasses and other disordered systems. Interestingly, these configurations may also be important for the explanation of
  the information  paradox of quantum black holes.
  Motivated by these developments, we investigate the thermodynamic properties of a $PT$-symmetric system composed of two random non-Hermitian Hamiltonians with no explicit coupling between them. After
  performing ensemble averaging, we identify numerically and analytically a robust first-order phase transition in the free energy of two models with quantum chaotic dynamics: the elliptic Ginibre ensemble of random matrices and a non-Hermitian Sachdev-Ye-Kitaev (SYK) model. The free energy of the Ginibre model is temperature-independent in the low-temperature phase. The SYK model has a similar behavior for sufficiently low temperature, then it experiences a possible continuous phase transition to a phase with a temperature-dependent free
  energy before the first-order transition takes place at a higher temperature.
 We identify the order parameter of the first-order phase transition and obtain analytical expressions for the critical temperature.
 The mechanism behind the transition is the existence of replica symmetry breaking
 configurations coupling Left and Right replicas
  that control the low-temperature limit of the partition function. We speculate that quantum chaos may be necessary for the observed dominance of off-diagonal replica symmetry breaking configurations in the low-temperature limit.  

\end{abstract}

\maketitle
\tableofcontents
\section{Introduction}
The replica trick \cite{edwards1975} is a powerful tool in the study of disordered systems. It consists of replicating the action $n$ times which facilitates the explicit calculation of the average over disorder. The resulting $n$-dependent action, describing the ensemble-averaged system, 
%and defined in an abstract %replica space, 
is then, in most cases, solved in the mean-field limit by the saddle-point method. In the last step of the calculation, the value of $n$ is set to a value that depends on the observable of interest (typically $0$ or $1$). 

The replica trick has been employed in a broad variety of problems in different research fields including disordered spin systems \cite{parisi1979,parisi1983}, quantum disordered conductors \cite{wegner1979}, random matrix theory \cite{mezard1999}, QCD \cite{stephanov1996,Akemann:2004dr} and the development of error correction codes \cite{nishimori1999}.  
For instance, in the context of disordered spin systems describing certain magnetic alloys, the replica trick plays a pivotal role in the physical description of the low-temperature spin-glass phase characterized by an energy landscape with multiple local minima and a splitting of the Gibbs measure into separate components (called pure states), which is a signature of breaking of ergodicity \cite{mezard1984,mezard1985}.

It was also found \cite{parisi1979,parisi1983,mezard1984,mezard1985} that replica symmetry breaking solutions of the Sherrington-Kirkpatrick model \cite{sherrington1972}, a model for these disordered spin systems, describe the low-temperature spin-glass region, while replica symmetric configurations are dominant for higher temperatures. Replica symmetry breaking (RSB) refers to solutions of the saddle point equations which couple different replicas and that,  superficially, should be subleading in the mean field limit. These RSB solutions
have a precise physical meaning for spin glasses \cite{parisi1983,mezard1984}: they represent the overlap of probability among pure states which is directly related to the order parameter of the transition.

A different type of RSB is found
in the context of disordered systems \cite{wegner1979}
and random matrix theory \cite{mezard1999,kanzieper2002,nishigaki2002}.
In this case, the replica symmetry between the advanced and retarded sectors
of the Green's function is broken leading to Goldstone's modes that dominate
the partition function. These configurations give non-perturbative contributions to spectral correlators that provide information on the dynamics for scales on the order of the Heisenberg time.
Indeed, fully accounting for all RSB solutions it reproduces \cite{mezard1999,kanzieper2002,nishigaki2002} the exact random matrix theory result for the two-level correlation function. 
 
Another model that has recently been intensively studied by means of the replica trick is the Sachdev-Ye-Kitaev model \cite{french1970,bohigas1971,kitaev2015,sachdev1993,maldacena2016,jensen2016}: a model describing $N$ Majorana fermions with infinite-range random interactions in Fock space. Variants of this model with complex fermions were originally introduced  \cite{french1970,french1971,bohigas1971,bohigas1971}
and studied \cite{brody1981,Verbaarschot:1985jn,Flores:2000ew,Benet:2002br,Zelevinsky:2003pi,Kota:2022lth}
  in the context of nuclear physics and quantum chaos over half a century ago.

  The renewed interest in this model is due to intriguing similarities with Jackiw-Teitelboim (JT) gravity \cite{teitelboim1983,jackiw1985}, a two-dimensional theory of gravity that describes almost extremal black holes in near AdS$_2$ backgrounds \cite{jensen2016,maldacena2016a,engels2016}. 
 In the infrared limit, both models share the same action: a Schwarzian whose path integral can be evaluated exactly \cite{stanford2017}. The resulting spectral density \cite{Cotler:2016fpe,garcia2017}, which grows exponentially for excitations close to the ground state, is consistent with that of quantum black holes.
 The dynamics is quantum chaotic \cite{kitaev2015} with spectral correlations given by random matrix theory predictions
 \cite{garcia2016,Cotler:2016fpe},
 classified according to the global symmetries of the system \cite{you2016,garcia2018a}. Likewise, a weakly coupled two-site SYK model, which is also quantum chaotic
 \cite{Garcia-Garcia:2019poj,Fremling:2021wwy,Cao:2021xcq,Caceres:2021nsa}
for sufficiently high energies, reproduces the physics of the transition from a traversable wormhole to a two-black-hole configuration in near-AdS$_2$ backgrounds with Lorentzian signature \cite{maldacena2018,gao2016}. 
 
On the gravity side, it may seem that disorder, and therefore any non-trivial structure in replica space, plays no role and that these similarities with the SYK model, where the replica symmetric solution is typically chosen, are unrelated to the fact that the SYK model is a disordered system. However, recent results in the gravity literature put in doubt this prediction.
In a recent work by Saad, Shenker and Stanford~\cite{Saad:2019lba}, it was found that the dual theory of JT gravity was exactly given by a random matrix theory
in a certain scaling limit which suggests that the gravitational path integral involves an average over different theories. 
 Moreover, a replica calculation \cite{engelhardt2020} of the free energy in JT gravity identified a range of parameters where the contribution of RSB configurations, called replica wormholes in this context, are dominant compared to replica symmetric configurations. 
Similarly, the calculation \cite{Almheiri:2019qdq,Almheiri:2020cfm,Penington:2019npb} of the evolution of the von Neumann entropy in JT gravity plus additional matter, modeling the black hole evaporation process, showed that for late times the growth stops due to additional RSB saddle points in the gravitational path integral, which represent wormholes connecting different copies of black holes. This behavior is in agreement with that expected for Hermitian systems \cite{page1993}. The ``information  paradox'' is therefore avoided.

However, these results also raise some fundamental issues. 
It seems that the 
gravitational path integral represents an ensemble over theories, 
something that is not yet well understood. Moreover, at least in field theories 
with a gravity dual, Euclidean wormholes raise the so-called factorization puzzle,
 namely,  the field theory dual
to wormholes connecting two boundaries should be related to a field theory 
partition function that does not factorize \cite{maldacena2004,Saad:2021uzi,Belin:2021ibv,Johnson:2022wsr,Berkooz:2022fso,Schlenker:2022dyo,Goto:2021wfs} but it is unclear 
how exactly to define such an object. Another problem is that these 
Euclidean wormholes, at least in JT gravity without additional matter, are not 
solutions of the classical equations of motion \cite{Saad:2019lba,Gao:2021tzr} 
so their interpretation as RSB saddle solutions is not
 straightforward. In the simplest case of two replicas, it was possible
 \cite{Garcia-Garcia:2020ttf} to find wormhole solutions of the classical JT gravity 
  equations provided that complex sources were added. The system undergoes 
  a first-order wormhole-black hole transition where the wormhole phase is 
  characterized by a free energy that depends only weakly on the temperature until
  a possible second continuous phase transition occurs, below which the free energy becomes
  temperature independent.
  
Given these recent advances, an interesting question to ask is whether it is possible to find field theories whose dominating saddle points are RSB configurations and whether their role is qualitatively similar to that of wormholes in gravity theories. A positive answer to this question may shed some light on the factorization and information loss puzzle mentioned above and, more generally, on the role of wormholes in holography and quantum gravity. Even putting aside any gravitational interpretation, it is a problem of fundamental interest to determine the conditions for the dominance of off-diagonal replicas in disordered and strongly interacting quantum mechanical systems. 
 
  The main goal of this paper is to address this problem by studying several 
  random non-Hermitian but $PT$ symmetric two-site systems with no explicit coupling
   between them. Among others, we investigate the elliptic Ginibre ensemble of random matrices
    and the non-Hermitian SYK model \cite{Garcia-Garcia:2021elz}. By downgrading the Hermiticity of the SYK model to just $PT$ symmetry 
    \cite{bender1998}, so that the model still has a real positive partition function, 
 we identify RSB configurations that control the free energy 
 in the low-temperature limit. The restoration of replica symmetry at higher 
 temperature triggers a first-order thermal phase transition. 
If the imaginary part of the SYK model is large enough, we have indications of the existence of an additional continuous phase transition at a temperature below the one at which the first-order transition takes place.    
 Moreover, we obtain explicit expressions for the critical temperature, the ground state energy and the order parameter that characterizes the RSB phase. Our results are qualitatively similar to those of a gravitational system
 \cite{Garcia-Garcia:2020ttf} and also largely universal provided that the dynamics is quantum chaotic \cite{Garcia-Garcia:2021elz}.

 We note that the role of RSB configurations has already been the subject of different studies \cite{arefeva2018,wang2018} for the SYK model with real couplings. Although there is not yet consensus in the literature, it seems that in these cases most of the features of the model, which are also present in JT gravity, do not involve any RSB. 
  
 The paper is organized as follows: in section \ref{sec:qualitative}, we qualitatively explain why we expect a universal thermal phase transition due to RSB configurations in a non-Hermitian random quantum system. This is illustrated in section \ref{sec:ginibre} by an analytical solution
 of a non-Hermitian random matrix model with $PT$ symmetry which roughly corresponds
 to the two-site non-Hermitian SYK model with a $q$-body ($q > 2$) interaction.
 In section \ref{sec:sykchaos},
 by an explicit solution of the Schwinger-Dyson (SD) equations and also by the numerical calculation of the free energy from the eigenvalues of the SYK Hamiltonian, we show that a $q=4$ two-site non-Hermitian SYK model with $PT$ symmetry and no explicit coupling between the  two sites, also undergoes a first-order phase transition induced by RSB configurations. We close with concluding remarks and a list of topics for further research in section \ref{sec:outlook}. Technical details are worked out in
 six appendices. Some of the results of this paper were announced in a recent letter
 \cite{Garcia-Garcia:2021elz}.
 \section{Replica symmetry breaking in random non-Hermitian, $PT$-symmetric systems}\label{sec:qualitative}
In this section, we aim to give a qualitative argument for the existence of a rather universal phase transition for the free energy of a $PT$-symmetric system composed of two random disconnected non-Hermitian Hamiltonians. This can be viewed as a replicated version (with two replicas) of a single-site non-Hermitian Hamiltonian.
The low-temperature phase is dominated by RSB configurations whose effect is strikingly similar to that of Euclidean wormholes in AdS$_2$ gravity.
In later sections, we discuss examples including a two-site non-Hermitian SYK model where an explicit replica analysis is possible. 
 
 We argue below that for the two-site non-Hermitian systems we are interested in, the replica trick gives correct results. We will also see that for these systems the quenched and annealed free energies are identical in the thermodynamic limit. This justifies using annealed averaging to obtain quenched free energies, which we will do for the Schwinger-Dyson calculation of the free energy. 
 
 In the second part of this section,  we show that when
eigenvalues
 have the universal characteristics of quantum chaotic systems,
 the connected two-level correlation function corresponding to RSB
 configurations,
 contributes  to the free energy at leading order.
 Moreover, we argue that these contributions indeed
 control the low-temperature limit of the free energy.
 In section \ref{sec:equations}, an analysis of the Schwinger-Dyson (SD) equations for the
 one-replica SYK model will show more explicitly that RSB
 configurations are directly responsible for the phase transition which mimics that observed for Euclidean wormholes in JT gravity \cite{Garcia-Garcia:2020ttf}.   
 
 \subsection{Quenched free energy by the replica trick} 
\label{sec:IIa}
 We consider the  partition functions of two-site Hamiltonians
of the form
 \be\label{eq:hami}
 H=H_L \otimes 1 + 1 \otimes H_R.
 \ee
 We are mostly interested in the case where $H_L = H_R^\dagger$, and
 in general $H_L, H_R, H$ are non-Hermitian but $H$ is $PT$-symmetric \cite{bender1998}, namely,
 \be
 [PT, H]=0, \nn
 \ee
 with $P$ a permutation matrix that interchanges the $L$ and $R$ Hilbert spaces and the anti-unitary operator is
 $T =CK \otimes C K$. Here $C$ is some charge conjugation matrix and $K$
 the complex conjugation operator.
 If the $D$ eigenvalues of the complex $D\times D$
 matrix $H_L$ are denoted by $E_k$, then the $D^2$ eigenvalues of $H$ are given by $E_k+E_l^*$. The eigenvalues with $k=l$ are real while the other eigenvalues
 come in complex-conjugate pairs, consistent with the existence of $PT$ symmetry.
 
 The partition function of this Hamiltonian (before averaging over the disorder) is given by
 \be
 Z(\beta)=\Tr e^{-\beta H} =  Z_L Z_R=|\Tr e^{-\beta H_L}|^2,
 \ee
where we have defined
\begin{equation}
Z_L\equiv\Tr e^{-\beta H_L}, \ Z_R\equiv\Tr e^{-\beta H_R},
\end{equation}
 and obviously $Z_L =Z_R^*$. If $\rho(z)$ is the eigenvalue density of $H_L$ then
 \be
 Z(\beta)= \int d^2z_1 d^2z_2 \rho(z_1)\rho(z_2) e^{-\beta(z_1+z_2^*)}.
 \ee
 
 The quenched free energy must be computed by a quenched 
 average $-\beta \langle F \rangle  = \langle \log Z \rangle $ where $\beta$ is the inverse of temperature $T$. A direct analytical calculation of the quenched disorder average is in general technically demanding. 
 The replica trick was introduced \cite{edwards1975} to circumvent these difficulties by using that 
 \be \label{eqn:two-siteReplicaLimit}
 \langle  \log Z \rangle = \lim_{n\to 0} \frac {\left \langle Z^n\right \rangle -1}n = \lim_{n\to 0} \frac {\left \langle (Z_L Z_L^*)^n\right \rangle -1}n .
 \ee
 The average on the right-hand side is much easier to evaluate analytically by replicating $n$ times the original action, carrying out the averages analytically and taking the limit $n \to 0$ at the end of the calculation.   
 
 However, a word of caution is in order: it is well-documented that the replica trick may give incorrect results if
 applied naively \cite{verbaarschot1985,zirnbauer1999another}. An example is the Sherrington-Kirkpatrick model mentioned earlier, where the entropy is negative for sufficiently low temperature if the replica trick is naively applied \cite{sherrington1972}. A number of
 fixes have been introduced \cite{kanzieper2002,mezard1999,parisi1983,splittorff:2003cu,sedrakyan2005toda}
 including the supersymmetric method that avoids the replica trick
 altogether
 \cite{efetov1983supersymmetry,wegner1983,verbaarschot1984,verbaarschot1984a,sedrakyan2020supersymmetry}.
 However, in many situations there are no realistic alternatives so it is necessary to understand under which conditions the trick is applicable.
 The replica trick is premised on Carlson's theorem \cite{carlson1914} which states that
 if a holomorphic function $f(z)$ on ${\rm Re}(z)> 0$ vanishes for all positive integers $n$, it also
 vanishes on the right half-plane, provided that $|f(z)|< C \exp(\pi|z|)$ on the imaginary
 axis and grows no faster than an exponential elsewhere on the right half plane.    
 For a non-Hermitian Hamiltonian such as $H_L$, $\log Z_L$  in general has a nonzero imaginary part, and therefore it is unclear whether the conditions of Carlson's theorem are satisfied in the low-temperature limit. 
 Hence, if we were interested in the free energy of the one-site model,
 the naive replica trick 
 \be \label{eqn:one-siteNaiveReplicaLimit}
 \langle  \log Z_L \rangle = \lim_{n\to 0} \frac {\left \langle Z_L^n\right \rangle -1}n
 \ee
 is likely to give incorrect results.

 The average of the one-site free energy can be expressed as
 \be
 \left\langle \log Z_L \right \rangle =  
 \left\langle \log |Z_L| \right \rangle +\left\langle i {\rm arg} Z_L \right \rangle.
   \ee
   For a non-Hermitian Hamiltonian, the phase of $Z_L$ is expected to
   oscillate rapidly so that the average of the second term vanishes.
If that is the case, we have 
   \be
   \left\langle \log Z_L \right \rangle =  
   \left\langle \log |Z_L| \right \rangle =
   \frac 12 \left\langle \log Z_L Z_L^* \right \rangle.
\label{quench}
   \ee
   This shows that the quenched average free energy is
   necessarily given by the
   quenched free energy of a replica and a conjugate replica (in the sense of the one-site model).
   Because $\log (Z_L Z_L^*)$ is real, we have that
 $\langle \exp n\log Z_LZ_L^* \rangle$ is bounded for imaginary $n$
 so that there is a chance we can apply Carlson's theorem to validate the replica trick. We thus have
\be
   \label{eqn:one-siteCorrectReplicaLimit}
 \langle  \log Z_L \rangle = \lim_{n\to 0} \frac 1{2}
 \frac {\left \langle( Z_LZ_L ^*)^n\right \rangle -1}n.
 \ee
 Notice that this is exactly half of   \eqref{eqn:two-siteReplicaLimit}, therefore the correct replica description of a non-Hermitian one-site model naturally involves the conjugate replicas. This procedure  is actually well known for quenched averages (now understood as
   ignoring the fermion determinant)
   of a similar quantity, namely  the resolvent $G(z) = {\rm Tr}(H_L-z)^{-1}$ 
    \cite{girko2012theory,stephanov1996,nishigaki2002a}. For a non-Hermitian Hamiltonian, the quenched resolvent is given by the replica limit,
 \be
 G(z) = \lim_{n\to 0} \frac 1{2n}  \frac{d}{dz}
 \left \langle {\det}^n(H_L+z)   {\det}^n(H_L^\dagger+z^*)\right \rangle,
 \ee   
 which is sometimes referred to as Hermitization \cite{feinberg1997non,girko2012theory,stephanov1996,Janik:1996xm}.

 More importantly,
 we will study the two-site system using the mean field approximation. We do not expect
 RSB to occur for the replication of the two-site
 system,  namely we expect replica diagonal behavior
 \be
 \langle( Z_L Z_L^*)^n \rangle = \langle Z_L Z_L^*\rangle^n.
 \ee
 Then the replica limit \eref{eqn:one-siteCorrectReplicaLimit} is given by
\be
 \langle \log Z_L \rangle=\lim_{n\to 0} \frac 1{2}
 \frac {\left \langle( Z_LZ_L ^*)^n\right \rangle -1}n
 =
 \lim_{n\to 0} \frac 1{2}
 \frac {\left \langle Z_LZ_L ^*\right \rangle^n -1}n
 =\frac 12\log  \left \langle Z_L Z_L^* \right\rangle.
 \ee
 We conclude that in the thermodynamic limit, the quenched free energy of
 $Z_L$ is given by half the annealed free energy of $ Z_L Z_L^* $.
 The latter, for non-Hermitian theories, is generally different from the annealed free energy of $Z_L$.

 At this point it is useful to clarify a potentially confusing semantic point of
 our notion of RSB which is different
 from that in spin glasses. It is reminiscent to RSB
 in disordered systems where  RSB happens between $n$
 retarded and $n$ advanced Green's functions, \textit{i.e.} 
 $U(2n) \to U(n) \times U(n)$. In the present case 
 we have RSB
 between replicas and conjugate replicas of the partition function.
 In the replica symmetric phase, the replicas remain uncoupled
 after averaging so that
 \be
 \langle( Z_L Z_L^*)^n \rangle = \langle Z_L\rangle^n \langle Z_L^*\rangle^n.
 \ee
 When
 replica symmetry is broken, this factorization no longer holds: 
 \be
 \langle( Z_L Z_L^*)^n \rangle \ne \langle Z_L\rangle^n \langle Z_L^*\rangle^n.
 \ee
 but we still have that
\be
 \langle( Z_L Z_L^*)^n \rangle = \langle Z_L Z_L^*\rangle^n.
 \ee
So from the two-site model perspective, $ \langle Z_LZ_L^* \rangle$
is the one-replica partition function of the two-site Hamiltonian $H$, which
is not expected to bring about any further RSB. On the other hand, this can be viewed as the two-replica partition function of the single-site Hamiltonian $H_L$. In that case, one can legitimately talk about RSB. However, the two perspectives are mathematically equivalent. 

 For a characterization of the conditions to observe dominant RSB configurations is important to split the partition function into a connected and a disconnected piece:
 \be
 \langle Z \rangle =  \langle Z_LZ_L^* \rangle =   \langle Z_LZ_L^* \rangle_c
+ \langle Z_L\rangle \langle Z_L^* \rangle.
 \ee
The first term receives contributions from
 the connected two-point function while the second term is
 determined by the one-point function. Because of the non-Hermiticity,
 $\langle Z_L(\beta)  \rangle$ may actually be exponentially
 suppressed so that the connected part of the partition function may
 become dominant.  As discussed in the previous paragraph,
 we will refer to this situation as RSB.
 In a field theory formulation of the partition function,
 the corresponding saddle-point configuration of the action connects 
 {\it different} replicas. In this paper, we will not pursue an explicit gravitational
 interpretation of these results. However, the analogy with gravity is evident:
 RSB configurations are the field theory analogue of
 Euclidean wormhole solutions in gravity, which are tunneling
 geometries connecting two or more otherwise disconnected space-time regions.
 However, we note that if this analogy applies, the relation with a
 gravitational system is not at the level of the microscopic Hamiltonian but
 rather at the level of the effective action resulting from the replica trick after ensemble averaging.
 
 The remainder of the paper is devoted to a better understanding of both the circumstances for which the connected part becomes relevant so that RSB configurations control the free energy of the system and the effect of RSB on the thermodynamic properties of the system.  
  
 \subsection{Existence of a phase transition induced by RSB configurations}
\label{sec:IIb}
 
 For the random Hamiltonian   (\ref{eq:hami}), we calculate the expectation value of the
 partition function as
 \be
 \left \langle Z(\beta) \right \rangle=
 \int d^2z_1 d^2z_2 \int\left \langle \rho(z_1)\rho(z_2)\right \rangle
 e^{-\beta(z_1+z_2^*)},
 \ee
where $\beta \equiv \frac 1T$ is the inverse temperature and the level density is given by
 \be
 \rho(z) = \sum_k \delta^2(z-z_k).
 \ee
The two-point correlation function can be decomposed as 
 \be
 \left \langle\rho(z_1)\rho(z_2)\right \rangle 
 =\left \langle\rho(z_1)\right \rangle\left \langle \rho(z_2)\right \rangle 
 +\delta^2 (z_1-z_2) \langle \rho(z_1) \rangle + R_{2,c}(z_1,z_2)
 \label{twop}
 \ee
 with 
 \be
 R_{2, c}(z_1,z_2)\equiv \sum_{k\ne l}\left [
 \left \langle \delta^2 (z_1-z_k)\delta^2(z_2-z_l) \right \rangle
 -\left\langle \delta ^2(z_1-z_k)\right \rangle
 \left \langle \delta^2(z_2 -z_l)  \right \rangle
 \right ]
 \ee
 which is the two-point correlator without self-correlations. Because of the normalization of the level density, we have the
 sum rule obtained by integrating   \eref{twop} over $z_1$:
  \be
 \int d z_1 R_{2,c}(z_1,z_2) = - \langle \rho(z_2) \rangle
 \label{sum-rule1}.
 \ee
 The decomposition of the partition function corresponding to   \eref{twop}
 is 
 \be
 \llangle Z(\beta) \rrangle =  |\llangle Z_L(\beta)\rrangle|^2
 + \int d^2z \rho(z)e^{-\beta(z+z^*)} + 
  \int d^2z_1 d^2 z_2 R_{2,c}(z_1,z_2)e^{-\beta(z_1+z^*_2)}. 
 \label{part-dec}
 \ee
Notice that, from   \eqref{part-dec} on, we will no longer make a notational distinction between $\rho(z)$ and $\langle \rho(z) \rangle$ when the context is free of confusion.  Because of the sum rule  \eref{sum-rule1}, 
 the second term in   \eqref{part-dec} cancels the third
 term for $\beta = 0$. Therefore, there is no RSB in the infinite temperature limit. We shall see that for sufficiently low temperature the situation is different.  

 To simplify the argument, for now we assume a rotationally invariant eigenvalue distribution, so that
 \be
 \int d^2z \rho(z) e^{-\beta z} = D
 \ee
 with $D$ the number of eigenvalues of the one-site Hamiltonian. We note that this is a realistic situation which can occur for instance for $H$ given by two copies of the Ginibre ensemble \cite{ginibre1965} of complex random matrices. 

 We next estimate the connected part of the partition function. To do
 that we need to make three  assumptions on the two-point correlations of the
 eigenvalues:
 \begin{itemize}
 \item[1.] The correlations are isotropic and
   only depend on the distance of the eigenvalues so
   that
   \be
   R_{2,c}(z_1,z_2) = -\rho(\bar z)^2 F(|z_1-z_2|/ \lambda),
\label{R2}
   \ee
   where $\bar z$ is the center of mass coordinate $\bar z =(z_1+z_2)/2$.
\item[2.] The correlation length is a function of $\rho(\bar z)$ only.
\item[3.] The average spectral density $\rho(\bar z)$ does not appreciably vary on the scale of the correlation length.  

 \end{itemize}
 These assumptions are expected to hold for the universal correlations
 of non-Hermitian quantum chaotic systems.
 In terms of the integral over the center of mass (i.e. $(z_1+z_2)/2$) and the differences of
 the eigenvalues (i.e. $z_1-z_2$), the sum rule reads
 \be
 -\rho(\bar z)^2 \int  F(|z_1-z_2|/\lambda) d^2(z_1-z_2) = - \rho(\bar z). 
 \ee
 This requires that $\lambda \sim 1/\sqrt{\rho(\bar z)}$ and
\be
 \rho(\bar z)\int d^2(z_1-z_2) F( |z_1-z_2|\sqrt{\rho(\bar z)}) = 1.
 \ee
 Since $\rho(\bar z) \sim D$ we have that the length scale of the
 eigenvalue correlations is $1/\sqrt D$.
 
For the connected part of the partition function we then obtain 
\be\label{eqn:z2cIntermediate}
Z_{2,c} = -\int d^2 \bar z  \int d^2(z_1-z_2) \rho(\bar z)^2  F(|z_1-z_2|\sqrt{\rho(\bar z)})
e^{-\beta[ \bar z +\bar z^* +i{\rm Im}(z_1-z_2)]}
+ \int d^2 \bar z  \rho(\bar z) e^{-\beta(\bar z+\bar z^*)}.\nn\\
\ee
Since the correlations are short-ranged, we can Taylor expand $\exp(i{\rm Im}(z_1-z_2))$. The first term in the Taylor expansion cancels with the second integral in equation \eqref{eqn:z2cIntermediate} because of the sum rule, the second term of the Taylor expansion vanishes
because the correlations are even in $y_1-y_2$. So to leading non-vanishing order in
$1/\rho(\bar z)$ we obtain
\be\label{eqn:z2cUniv}
Z_{2,c} = \frac {\beta^2 }2
\int d^2 \bar z  \int d^2(z_1-z_2) \rho(\bar z) ^2 F\left(|z_1-z_2|\sqrt{\rho(\bar z)}\right)
( {\rm Im}(z_1-z_2))^2e^{-\beta( \bar z +\bar z^*)}.
\ee
We can scale $\rho(\bar z)$ out of the $z_1-z_2$ integrations. Then, the integral
factorizes and the integral over the difference is just a constant which we
will denote by $\langle \zeta^2 \rangle$. The connected part of the partition
function is thus given by 
\begin{align}\label{eqn:z2cCircular}
Z_{2,c} &= \frac {\beta^2 \langle \zeta^2 \rangle }2
\int_{|\bar z|<E_0} d^2 \bar z e^{-\beta( \bar z +\bar z^*)}\nn\\
&= \frac {\pi}{2} \beta E_0  \langle \zeta^2 \rangle  I_1(2\beta E_0) 
\end{align}
where $E_0$ is the radius of the support of $\rho(\bar z)$, $I_1$ is a modified Bessel function of first kind  and
\be\label{eqn:zetaUniv}
\langle \zeta ^2 \rangle = \int d^2s F( |s|) ( {\rm Im} \, s)^2,
\ee
with $s = \sqrt{\rho(\bar z)} (z_1-z_2)$. The annealed free energy is thus given by
 \be
 -\beta F = \log \left[  c \beta I_1(2\beta E_0) +D^2 \right ].
 \ee
 Since $E_0>0$ the first term becomes dominant at low temperatures.
 More specifically, a genuine phase transition at finite temperature 
 can occur provided that $\log D$ and $E_0$ scale linearly with the system size $N$. 
 This is indeed the case for interacting fermionic systems such as the (two-site) SYK model, where the ground state energy $E_0 = N e_0/2$ (each site has $N/2$ Majoranas) with $e_0$
 the ground state energy per particle that does not depend on $N$
and $ D = 2^{N/4}$. In the thermodynamic limit, we then can use the asymptotic
limit of the modified Bessel function while the prefactors are irrelevant.
The free energy
 is given by
 \be\label{eq:fheuristic}
 F/N =- \theta(T_c-T) e_0 -\theta(T-T_c) T \log 2/2
 \ee
 with
 \be
 T_c = \frac {2e_0}{\log 2}.
 \ee
 This argument for the transition is based on the existence of the above large $N$-scaling behavior, and the scaling properties of the two-point correlation function. In the next section we will study an example where these conditions are met.   
  The rotational invariance of the spectrum is in fact not essential.  In particular, from the universality arguments and equations \eqref{eqn:z2cUniv}-\eqref{eqn:zetaUniv},  we see that the connected part of the partition function depends on $\rho(z)$ only through the shape of its support. Moreover, the exponential dependence on $2 \beta E_0$ is largely independent of the shape of the support of $\rho(z)$: consider the integral in equation \eqref{eqn:z2cCircular} with a generic support, namely,
 \begin{equation}
\int_{\rm{supp } (\rho)} d^2 \bar z e^{-\beta( \bar z +\bar z^*)}  =\int_{\rm{supp } (\rho)} e^{-2\beta x} dxdy. 
 \end{equation}
If the support of $\rho$ on the real axis has a projection $[-E_0, E_{\rm max}]$, and for each $x$ the slice of the support has a length $L(x)$ along the $y$ direction, the integral becomes 
\begin{equation}
\int_{-E_0}^{E_{\rm max}} e^{-2\beta x} L(x) dx .
\end{equation}
Since $E_0$ is a large parameter ($E_0 \sim N$), the integral on the right-hand side localizes at
the maximum of the integrand, namely, at $\widetilde x =-E_0$ as long as $L(x)\ll e^{x}$. This establishes that $Z_{2,c} \sim e^{2\beta E_0}$.

  \section{Free energy for the Elliptic Ginibre Model}
 \label{sec:ginibre}
 In this section, we evaluate the free energy for the elliptic Ginibre model. In the large $D$ limit, the elliptic Ginibre model  has a constant level density inside an ellipse, while the Ginibre model has a constant level density inside a circle. This model can be seen as a representative of the
 universality class of non-Hermitian Hamiltonians which do not necessarily have a rotationally invariant
 level density. Our calculations will be framed in general terms and can be applied  
to the non-Hermitian SYK model with slight modifications. 
 Using the general arguments of section \ref{sec:IIa},
 the quenched free energy of the elliptic Ginibre model is equal to the annealed free energy
of the model with one replica and one conjugate replica.
 In the case of the ordinary Ginibre model 
the two-site partition function can be evaluated analytically for finite $D$ 
(see appendix \ref{a:ginibre}). In that case, the one-site partition function does not
  depend on $\beta$.
  This is related directly to the fact that the eigenvalue density
  is rotationally invariant. We now show that the derivation of the previous section
  is also valid for the universality class of the elliptic Ginibre ensemble.

  The Hamiltonian of the elliptic Ginibre ensemble
  \cite{fyodorov1997,fyodorov2003} is given by
   \be
 H_L=\frac { H_1 + ik H_2}{\sigma (k)}
\label{H-ellip}
\ee
here $ \sigma(k)$ is a scale factor which can be $k$-dependent. 
In this model, $H_1$ and $H_2$ are independent matrices extracted from the Gaussian Unitary Ensemble 
according to the
probability distribution 
\be\label{eqn:ellipticGinibreNormalization}
 e^{-\frac N{\sigma^2_0} \left( \Tr H_1^2 +\Tr H_2^2 \right)}.
 \ee 
Note $\sigma_0$ is a parameter that is  unrelated to $\sigma(k)$. In most of
this paper we choose $\sigma(k) = 1$, but we point out another interesting
 possibility $\sigma(k)=\sqrt {2(1-k^2)}$ such that the variance of the
 real part of the eigenvalues is independent
 of $k$. This guarantees, as we will see below, that the disconnected
 partition function, dominant in the high-temperature phase, is $k$-independent.
  This could be of interest for quantitative comparisons with the 
  gravity picture \cite{Garcia-Garcia:2020ttf} where the high-temperature phase 
  corresponds to two decoupled black holes. 
 
 We can choose (\ref{eq:hami}) as the definition of our total Hamiltonian where $H_L = H_R^\dagger$, but unlike the SYK model this would not give a $PT$-symmetric Hamiltonian. We can instead let $H_L = H_R^*$, then we have a Hamiltonian which is $PT$-symmetric. In any case, both choices give the same spectrum and hence the same partition function. 
 We next evaluate the two-site partition function with $H_L$ given by \eref{H-ellip}.

 The calculation of the partition function can actually be carried out
at finite $D$
 by integrating the expressions for the spectral density and two-level correlation functions of the elliptic Ginibre ensemble which were first obtained in  Ref.~\cite{fyodorov1997}.  Here we are interested in the large $D$ limit 
 and only need the asymptotic form of the spectral density which is
 constant inside an ellipse with long axis length $2 E_0$ and short axis length $2 y_0$:
  	\be
\rho(z) 	=\frac D{\pi E_0 y_0} \theta \left(1-\frac {x^2}{E_0^2} -\frac {y^2}{y_0^2}\right)
\label{rho-gin}
        \ee
        with
        \be
        E_0 = \frac{ \sigma_0}{\sqrt{1+k^2}\sigma(k)},
        \qquad y_0 = \frac{ \sigma_0 k^2}{\sqrt{1+k^2}\sigma(k)},
        \label{Ey} 
        \ee
       and $z=x+iy$.
We do not need the specific form of the two-point correlations other than that they are isotropic and short-range with a range that scales as $1/\sqrt D$ as given by the general form \eref{R2}.
        
To obtain the disconnected part of the partition function we need to evaluate the one-site partition function which can be easily computed by the following parameterization of the energy integration variable:
         	\be
 	z= E_0 r \cos\phi +i y_0 r \sin\phi 
\label{zpar}
 	\ee
 	with $r \in [0,1]$ and $\phi \in [-\pi,\pi]$.
 	Using that the Jacobian of this transformation is $r E_0 y_0$,
	we find the
 	partition function 
 	\be
 	Z_L(\beta) &=& \int d^2z e^{-\beta z}\rho(z)\nn\\
 	&=&
 	\frac D \pi  \int_0^1 rdr \int_{-\pi}^\pi d\phi e^{-\beta r(E_0\cos\phi + i y_0\sin\phi)}\nn\\
 	&=& \frac {2D}{\beta \sqrt{E_0^2-y_0^2}} I_1\left(\beta \sqrt{E_0^2-y_0^2}\right).
 	\ee
 Using that $E_0^2-y_0^2 =\sigma^2_0(1-k^2)/\sigma^2(k)$ we obtain the disconnected piece of the two-site partition function
 	\be
 	Z_{2,{\rm dis}}(\beta)=   \frac {4D^2\sigma^2(k)}{\beta^2\sigma^2_0(1-k^2)}
 	[  I_1(\beta \sigma_0\sqrt{1-k^2}/\sigma(k)) ]^2.
 	\ee
 	When we have $\sqrt{1-k^2}/\sigma(k)\to 0$ for $k\to 1$, as is the case for
        our generic choice of $\sigma(k) =1$, the disconnected contribution to the  free energy becomes temperature
        independent for $k=1$. 
        The connected part of the partition function is given by
        \be
        Z_{2,c} = \int d^2 z_1 d^2z_2 \rho_{2c}(z_1,z_2) e^{-\beta(z_1+z_2^*)}
\label{zcon}
        \ee
        where the connected two-point correlation function $ \rho_{2c}(z_1,z_2)$
        is given by the sum
         	\be
 	\rho_{2,c}(z_1,z_2) = R_{2,c}(z_1,z_2) + \delta^2(z_1-z_2) \rho(z_2).
 	\ee
        The first term represents the true two-point correlations involving two
        different eigenvalues while the second term is due to  self-correlations.
        To evaluate the connected partition function we can use the general
        argument given in section \ref{sec:IIb} but now we have an explicit
        expression for the two-point correlation function which satisfies
        the conditions used in that section. In particular, the two-point
        correlation function is given by
        \be
        R_{2,c}(z_1,z_2)= -\rho(\bar z)^2 F_{\rm unv}(\sqrt{ \rho(\bar z)} | z_1- z_2| ).
        \ee
        where $\bar z= (z_1+z_2)/2$
        and $F_{\rm unv}$ is a universal function
        that is given by the large $D$ result for the Ginibre ensemble \cite{ginibre1965}:
        \be
        F_{\rm unv}(s) = e^{-\pi s^2}.
        \label{unicor}
        \ee
It satisfies the sum rule
         	\be
 	\int d z_1 R_{2,c}(z_1,z_2) = - \rho(z_2)
 	\label{sum-rule}
 	\ee
        and that the eigenvalue correlations are short-ranged on the scale of $1/\sqrt D$. 
        The two-point function \eref{unicor} can also be derived rigorously for
        the elliptic Ginibre ensemble \cite{fyodorov1997}.

        We can now proceed in exactly the same way as in section \ref{sec:IIb} but with an
        explicit expression for the two-point correlation function.
        Let us expand the Boltzmann factor in powers of $y_1-y_2$, the imaginary part of $ z_1-z_2$:
         	\be
 	e^{-\beta(z_1+z_2^*)} = e^{-\beta ( \bar z+\bar z^*) } \left(1 -i\beta (y_1-y_2)
 	-\frac 12 \beta^2 (y_1 -y_2)^2 +\cdots \right),
 	\ee
        where $\bar z = (z_1+z_2)/2$.
      As mentioned in section \ref{sec:IIb}, the contribution due to the first term vanishes because of the sum rule \eref{sum-rule},
        and the second term does not contribute because the integral is even in
        $y_1-y_2$. We thus find
        \be 
        Z_{2,c} = \frac{\beta^2}2 \langle (y_1-y_2)^2\rangle \int d^2\bar z
       e^{-\beta(\bar z+\bar z^*)}\theta \left(1-\frac {\bar x^2}{E_0^2} -\frac {\bar y^2}{y_0^2}\right),
       \ee
       where
       \be
       \langle (y_1-y_2)^2\rangle =\int (y_1-y_2)^2
       \rho(\bar z)^2 e^{-\pi |z_1-z_2|^2{\rho(\bar z)}}d^2(z_1-z_2)=\frac 1{2\pi}.
       \ee
       Using the parameterization \eref{zpar}
          we obtain 
          \be
          Z_{2,c} = \frac{\beta y_0}{4} I_1(2\beta E_0).
          \label{z2c}
          \ee
          To derive this result we have interchanged the large $D$ limit and the
          integrations over the spectral density and spectral correlations. In Appendix
          \ref{a:ginibre} we show that this misses additional 
          corrections which change the prefactor in    \eref{z2c}. Since these corrections
          do not change the exponential $D$ dependence of the contribution, they do not affect the
          free energy in the thermodynamic limit.

 The total partition function is  given by
 \be
 Z= Z_{2,{\rm dis}}  + Z_{2,c}. 
 \ee
 Taking only the leading non-vanishing terms in the thermodynamic limit, we simplify the free energy to
 \be
 F(T) = - T \log \left ( e^{2\frac{E_0}{T}} + D^2 e^{\frac 2T \sqrt{E_0^2 -y_0^2}} \right).
\ee
In order to mimic forthcoming results for the two-site SYK model and more generically of interacting fermionic systems, we set $D = 2^{N/4}$
and $ \sigma_0 = e_0 N/2$, where $e_0$ is a size independent microscopic energy scale and we stress that $N$ is not the number of eigenvalues of the Ginibre Hamiltonian. With these choices, the free energy per particle can be written as
\be
\frac {F}N =-\theta(T_c-T)\frac {e_0}{\sqrt{1+k^2} \sigma(k)}
- \theta(T-T_c)\left(T \frac {\log 2}2  +\frac{e_0 \sqrt{1-k^4}}{\sqrt{1+k^2} \sigma(k)}\right )
\ee
where the critical temperature of the first-order phase transition is given by 
 \be
 T_c = \frac {2e_0}{\sigma(k) \sqrt{1+k^2}\log 2}\left(
1-\sqrt{1-k^4}\right )
 \ee
 which for small $k$ scales as $k^4$. These results are fully consistent with the universal expression (\ref{eq:fheuristic}). Indeed, the free energy for the elliptic Ginibre model are qualitatively similar as those of the ordinary (circular) Ginibre case:  in both cases, there is a first-order phase transition separating a low-temperature region where the free energy is dominated by RSB configurations. We now explore whether this first-order 
 transition is a feature of more realistic fermionic systems such as a non-Hermitian SYK model,  where there are $N$ Majoranas in zero spatial dimension with infinite-range interactions and random complex couplings. Therefore,  we do not expect that any artificial choice of scaling is necessary to observe the transition. 
For $k =1$ the elliptic Ginibre model reduces to
 the ordinary circular Ginibre model.
  Its partition function can be evaluated exactly at finite
 $D$, and up to a prefactor, the large $D$ limit of this result is in agreement with the results
 derived in this section. The details are worked out in Appendix \ref{a:ginibre}.
 
 \section{Free energy and RSB for the $PT$-symmetric SYK Model}\label{sec:sykchaos}

 We now turn to the study of the Hamiltonian (\ref{eq:hami}) with 
 $H_L$ and $H_R$ given by a $q = 4$ SYK model with complex couplings:
 \begin{align}
 \label{eq:hamiltonian_antonio}
 & H = \sum_{i < j < k < l}^{N / 2} \left( J_{ijkl} + i \, k M_{ijkl} \right) \psi_L^i \psi_L^j \psi_L^k \psi_L^l + \sum_{i < j < k < l}^{N / 2} \left( J_{ijkl} - i \, k M_{ijkl} \right) \psi_R^i \psi_R^j \psi_R^k \psi_R^l \ ,
 \end{align}
 where the variances of the couplings are 
 \begin{align}
 \label{eq:variances_antonio}
 & \langle (J_{i_1  \cdots i_q})^2 \rangle = 
 \langle (M_{i_1  \cdots i_q})^2 \rangle =
 \frac{2^{q-1} (q-1)!}{q (N/2)^{q-1}}v^2 \ ,
 \end{align}
 and $v$ sets the physical scale.  The strength of the complex deformation
resulting in a non-Hermitian Hamiltonian
is controlled by the parameter $k$. The Majorana fermions satisfy the Clifford algebra
\be
\{ \psi^i_L, \psi^j_L\} = \{ \psi^i_R, \psi^j_R\} =\delta^{ij}, \quad  \{ \psi^i_L, \psi^j_R\}=0.
\ee
We have also studied variations of this non-Hermitian SYK model. For example,
a model where the couplings are not complex conjugated. However, the partition
function of this model is not positive definite. A more interesting possibility
is to include an explicit coupling term between the two sites. This model has the
remarkable property that all eigenvalues become real beyond a critical
value of the coupling. Below, we will see that we will have to
add an infinitesimal explicit coupling term to break the symmetry between the
Left and Right replicas. The effect of a finite coupling will be studied in
detail in \cite{Garcia-Godet-2022}.

This section is divided into two parts. First, we provide theoretical 
arguments, supported by numerical results obtained by exact diagonalization of the Hamiltonian, which show that the free energy of this SYK model is quantitatively similar to that of the elliptic Ginibre model. In the second part,
 we confirm this conclusion by
 explicitly calculating the free energy from the solution of the Schwinger-Dyson (SD) equations. These  equations are the  saddle point equations  derived for one replica and one conjugate replica  and give  the large $N$ limit
 of the free energy \cite{maldacena2016}. We will see the free energy obtained from the SD equations  agrees with the Ginibre prediction. 
 
 For the numerical calculations, we diagonalize the {\it one-site} Hamiltonian with up to $N/2 = 34$ 
  Majoranas. In this case, we can directly calculate the
  quenched free energy which is equal to  half the free energy of the two-site model (see equation \eref{quench})
  and there is no need to use the replica trick. We have 
 found that 
 the ensemble fluctuations of $\log |Z_L|^2 $ are small and this quantity seems to
 be self-averaging for large $N$. The annealed average $\langle |Z_L |^2 \rangle$, which corresponds to one replica and one conjugate replica, shows much
 stronger fluctuations, and it is not clear if it is self-averaging.
 For comparison with theoretical predictions it is necessary
 to eliminate the fluctuations by averaging
 about many disorder realizations.
  We shall see that indeed, after averaging, annealed and quenched averages 
  lead to similar results by comparing the quenched free energy from exact 
  diagonalization with that obtained from the solution of the Schwinger-Dyson 
  equations that assumes annealed averages in its derivation.

   \begin{figure}[t!]
 	\centerline{ \includegraphics[width=8cm]{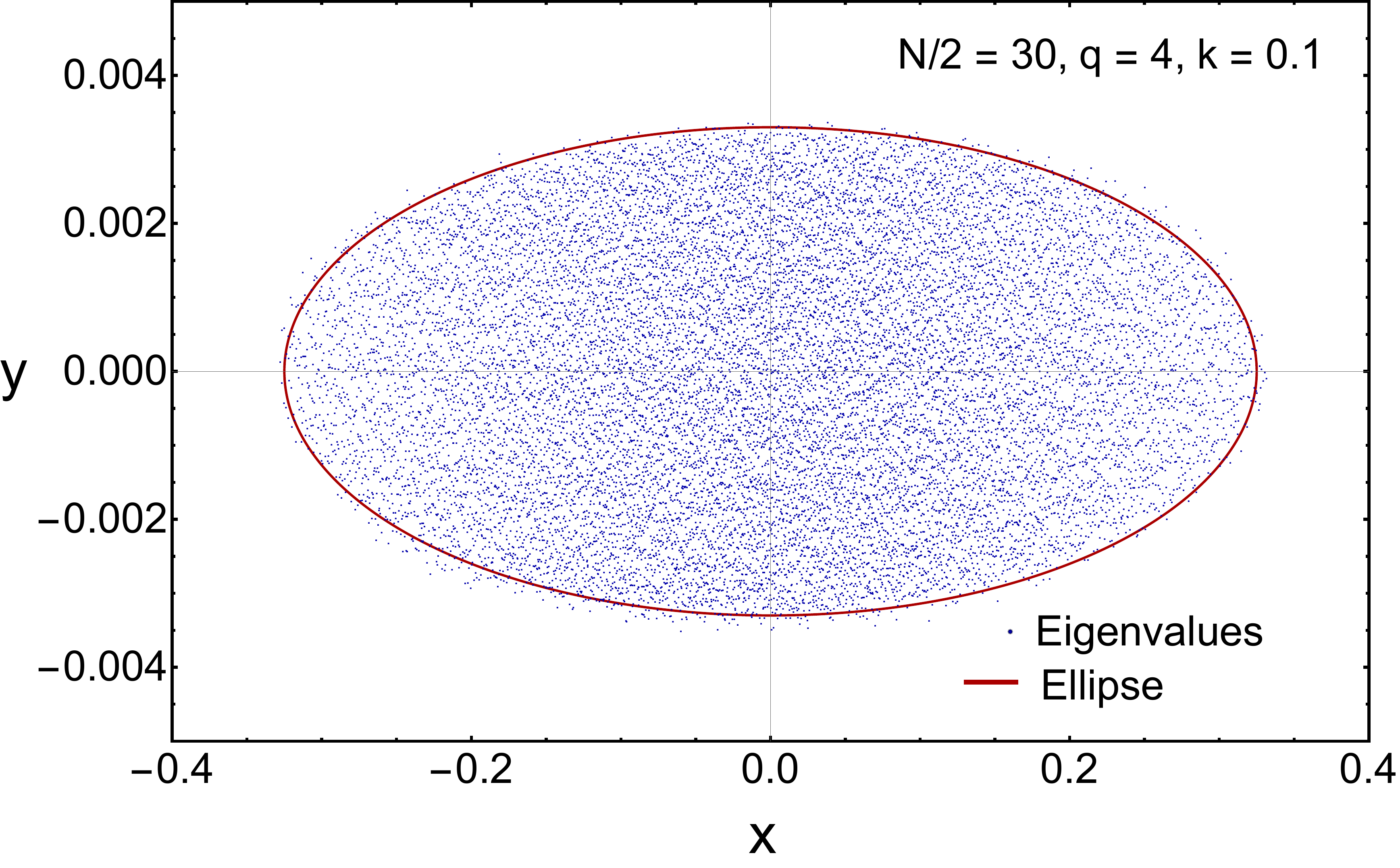}
 		\includegraphics[width=8cm]{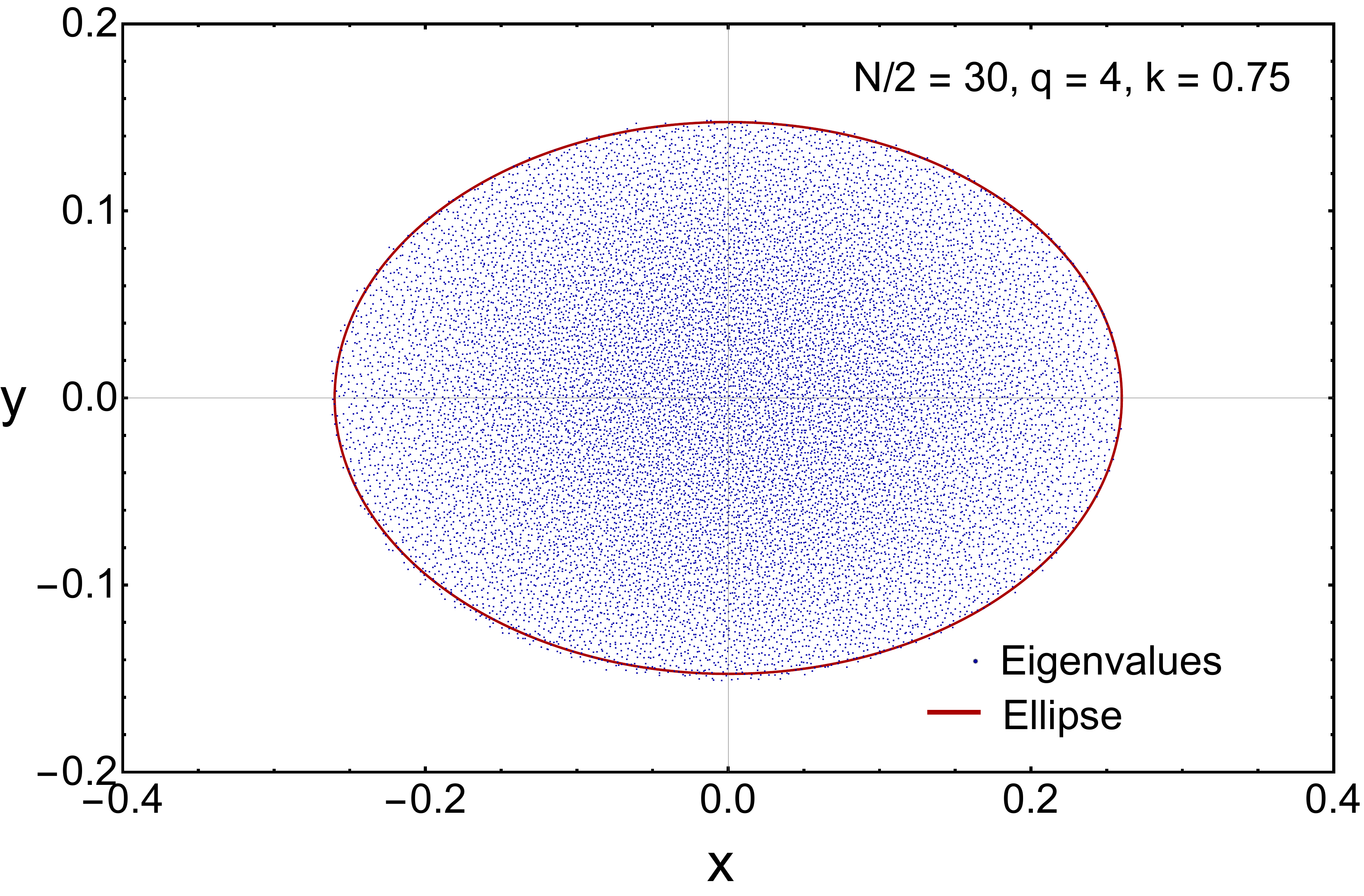}}
 	\caption{Eigenvalue distributions of a single realization of the non-Hermitian one-site
 		SYK model compared to the elliptical eigenvalue distributions
 		with non-Hermiticity parameters $k=0.1$ (left) and
 		$k=0.75$ (right).  Note the different scales of the $y$-axes in the two plots. 
 	\label{fig:evdisk}}
 \end{figure}

\begin{figure}[t!]
 	\centerline{ \includegraphics[width=8cm]{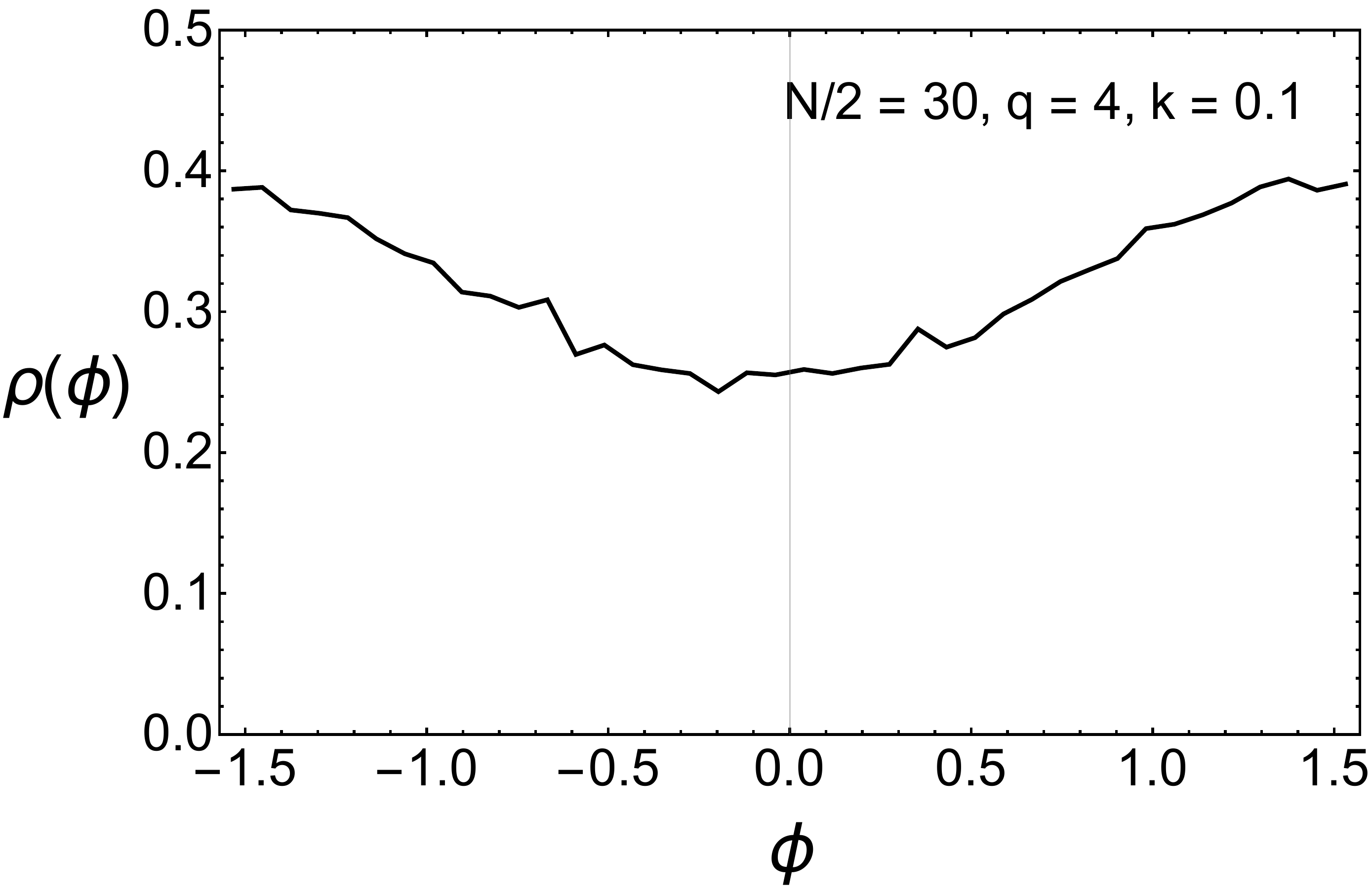}
 		\includegraphics[width=8cm]{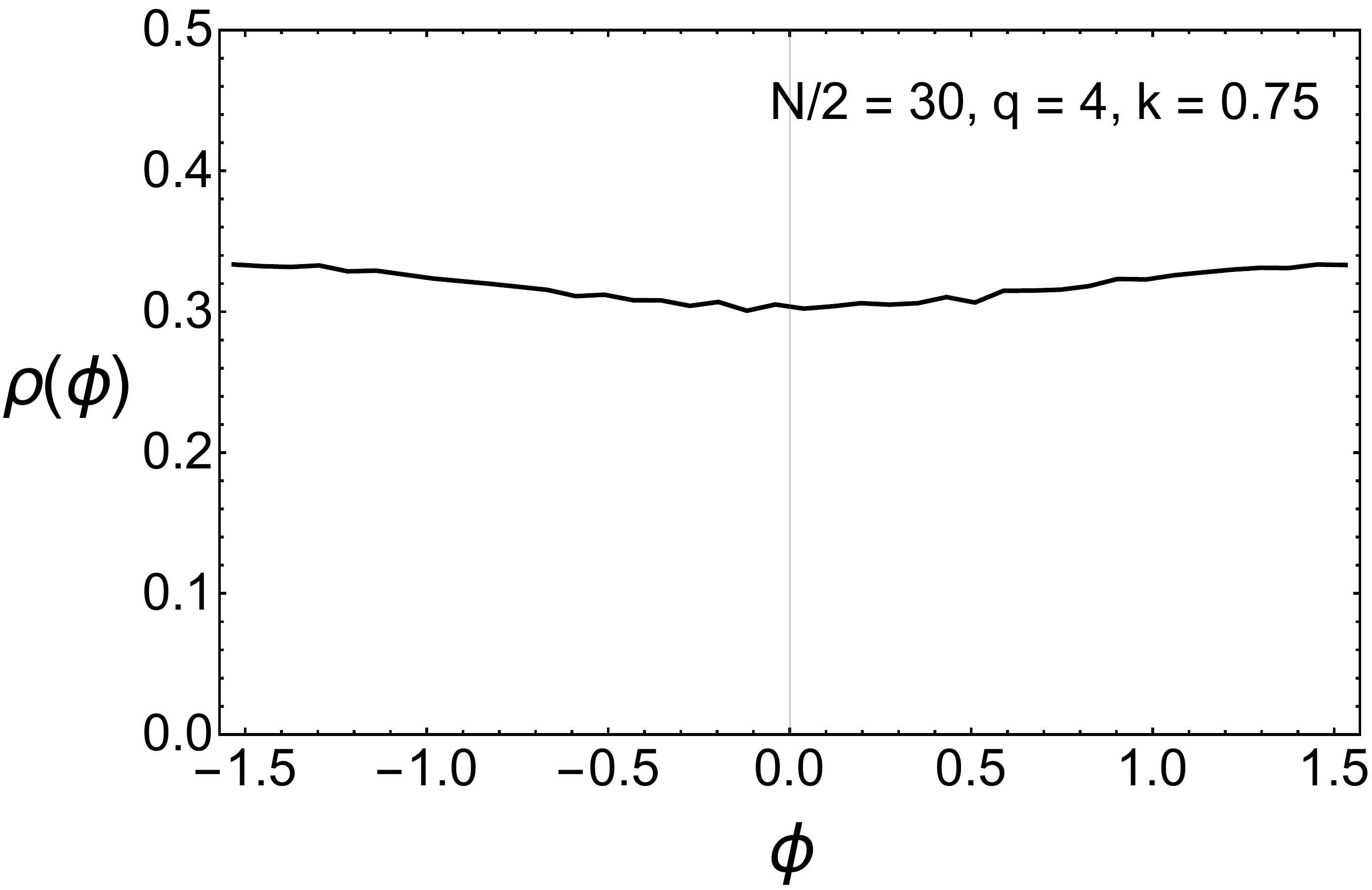}}
 	\caption{ Distribution of the phases of the eigenvalues after linearly
 		rescaling the spectrum from an elliptical shape to a circular
 		shape. This distribution is uniform for $k=1$ and becomes
 		less uniform for smaller values of $k$. This is illustrated for $k=0.1$ (left) and for
 		$k=0.75$ (right).
 	\label{fig:evphase}}
 \end{figure}

 We start with the analysis of the distribution of complex eigenvalues. 
 In Figure \ref{fig:evdisk}, we depict the distribution of the eigenvalues
 of a one-site SYK model with $N/2 = 30$ Majoranas and  compare it with the
 ellipse (red curve) given by
 \be
 \frac {x^2}{E_0^2(k)}+\frac {y^2}{y_0^2(k)} = 1
 \ee
 with $E_0(k)$ and $ y_0(k)$ fitting parameters. The quality of the fit
of  the support of the spectrum 
 is comparable to that of the elliptic Ginibre model, but contrary
 to the elliptic Ginibre model, the eigenvalue distribution is not
 completely uniform. In Figure \ref{fig:evphase}, we plot the distribution
 of
 the phase
 $\phi_n$  of the rescaled eigenvalues
 \be
 |\tilde E_n|  e^{i\phi_n} = \frac {\text{Re} (E_n)}{E_0(k)}+ i  \frac {\text{Im} (E_n)}{y_0(k)}
 \ee
 for $k=0.1$ (left) and $k=0.75$ (right). For $k = 1$, the distribution of the phase
 is uniform but becomes less uniform for smaller values of $k$. However, the deviation
 from uniformity is well fitted by a $\cos 2\phi $ dependence.

\begin{figure}[t!]
 	\centerline{\includegraphics[width=8cm]{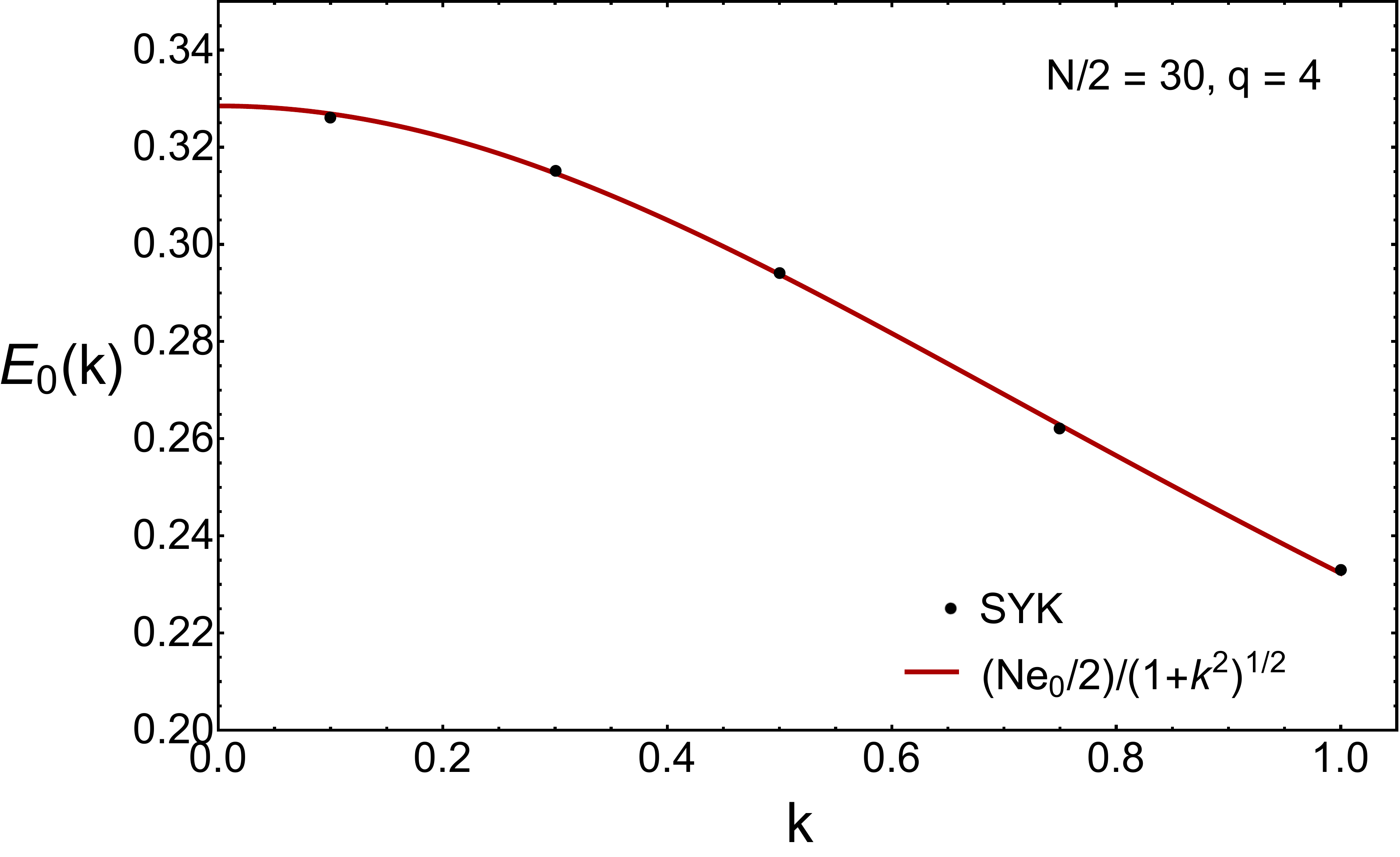}
 		\includegraphics[width=8cm]{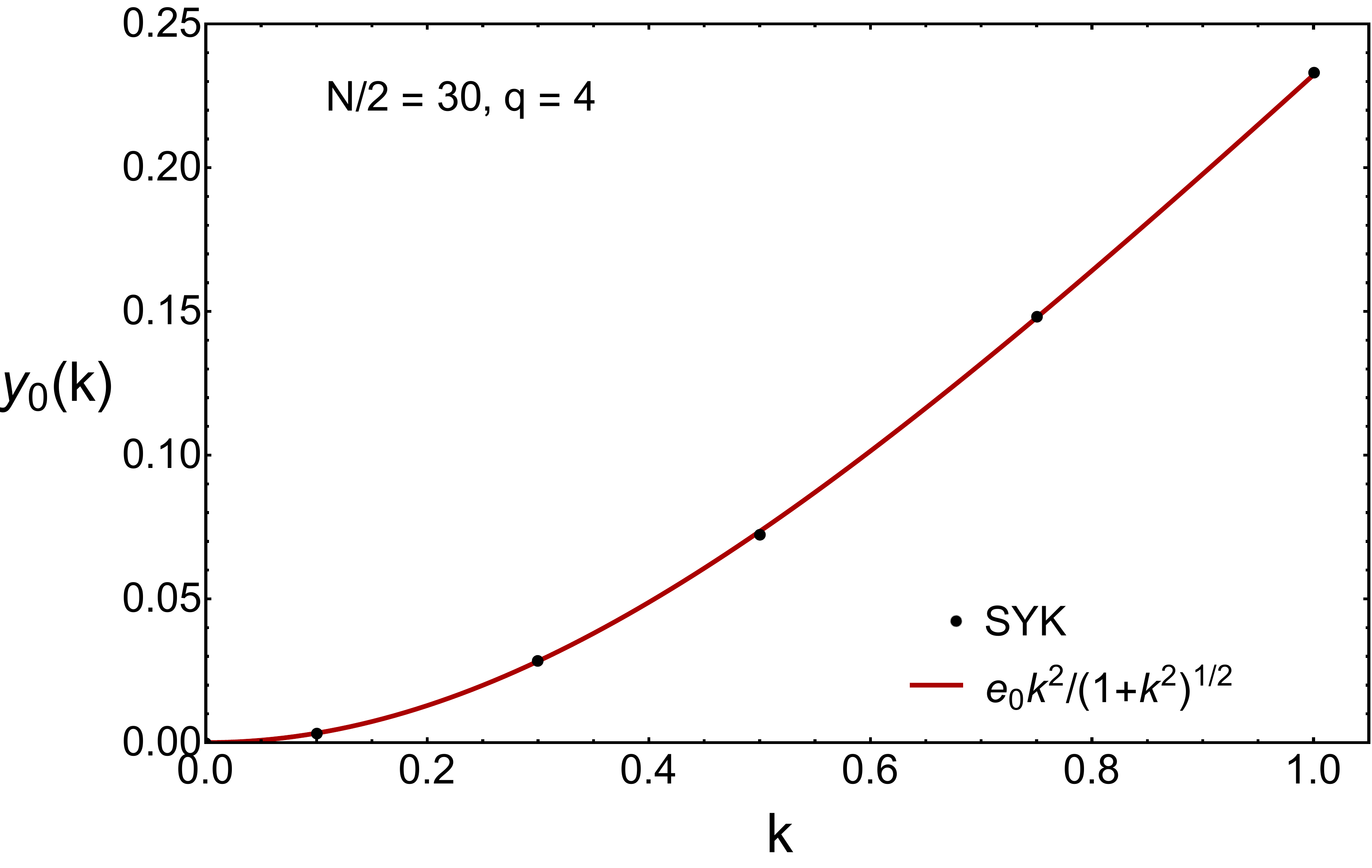}}
 	\caption{The $k$-dependence of the long axis, the ground state energy $E_0(k)$, and the short axis
 	  $y_0(k)$ that define the elliptical support of the eigenvalues of a one-site non-Hermitian SYK model with $N/2=30$. The
          $k$-dependence is compared to the prediction for
 		the elliptic Ginibre model.\label{fig:e0y0com}}
 \end{figure}

 Finally, in Figure \ref{fig:e0y0com}, we compare the fitted values of $E_0(k) $ and $y_0(k)$ to the 
 analytical functional dependence obtained for the elliptic Ginibre model,
namely,
 \be
{ E_0(k)} = \frac N 2 \frac{e_0 }{\sqrt{1+k^2}},\\
{ y_0(k)} =\frac N2 \frac{e_0 k^2 }{\sqrt{1+k^2}},
\label{e0k}
 \ee
 with $e_0$ the ground state energy per particle for $k=0$
 (for $N/2 =30$ we obtain  $e_0 \approx 0.011$, see equation \eqref{eqn:groundEnSYK}).
 The agreement is excellent which strongly suggests that indeed the two models have very similar spectral properties. We now turn to the study of the free energy.  
 
We recall two of the main features of the free energy of the elliptic Ginibre model.  First, because of the non-Hermiticity,
 the disconnected part of the partition function is exponentially suppressed for $k \ne 0$ which
 makes it possible for its magnitude to be comparable to that of the
connected part. Second, because the spectral correlations are
short-range, the details of these correlations are irrelevant.
As a consequence of a spectral sum rule, both the leading contribution due to the
self-correlations and those due to the genuine two-point correlations are of the same
magnitude but with an opposite sign and
cancel at leading order for large $N$.
 A phase transition induced by RSB can only occur if after this  leading-order  cancellation, the remaining two-point piece is comparable with the disconnected part. We can get an estimate of the critical temperature by assuming
that the eigenvalues are distributed uniformly
inside a ellipse with long axes $2E_0$ and short axis $2y_0$, in other words,
they are given by the distribution of the elliptic Ginibre ensemble. Using the
results of the previous section we find the critical temperature
 \be\label{tceli}
 T_c= \frac{E_0(k)-\sqrt {E_0^2(k) -y_0^2(k) }}{\log 2}\frac{4}{N}
 \ee
 with the $k$-dependence of $E_0(k)$ and $y_0(k)$ given by the results for the elliptic
 Ginibre model \eref{e0k}.
 This gives a critical temperature
 \be                              
 T_c = \frac {2 e_0}{\log 2} \frac{1- \sqrt{1-k^4}}{\sqrt{1+k^2}},
 \label{tc}
 \ee
 which behaves as $k^4 $ for small $k$.
The free energy in terms of $e_0$ is given by
 \be
  -\frac{F}{N/2} = \theta(T_c-T)\frac {2 e_0}{\sqrt{1+k^2}} +\theta(T-T_c)( T \log 2 + 2 e_0\sqrt{1-k^2}).
\ee
The energy $e_0$, the ground state energy per particle for the $k = 0$ SYK model, is given by \cite{Cotler:2016fpe,garcia2017}
 \be\label{eqn:groundEnSYK}
 (e_0 N/2)^2 = \frac {4M_2}{1-\eta}
 \ee
 where $M_2$ is our choice for the second moment of the one-site $k=0$ SYK model
\be
M_2 = {N/2 \choose 4} \vev{J_{ijkl}^2}=
      {N/2 \choose 4} \frac 1{6 N^3},
 \ee
 and
 \be \label{eqn:etaDef}
 \eta ={N/2\choose q}^{-1}\sum_{m=0}^{q}(-1)^m{q\choose m}{N/2-q\choose q-m}\sim 1- \frac{4q^2}{N}
,
 \ee
 where we have chosen $q = 4$ and $v =\frac{1}{24}$ in    (\ref{eq:variances_antonio}). This choice is the one employed in the numerical calculations.
 For $N/2=30$, we find $e_0 N/2 = 0.332$ while from Figure \ref{fig:e0y0com} we can
 read off a value of $0.328$ which is only slightly lower. The analytical result for the critical temperature
 using equation \eref{tceli} and $e_0 N/2 = 0.332$ is equal to $T_c =0.0225$ for $k=1$ which is also
 close to the result from exact diagonalization which is approximately $0.22$.
 As will be discussed later in
this section,
an independent calculation of $T_c$ and $E_0$ by exact diagonalization is in 
 agreement with these results.

 The analytical results  for the Ginibre model are largely  based on the
uniformity of   the distribution of  the eigenvalues.
 However, in the SYK case the phase  is only uniform for $k=1$  while the radial distribution
 is never uniform.
 For $k<1$, we have found that the $\phi$ dependence of the
 spectral density is  well fitted by (see Figure \ref{fig:evphase})
 \be
 \rho(\phi) =\frac 1{2\pi}( 1 - \alpha \cos2\phi).
 \ee
 The angular integral of the disconnected part of the partition function then becomes 
 \be
 2I_0(\sqrt{2}\beta r \sqrt{1-1/k^2})-2\alpha \frac{k^4+1}{k^4-1}
 I_2(\sqrt{2}\beta r \sqrt{1-1/k^2}).
 \ee
 Therefore, the leading exponent is not affected. The same argument can be made for
 the self-correlations and genuine two-point correlations.
 The deviation of the radial distribution from uniformity also does not change
 the leading exponent.
 This implies that even for $k < 1$ we expect the same results as for
 the elliptic Ginibre model, namely, in the large $N$ limit there is a $k$-dependent first-order phase transition.

 So far, we have restricted our analysis to the $k\leq 1$ region.
 It is easy to see that for $k > 1$, the partition function is equivalent to that resulting from the transformation $k \to 1/k$ and
 $\beta \to i \beta$. The calculation of the free energy can be carried
 out along the line of the $k < 1$ calculation.  Details are worked out in appendix \ref{a:pfg1}.
 
 Another interesting question is whether the eigenvalue distribution
can be related to that of the SYK with real couplings. We have found, see appendix \ref{a:denSYKq} for details, that indeed the real and imaginary parts of the eigenvalues of the non-Hermitian
SYK are still well described by the $Q-$Hermite prediction \cite{erdos2014,garcia2017} though the fitted value of $\eta$ is no longer given by the analytical
estimate \eref{eqn:etaDef}. Notably, for $k<0.3$, the distribution of the imaginary part
of the eigenvalues is very close to semi-circular.
 These facts are not directly related to the physics of the RSB but illustrate the rather deep connections between the  models we are considering.

 \subsection{Free energy, ground state energy and gap of the SYK model from the SD equations}
 \label{sec:equations}
 We now compare the predictions of the Ginibre model
 with a calculation of the free energy from the solution of the Schwinger-Dyson (SD) equations for the same two-site $q=4$ non-Hermitian SYK model in the
 $\Sigma G$ representation \cite{maldacena2016,bagrets2016,bagrets2017}. This formulation is based
 on the replica trick for the quenched partition function, $\vev{\log| Z_L|^2}$. However, we assume that the  mean field calculation does not break the replica symmetry so
 that the free energy can be obtained from the one-replica calculation, i.e.
 from the annealed partition function $\langle |Z_L|^2\rangle$, see the end of section \ref{sec:IIa} where this terminology is introduced. 
 However, we
 do have RSB between $Z_L$ and $Z^*_L$ which are only
 coupled by the disorder. We will see that, 
 when the temperature is sufficiently low, the dominant
 solutions of the saddle point equations couple a replica and a conjugate
 replica.
 We refer to section \ref{sec:qualitative} for a justification of
 both the correctness of the replica trick in this non-Hermitian case and
 the equivalence of quenched and annealed averages. This is important as
 by design the free energy from the SD equations involves an
 annealed average but we are interested only in quenched averages.

Following the standard procedure \cite{kitaev2015,maldacena2016}, we obtain the SYK action in Euclidean time as a simple variation of the action considered in \cite{maldacena2018}:
\be
	\label{eq:MQ_model_action}
        - \frac{2 S_E}{N} &=& \log \mathrm{Pf}(\partial_t \delta_{ab} - \Sigma_{ab}) - \frac 12  \int d\tau_1 d\tau_2 \sum_{a , b} \left[\Sigma_{ab}(\tau_1 , \tau_2) G_{ab}(\tau_1 , \tau_2) - s_{ab} \frac{\mathcal J_{ab}^2}{2 q^2} [2 G_{ab}(\tau_1 , \tau_2)]^q \right]
        \nn\\
        &&-\frac i2\epsilon \int d\tau(G_{LR}(\tau,\tau) -G_{RL}(\tau,\tau))
        \ ,
\ee
where the indices $a , \, b$ can be equal to $L$ or $R$. The function $s_{ab}$ takes the values $s_{LL} = s_{RR} = 1$ and $s_{LR} = s_{RL} = (- 1)^{q/2}$ and the couplings $\mathcal J_{ab}$ are taken to be $\mathcal J$ when $a = b$ and $\mathcal{\tilde J}$ when $a \neq b$. 
They are related to the variance and co-variance of  the random
$L$ and $R$ couplings
by
 \begin{align}
\label{eq:variances_mine}
& \langle (J^L_{ijkl})^2 \rangle = \langle (J^R_{ijkl})^2 \rangle =  \frac{2^{q-1} (q-1)!}{q (N/2)^{q-1}}(1-k^2)v^2 \equiv \frac{2^{q-1} (q-1)!}{q (N/2)^{q-1}}\mathcal J^2 \ , \nn \\
& \langle J^L_{ijkl}   J^R_{ijkl}\rangle =  \frac{2^{q-1} (q-1)!}{q (N/2)^{q-1}}(1+k^2)v^2 \equiv \frac{2^{q-1} (q-1)!}{q (N/2)^{q-1}}\tilde{\mathcal J}^2\ ,
\end{align}
  where the left and right couplings are related to the couplings of
 the Hamiltonian \eref{eq:hamiltonian_antonio}                                                   by
\be
\label{eq:coupling_mine_antonio_mapping_definition}
& J^L_{ijkl} \equiv J_{ijkl} +  i \, k M_{ijkl} \ , \nn \\
&J^R_{ijkl} \equiv J_{ijkl} -  i \, k M_{ijkl} \ .
\ee
This gives
\begin{equation}\label{eqn:JvRelation}
\mathcal J^2 \equiv(1-k^2)v^2 , \quad \tilde{\mathcal J}^2  \equiv(1+k^2)v^2.
\end{equation}
Finally, $G_{ab}$ denotes the fermion bi-linear  defined via the equations
 \begin{equation}
 \label{eq:propagators_definitions}
 G_{ab}(\tau_1,\tau_2) \equiv \frac 2N \sum_{i = 1}^{N / 2}  \psi^i_a (\tau_1) \psi^i_b (\tau_2)\ ,
 \end{equation}
 (we are assuming $0 < \tau < \beta$), while $\Sigma_{ab}$ are the Lagrange
 multipliers that implement this constraint. They can also be interpreted
 as the self-energies of the fermions, and the expectation value of
 $G_{ab}(\tau_1,\tau_2)$
 is the Green's function.  Note the $i\epsilon$ term in the action \eqref{eq:MQ_model_action} would have corresponded to a term 
 \be\label{eqn:iepsTerm}
  i \epsilon \sum_i \psi_L^i \psi_R^i
 \ee
in the Hamiltonian, which was not present in the original Hamiltonian \eqref{eq:hamiltonian_antonio}. 
However, as we will see in the next section  we will need this infinitesimal term added to detect the symmetry breaking whose order parameter is $G_{LR}$.

 \subsubsection{Symmetries of the Green's functions}
\label{symmetries}
 From the definition \eref{eq:propagators_definitions} we then obtain the
 symmetry relations
 \be
 G_{ab}(\tau_1,\tau_2) = - G_{ba}(\tau_2,\tau_1).
 \ee
 Assuming translational invariance as we will do in the remainder of
 this section, we have
 \be
 G_{ab}(\tau_1,\tau_2) \to G_{ab}(\tau_1-\tau_2),\qquad
 \Sigma_{ab}(\tau_1,\tau_2) \to \Sigma_{ab}(\tau_1-\tau_2).
 \ee
This results in
 \be
 G_{LR}(\tau) = -G_{RL}(-\tau),\qquad G_{LL}(\tau) = -G_{LL}(-\tau),
 \qquad  G_{RR}(\tau) = -G_{RR}(-\tau).
 \label{first-sym}
\ee

 For $\epsilon = 0$, the action is invariant under
 \be\label{eqn:chiralityL}
 \psi_L(\tau) \to  -\psi_L(\tau), \qquad  \psi_R(\tau) \to  \psi_R(\tau).
\ee
This symmetry can be implemented by the operator $\prod_{i=1}^{N/2} \psi_L^i$, when $N/2$ is even. Therefore we have that
\be
 G_{LR}(\tau) = \langle  \psi_L(\tau)  \psi_R(0)\rangle_\beta =0,
\ee
where $\vev{\cdot}_\beta$ denotes a thermal expectation value.  We take $\tau>0$ to avoid using the time-ordering symbol.
This symmetry is broken by a nonzero value of $\epsilon$. In the large
$N$ limit we shall see it is broken spontaneously for $\epsilon \to 0$, at sufficiently low temperature. Below we always
consider the limit
\be
\lim_{\epsilon \to 0} \lim_{N\to \infty} G_{LR}(\tau).
\ee
Also in terms of eigenvectors and eigenfunctions of the SYK model,
$G_{LR}(\tau)$ vanishes identically without the presence of $\epsilon$ term as one can easily check numerically for small values of $N$.

  Next we consider an anti-unitary $PT$ operation
 \be
 PT_1: \ \psi_L^i(\tau) \to  \psi_R^i(\tau), \qquad  \psi_R^i(\tau) \to  -\psi_L^i(\tau),\qquad
{i \to -i}.
\label{S4}
 \ee
This operation can be implemented by 
\begin{equation}
P = \exp\left(-\frac{\pi}{2}\sum_j \psi_L^j \psi_R^j\right)= \prod_j\frac{1}{\sqrt{2}}\left(1- 2 \psi_L^j \psi_R^j\right), \quad  T_1 = T_L \otimes T_R,
\end{equation}
where $T_L, T_R$ are just the conventional time reversals of a single-site SYK that leaves the fermions invariant and takes $i$ to $-i$. 
We note that $PT_1$ is a symmetry of the two-site Hamiltonian without the $i\epsilon$ term \eqref{eqn:iepsTerm},  but gets explicitly broken by the $i\epsilon$ term.  Fortunately,  if we compose $PT_1$ with the symmetry of equation \eqref{eqn:chiralityL},  we get another anti-unitary operation that is a symmetry even in the presence of the $i\epsilon$ term:
\begin{equation}
PT_2:   \ \psi_L^i(\tau) \to  \psi_R^i(\tau), \qquad  \psi_R^i(\tau) \to \psi_L^i(\tau),\qquad
{i \to -i},
\end{equation}
implemented by the same $P$ and 
\begin{equation}
T_2 = T_1  \prod_{i=1}^{N/2} \psi_L^i.
\end{equation}
This $PT_2$ symmetry ensures the reality of the partition function even in the presence of the $i\epsilon$ term.  The  $PT_2$ symmetry results in the identities
\be
G_{LR}(\tau) &=& \langle  \psi_L(\tau)  \psi_R(0)\rangle _\beta
= \langle  \psi_R(\tau)  \psi_L(0)\rangle^*_\beta =G_{RL}^*(\tau),\label{eqn:LRsymmFromPT2} \\
G_{LL}(\tau) &=& \langle  \psi_L(\tau)  \psi_L(0)\rangle_\beta
= \langle  \psi_R(\tau)  \psi_R(0)\rangle^*_\beta =G_{RR}^*(\tau),\label{eqn:LLsymmFromPT2}
 \ee
 where we have used the antiunitarity of $PT_2$.    Note that although the $P$ operator alone is not a symmetry,  it satisfies
 \begin{equation}
 P H P^{-1}  =  H^{\dagger} 
 \end{equation}
in the presence of the $i \epsilon$ term.    This means 
\begin{equation}
\psi_L(\tau)^\dagger = -P \psi_R(-\tau) P^{-1}, \quad \psi_R(\tau)^\dagger = P \psi_L(-\tau) P^{-1}.
\end{equation}
Hence, we have\footnote{We omit the normalization factor $\Tr(e^{-\beta H})$ in the denominator for this derivation since it is real and does not affect the reality property of Green's functions. Also note that $\tau$ is the Euclidean time.}
\be 
G^*_{LR}(\tau)&=&\left(\Tr e^{-\beta H}\psi_L(\tau)\psi_R(0)\right)^*=\Tr  \left(\psi_R(0)^\dagger \psi_L(\tau)^\dagger e^{-\beta H^\dagger}\right) \nn \\
&=&-\Tr  \left(\psi_L(0) \psi_R(-\tau) e^{-\beta H} \right)=-\Tr  \left( e^{-\beta H}\psi_L(\tau) \psi_R(0) \right) =-G_{LR}(\tau).
\ee
This is to say $G_{LR}(\tau)$ is purely imaginary.  Together with  the symmetries \eref{first-sym} and \eqref{eqn:LRsymmFromPT2} this gives
 \be
 G_{LR}(\tau) = G_{LR}(-\tau).
 \ee
  Since $G_{LR}(\tau +\beta) = - G_{LR}(\tau)$ we also find
\be
G_{LR}(\beta -\tau) = -G_{LR}(\tau).
  \ee
   Using the symmetries \eref{first-sym}  and the anti-periodicity of $G_{LL}(\tau) $ we obtain 
\be
  G_{LL} (\beta -\tau) = G_{LL}(\tau).
  \ee
  That is,  $G_{LR}(\tau)$ is odd about $\beta/2$ whereas $G_{LL}(\tau)$ is even about $\beta/2$.

  So far all the symmetry relations we worked out are true for each independent realization of the random couplings.   For
  the Hermitian Maldacena-Qi SYK model \cite{maldacena2018}, there is one more relation that holds:
  \be\label{eqn:LLrelationMQ}
  G_{RR}(\tau) = G_{LL}(\tau), \quad   G_{LL}(\tau) =   G_{LL}^*(\tau).
  \ee
 In our non-Hermitian model, there is not enough symmetry for the above to hold realization by realization.  Indeed as we can numerically verify,  for a generic realization $G_{LL}$ and $G_{RR}$ are complex and only $G_{LL} =G^*_{RR}$ (equation \eqref{eqn:LLsymmFromPT2}) holds.  However,  if we perform the ensemble averaging we would expect equation \eqref{eqn:LLrelationMQ} to hold for the non-Hermitian model,  because
 \begin{equation}
 H(J,M,\epsilon)^\dagger = H(J, -M, \epsilon)
 \end{equation}
 and the distribution of the disorder $M$ is an even function.  Let us summarize all the symmetry relations of the Green's functions in one place:
 \be\label{eqn:symmSummary}
&G_{ab}(\tau)= - G_{ba}(-\tau), \quad G_{LR}(\tau)= G_{LR}(-\tau)=-G^*_{LR}(\tau), \nn \\
& G_{LL} (\beta -\tau) = G_{LL}(\tau), \ G_{LR}(\beta -\tau) = -G_{LR}(\tau), \nn\\
&G_{LR}(\tau)= G^*_{RL}(\tau), \quad  G_{LL}(\tau)= G^*_{RR}(\tau),\nn \\
 &\langle  G_{RR}(\tau)\rangle =\langle  G_{LL}(\tau) \rangle,
 \quad   \langle G_{LL}(\tau)\rangle =  \langle G_{LL}^*(\tau)\rangle.
 \ee
  We stress all the above equations except the last line hold for each realization of the ensemble. It is also useful to note that the $LL$ Green's function satisfies
  \be
   G_{LL}(0) = \frac{\left \langle \frac{2}{N} \Tr \sum_k (\psi^k_L )^2 e^{-\beta H} \right \rangle}
{\left\langle \Tr e^{-\beta H}\right \rangle}
  = \frac 12.
  \ee

  The symmetries of $G_{ab}$ are inherited by $\Sigma_{ab}$.  This also  follows from the
 Schwinger-Dyson equations which will be discussed in the next subsection.

 \subsubsection{The Schwinger-Dyson Equations} 
Starting from the action \eqref{eq:MQ_model_action}, the stationarity of
 $\Sigma_{ab}$ gives the following set of saddle point equations for
 the Fourier components of $\Sigma_{ab}$ and $G_{ab}$,\footnote{The omitted correlators $G_{RL}$ and $G_{RR}$ are easy to obtain from $G_{LL}$ and $G_{LR}$,  thanks to the symmetry properties  \eqref{eqn:symmSummary}.}
 \be
 \bmat
 i\omega_n+\Sigma_{LL}(\omega_n) &\Sigma_{LR}(\omega_n)\\\Sigma_{RL}(\omega_n) &i\omega_n+\Sigma_{RR}(\omega_n)
 \emat
 \bmat
 G_{LL}(-\omega_n) & G_{RL}(-\omega_n) \\ G_{LR}(-\omega_n) & G_{RR}(-\omega_n) 
 \emat
 =\bmat 1 & 0 \\ 0 & 1 \emat.
\ee
Using that $G_{RR}(\omega_n) = G_{LL}(\omega_n)=-G_{LL}(-\omega_n)$ and
$G_{LR}(\omega_n) = G_{LR}(-\omega_n)=-G_{RL}(\omega)$, the saddle point equations can
be simplified to
\be
 \label{eq:SD_equations_general}
 &&  - (i\omega_n  + \Sigma_{LL}(\omega_n)) G_{LL}(\omega_n)
+  \Sigma_{LR}(\omega_n) G_{LR}(\omega_n) =  1 \ , \nn \\
 && (i \omega_n +  \Sigma_{LL}(\omega_n)) G_{LR}(\omega_n)
   + \Sigma_{LR}(\omega_n) G_{LL}(\omega_n) = 0 \ .
   \ee
 In \eqref{eq:SD_equations_general}, we have introduced the fermionic Matsubara frequencies
 \begin{equation}                                                          
 \label{eq:matsubara_freq_def}     
 \omega_{n} \equiv \frac{2 \pi}{\beta} \left(n + \frac 12\right) \ ,
 \end{equation}
 with $\beta$ being the inverse temperature. At the stationary points of
 the $G$ integral,
 the SD equations for 
 self-energies $\Sigma_{LL}$ and $\Sigma_{LR}$ in the time domain take the form of
 \be
 \label{eq:self_energies_general}
 \Sigma_{LL}(\tau) &=& \frac{\mathcal J^2}{q} (2 G_{LL}(\tau))^{q - 1} \ ,
 \qquad \Sigma_{LR}(\tau) = (-1)^{q/2}\frac{\tilde{\mathcal J}^2}{q}
 (2 G_{LR}(\tau))^{q - 1}-i\epsilon \delta(\tau)\ , \nn \\
 \Sigma_{RR}(\tau) &=&\frac{\mathcal J^2}{q} (2 G_{RR}(\tau))^{q - 1} \ ,
 \qquad \Sigma_{LR}(\tau) = (-1)^{q/2}\frac{\tilde{\mathcal J}^2}{q} (2 G_{LR}(\tau))^{q - 1} +i\epsilon \delta(\tau)\ .
 \ee
They can be rewritten as integral equations for the Fourier components
of $G_{ab}$ and $\Sigma_{ab}$, so that
 the SD equations constitute a set of coupled integral equations.

 Using the symmetry properties of $G_{ab}$ in equation \eqref{eqn:symmSummary}, we obtain the following relations
 (for even $q$):
 \be
 \Sigma_{LL}(\tau) &=& \Sigma_{RR}(\tau), \qquad  \Sigma_{RL}(\tau) = -\Sigma_{LR}(\tau),\nn \\
 \Sigma_{LL}(-\tau) &=& -\Sigma_{LL}(\tau), \qquad  \Sigma_{LR}(-\tau)
 = \Sigma_{LR}(\tau).
 \ee
 This again leads to the evenness of $\Sigma_{LL}(\tau)$  
 the oddness of $\Sigma_{LR}(\tau)$  about $\beta/2$.
   
Using the symmetry properties,
 the SD equations \eqref{eq:SD_equations_general} can be conveniently rewritten in the following form:
 \begin{align}
 \label{eq:SD_equations_final_form}
 & G_{LL}(\omega_n) = -\frac{i \omega_n + \Sigma_{LL}(\omega_n)}{(i \omega_n + \Sigma_{LL}(\omega_n))^2 + \Sigma_{LR}^2(\omega_n)} \ , \nn \\
 & G_{LR}(\omega_n) =  \frac{\Sigma_{LR}(\omega_n)}{(i \omega_n + \Sigma_{LL}(\omega_n))^2 + \Sigma_{LR}^2(\omega_n)} \ .
 \end{align}
\begin{figure} [t!]
	\centering
\includegraphics[width=8cm]{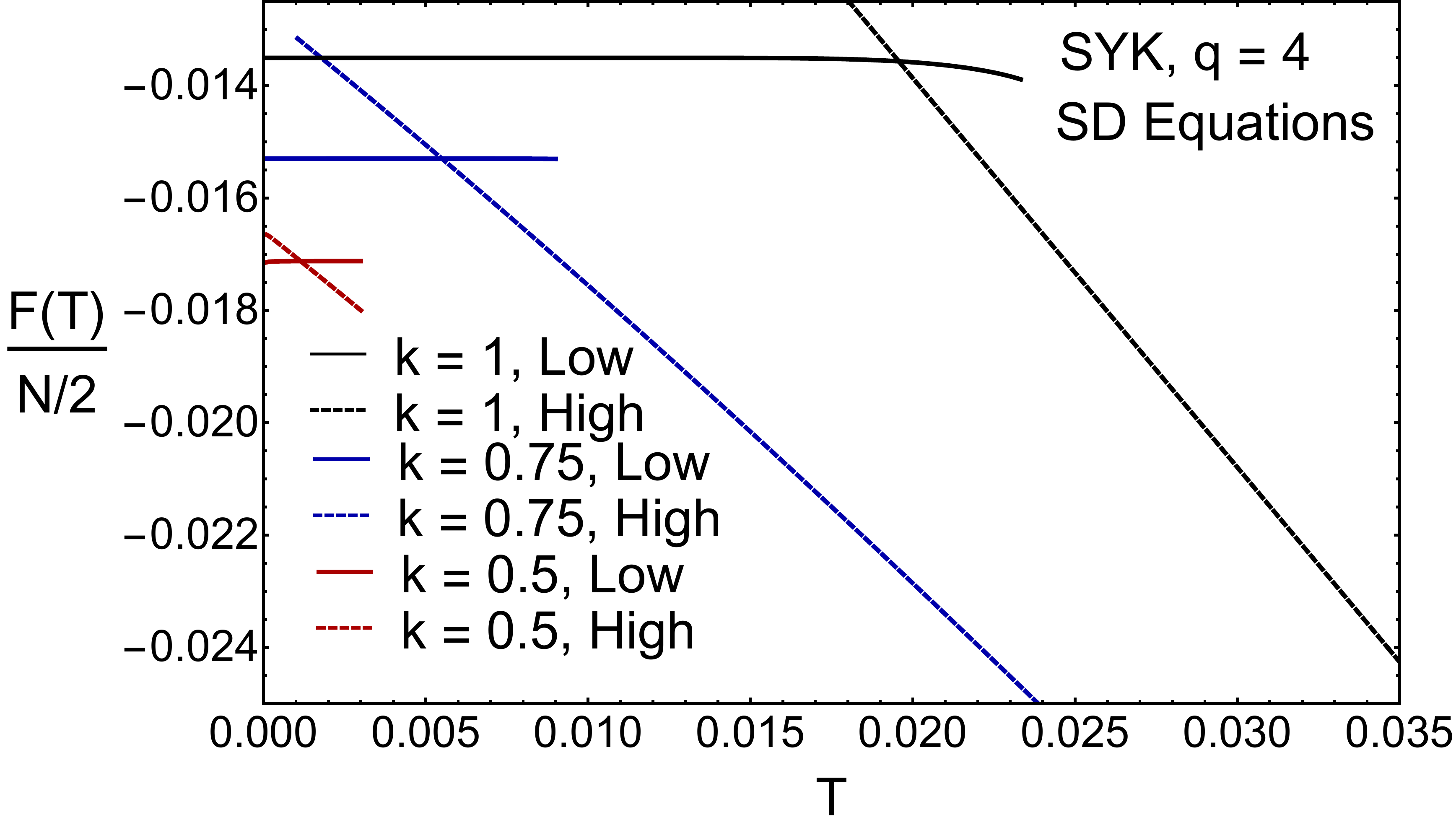}
\includegraphics[width=8cm]{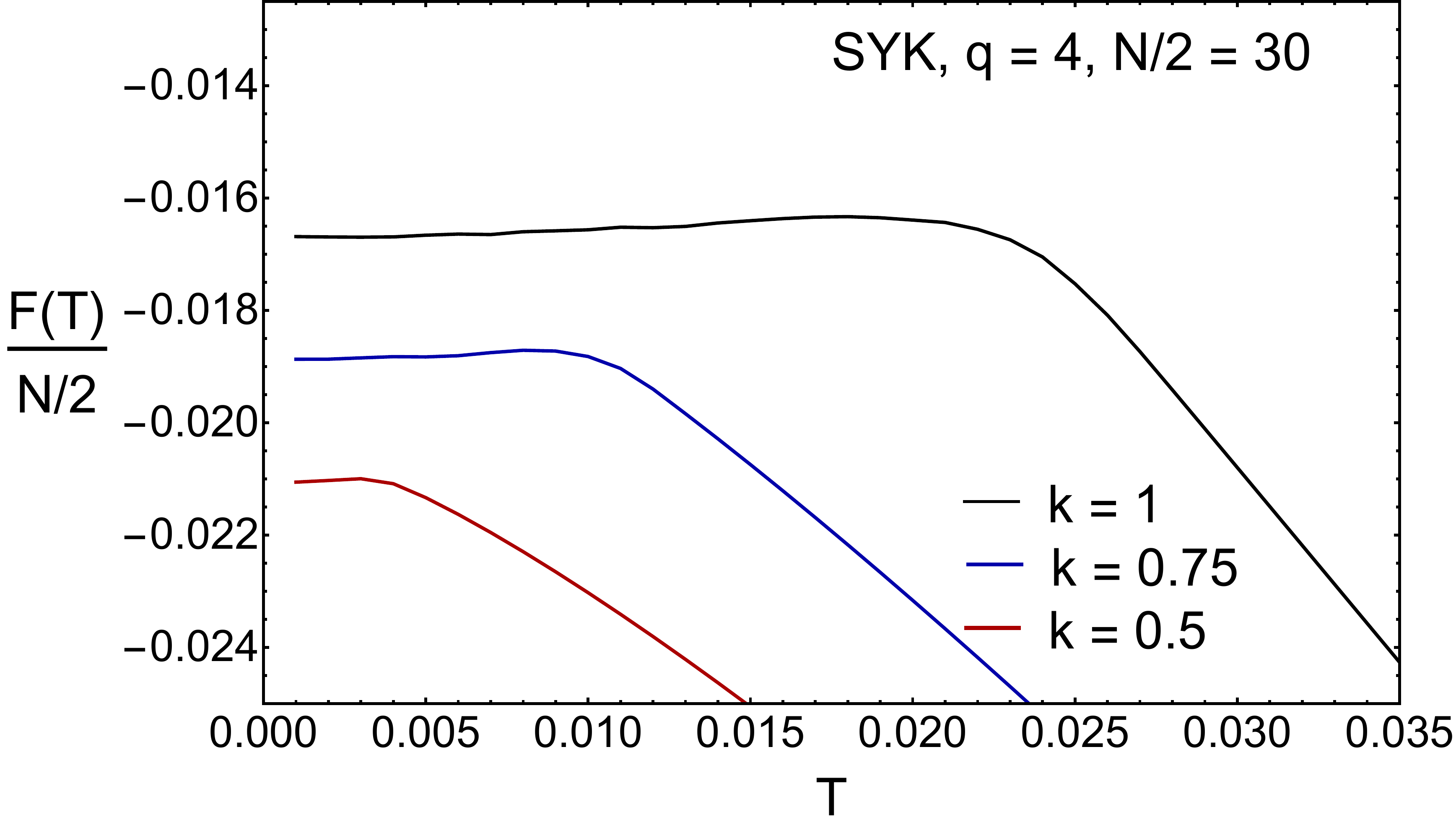}
\caption{Left: annealed free energy for different values of $k$ from the solutions of the SD equations.
Right: quenched free energy per particle from exact diagonalization
of the two-site SYK model for different values of $k$ and $N/2=30$.
\label{fig:freek}
}
\end{figure}
 Except for the special case $q = 2$, for which the self-energies are linear in $G_{ab}(\tau)$ and can be easily transformed to the frequency space, the exact solutions of the SD equations above have to be obtained numerically. Details of the numerical procedure are given in Appendix \ref{a:nSD}.
 
 Depending on the temperature, the saddle point equations may have more than
 one solution. The physical solution is the one with the lowest free energy
 subject to the condition that the steepest descent manifold (Lefschetz thimble)
 it lies on can be continuously deformed into the original integration manifold. This can be worked out explicitly for $q=2$
 where it turns out that the saddle point with the lowest free energy is
 not always the one that determines the physical free energy \cite{Jia:2022reh}.
 The phase transition occurs when the solution with the lower free
 energy switches to a different solution at the critical temperature. 
 The free energy depicted in the left panel of
 Figure~\ref{fig:freek} clearly illustrates this hysteresis mechanism.

In our case, we shall see that at low temperatures the system is dominated by the solution with a non-zero $G_{LR}$ while at higher temperatures the replica symmetric solution with a vanishing  $G_{LR}$ dominates.
 The two solutions intersect at a point where the system undergoes a first-order phase transition.
 Interestingly, we see that in the RSB phase the free energy is almost constant.
This is suggestive of the existence of a finite gap 
between the ground state and the excited spectrum of the effective theory
similar to the wormhole phase in the Maldacena-Qi model
\cite{maldacena2018,Garcia-Garcia:2019poj}. A comment is in order: from the gravity perspective,
it may seem strange that the high-temperature phase  depends on the strength $k$ of
the imaginary part of the coupling. However,  note that this $k$-dependence can be eliminated by an overall rescaling of the Hamiltonian by a function of $k$ as we did for the Ginibre case.
For the sake of simplicity, we stick with the Hamiltonian (\ref{eq:hamiltonian_antonio}).
\begin{figure}[t!] 
	\centering
\includegraphics[width=10cm]{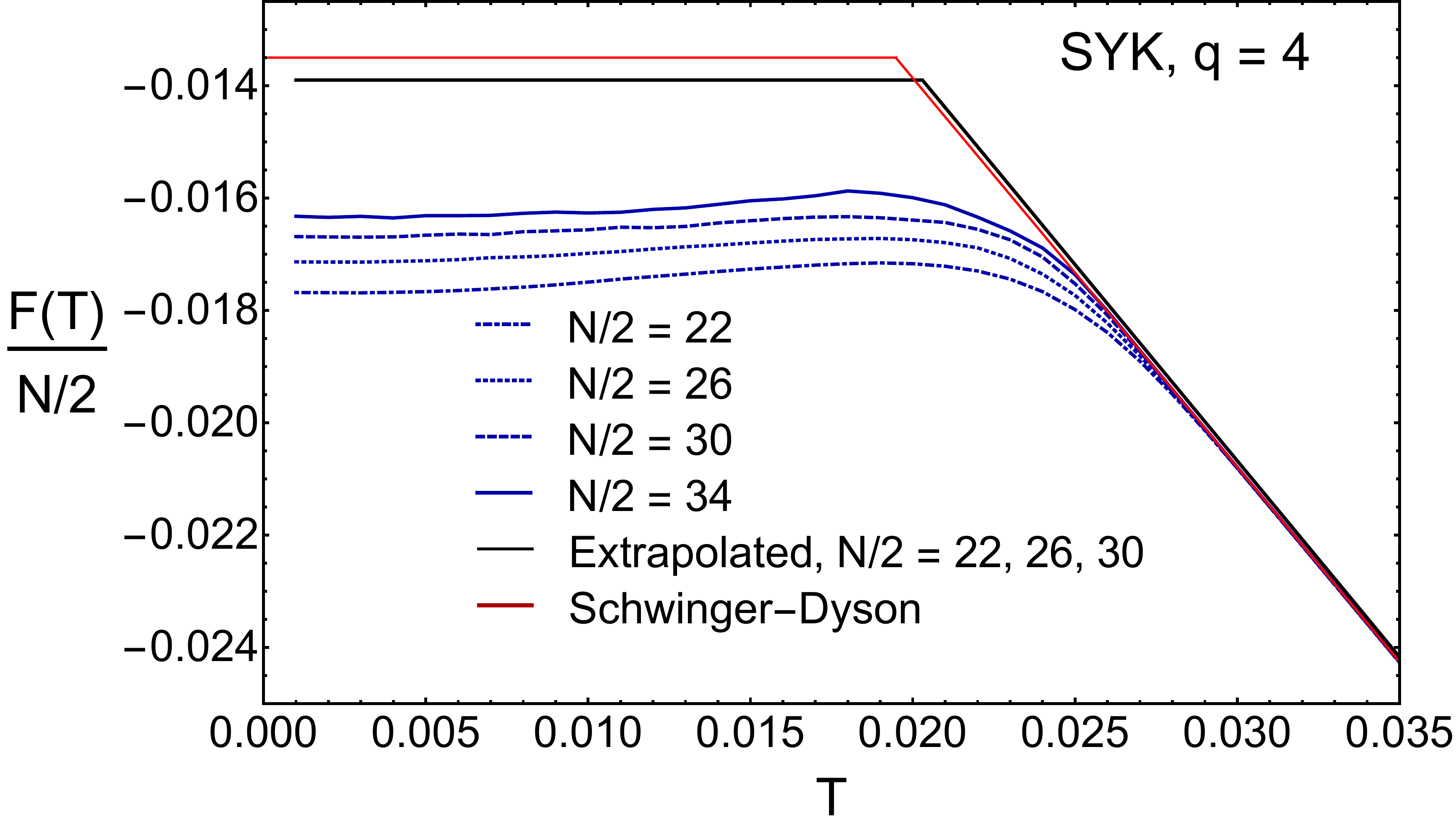}   
\caption{ Quenched free energy per particle obtained from exact diagonalization
  of the two-site SYK model for $k=1$ and different values of $N$.
  We also include the extrapolated, $N/2 \to \infty$, free energy resulting
  from a finite size  scaling analysis of the numerical results (black curve) together with the large $N$ prediction from the solution of the SD saddle point equations (red curve). 
 Agreement between the two results is excellent. See main text and Figure~\ref{fig:exp}
 for details and limitations
 of the finite size scaling analysis.
	\label{fig:free} }
\end{figure}

\begin{figure}[b!]
  \centering
       \includegraphics[width=5cm]{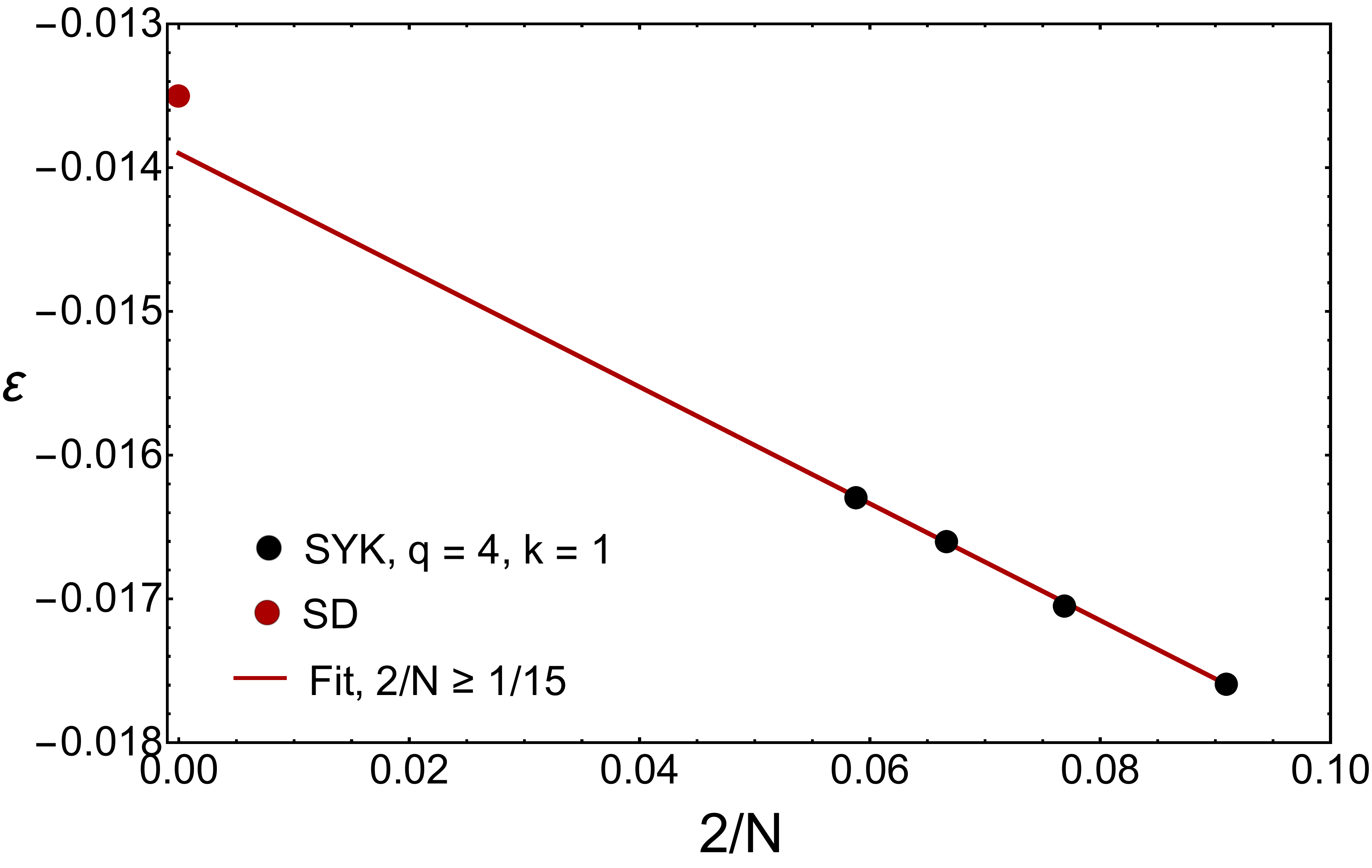}
       \includegraphics[width=5cm]{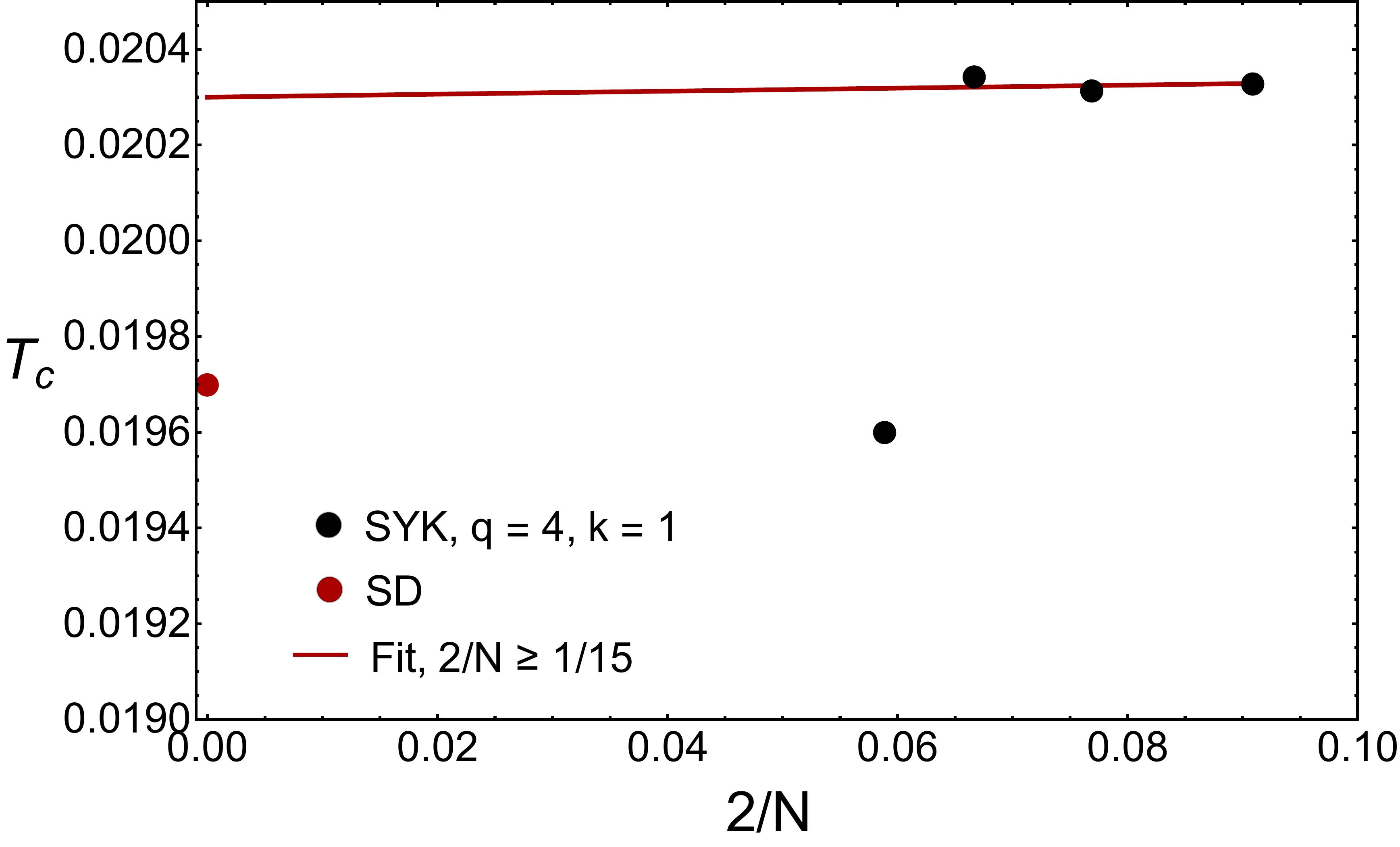}
       \includegraphics[width=5cm]{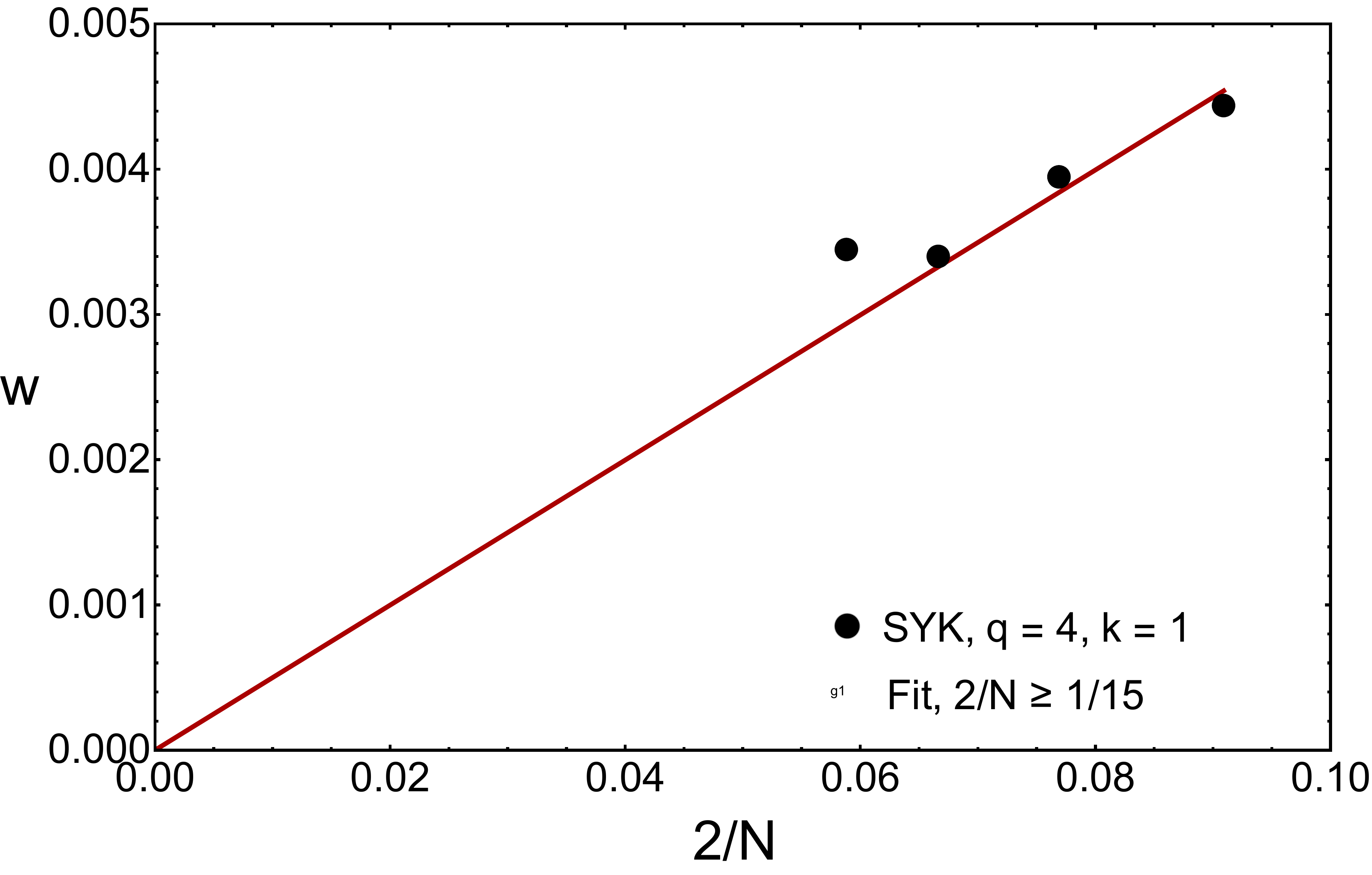}
       \caption{The fitting parameters $\varepsilon =\varepsilon(k)$ (left), $T_c$ (middle) and $w$ (right) as a function
       of $2/N$ for $N=22$, 26, 30 and 34.
  The linear dependence works well for all $N$
       for $\varepsilon $ but in case of $T_c$ and
         $w$, the $N=34$ values are aberrant. This is due to the limited size of the ensemble
         (40 realizations) in this case. The result obtained from the Schwinger-Dyson (SD) equations is
       indicated by the red dot at $2/N \to 0$.
       \label{fig:exp}
       }
     \end{figure}

 Finally, we study the differences between the annealed free energy, 
 obtained from the solution of the saddle point Schwinger-Dyson (SD)
 equations (see Figure \ref{fig:freek}, left), 
 and the quenched free energy  (see Figure~\ref{fig:freek} (right) and Figure\ref{fig:free}), which is accessible by an exact diagonalization of 
 the Hamiltonian.
 For the latter, due to technical limitations, 
 we have considered $N/2 \leq 34$.
 The number of disorder realizations is such that for any given $N$ and $k$, at least $10^6$ eigenvalues are obtained. It is well known \cite{garcia2016,Cotler:2016fpe} that the SYK model requires a relatively large number of fermions,  $N \geq 30$ in most cases,
 to approach the thermodynamic limit. For that reason, we have also carried
 out a finite size  scaling analysis with a fitting function (for $k=1$)
 \be
 F(\varepsilon, T_c, w) =\frac{\int_0^\infty ds e^{-(s-T)^2/2 w^2} \left[ \theta(T_c-s) \varepsilon - \theta(s-T_c) s \log 2\right ]}{\int_0^\infty ds e^{-(s-T)^2/2 w^2}},
\label{feps}
 \ee
which provides an excellent global fit for  $N \le 30$. The accuracy of the free energy data for
     $N=34$ reduces the quality of the fit, and we did not include it in the $N \to \infty$
     extrapolation. However, the $N=34$ data are within fluctuations of the
     extrapolation from $N= 22, \; N=26$ and $N=30$. In Figure \ref{fig:exp} we show the
     dependence of $\varepsilon$ (left), $T_c$ (middle) and $w$ (right) on $1/N$.

     The large
     deviation from linearity for $N=34$ are due to fluctuations close to the critical temperature, which
     can only be suppressed by increasing the size of the ensemble well beyond our computational
     resources.
     The extrapolated free energy agrees well with the result from solving
     the Schwinger-Dyson equations (compare the red and the black curves in Figure~\ref{fig:free}).
     Also, the extrapolated values of the physical parameters
     $\varepsilon(k=1) = -0.0139$ and $T_c = 0.0203$  are in agreement with
     results from the Schwinger-Dyson equation where  $\varepsilon(k=1) = -0.0135$ and  $T_c = 0.0197$.
     In the thermodynamic limit at fixed $q=4$ the values of these parameters are equal to
     $ \varepsilon(k=1) =-0.0147$ and $T_c = 0.0225$. The parameter
     $\varepsilon(k)$ is well determined
     by the global fit and can also be obtained from extrapolating at a single temperature well
     below $T_c$ where we can also include the $N=34$ data.
The critical temperature then follows from the intersection point with
the high-temperature curve which gives 
\begin{equation}
T_c=\frac{\gamma(k)-\varepsilon(k)}{S_0},
\end{equation}
with $-S_0$ the slope of the high-temperature curve, and $\gamma(k)$ and $\epsilon(k)$
the intercepts of the high-temperature curve and the low-temperature curve with the $T=0$
vertical axis, respectively.   Within the accuracy of our
calculations, this finite size scaling analysis gives the same result
as obtained from using the finite size scaling form \eref{feps}

 \subsection{The critical temperature and the ground state energy}
 \label{subsec:critical_temperature}

\begin{figure}[t!]
   \centerline{\includegraphics[width=8cm]{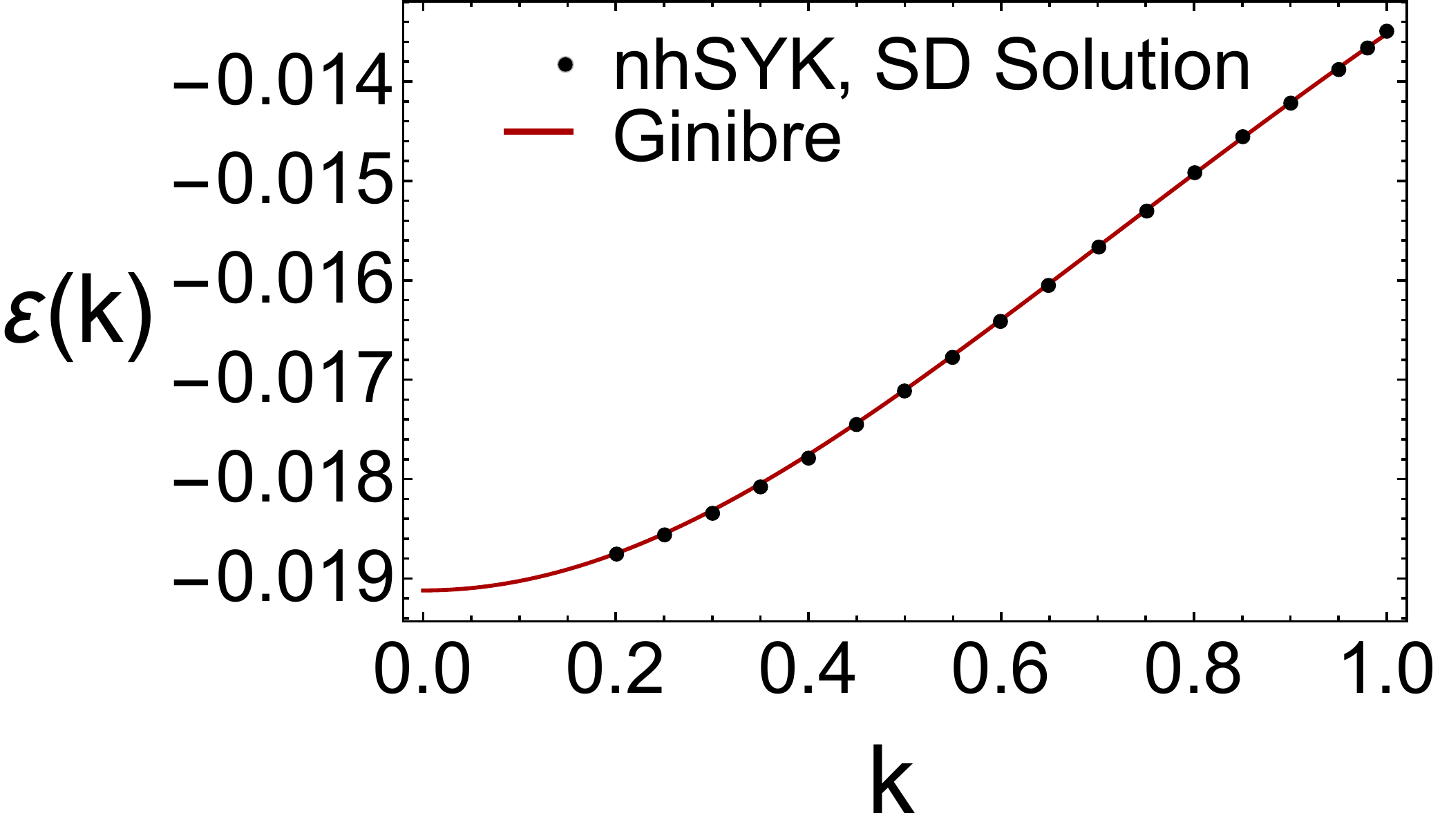}
   \includegraphics[width=8.5cm]{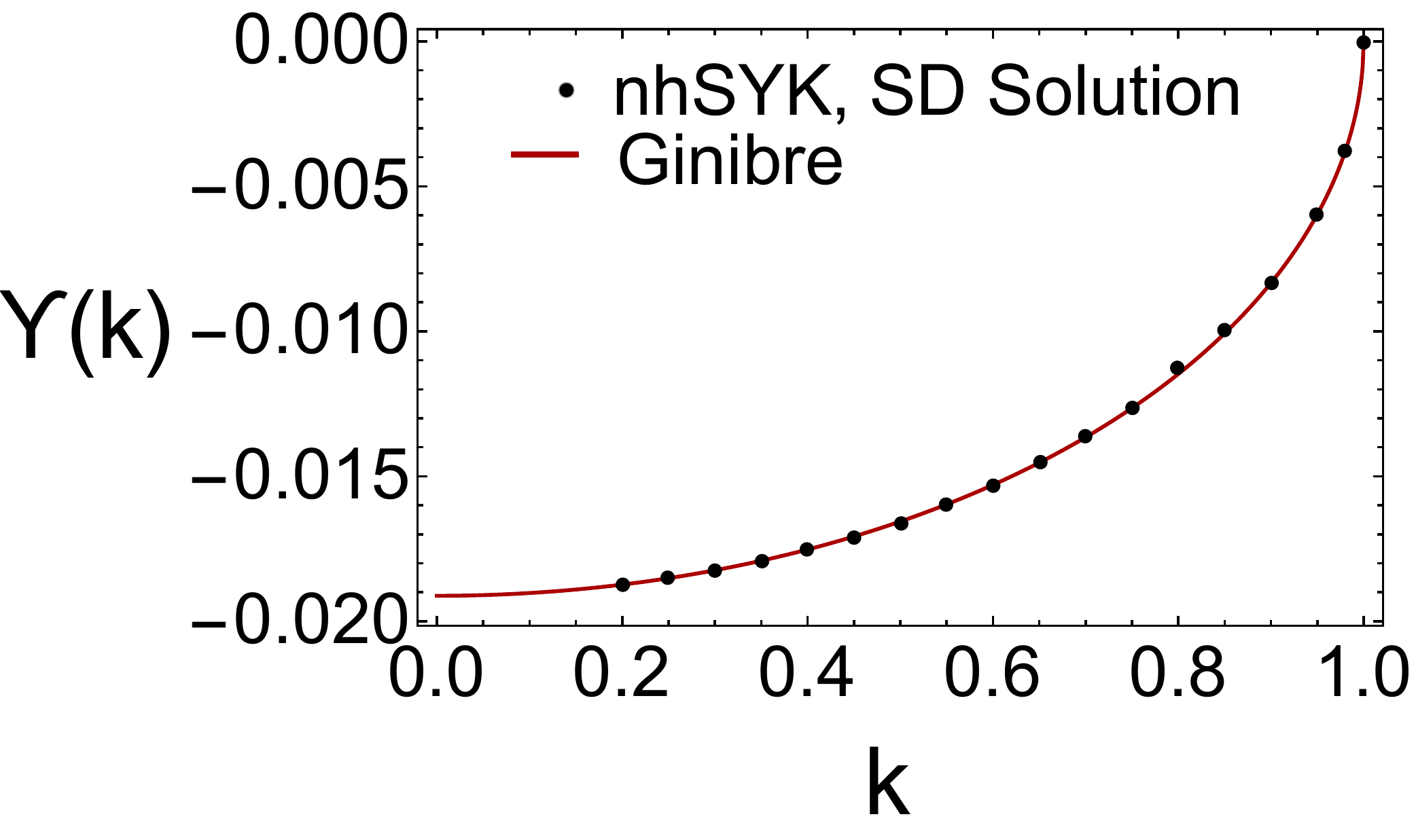}}
       \caption{The intercepts of the free energy with the $T=0$ axis for
         the low-temperature phase (left) and the high-temperature phase
         (right) both obtained from solving the Schwinger-Dyson equations
         of the non-Hermitian SYK model. The results are compared to
         analytical formulas for the Ginibre ensemble, where $\varepsilon(k)$
and $\gamma(k)$ are
           half of the long
          axis and the focal length of the ellipse containing the
          eigenvalues of the Hamiltonian. The curves are obtained without fitting
          with $\varepsilon(0)$ as the only parameter.
\label{fig:geometry}}
 \end{figure}

 Since the system develops a first-order phase transition, we can 
 study the free energy  of both phases  separately and
 use this to determine the critical 
 temperature $T_c (k)$ as a function of $k$. We start with the low-temperature
 phase. From Figure \ref{fig:exp} it is clear that the free energy is close
 to being temperature-independent.
   The intercept of the free energy of the low-temperature phase
   with the $T=0$ axis is well determined.
   In Figure \ref{fig:geometry}, we show the intercept $\varepsilon(k)$ versus
   $k$ (black points) and compare it to the $k$-dependence obtained
   for the elliptic Ginibre model:
\be
\varepsilon(k) =  \frac 2N E_0(k) = \frac{e_0}{\sqrt{1+k^2}}
\ee
with $e_0$ determined by the ground state energy of the $k=0$ SYK model.
From elementary considerations it is clear that $-\varepsilon(k) N/2$ is equal
to the smallest real part of the eigenvalues.
The excellent agreement of the $k$-dependence
shows that the two-site non-Hermitian SYK model is in the universality
   class of the elliptic Ginibre Model. 

\begin{figure}[t!]
   \centerline{\includegraphics[width=8cm]{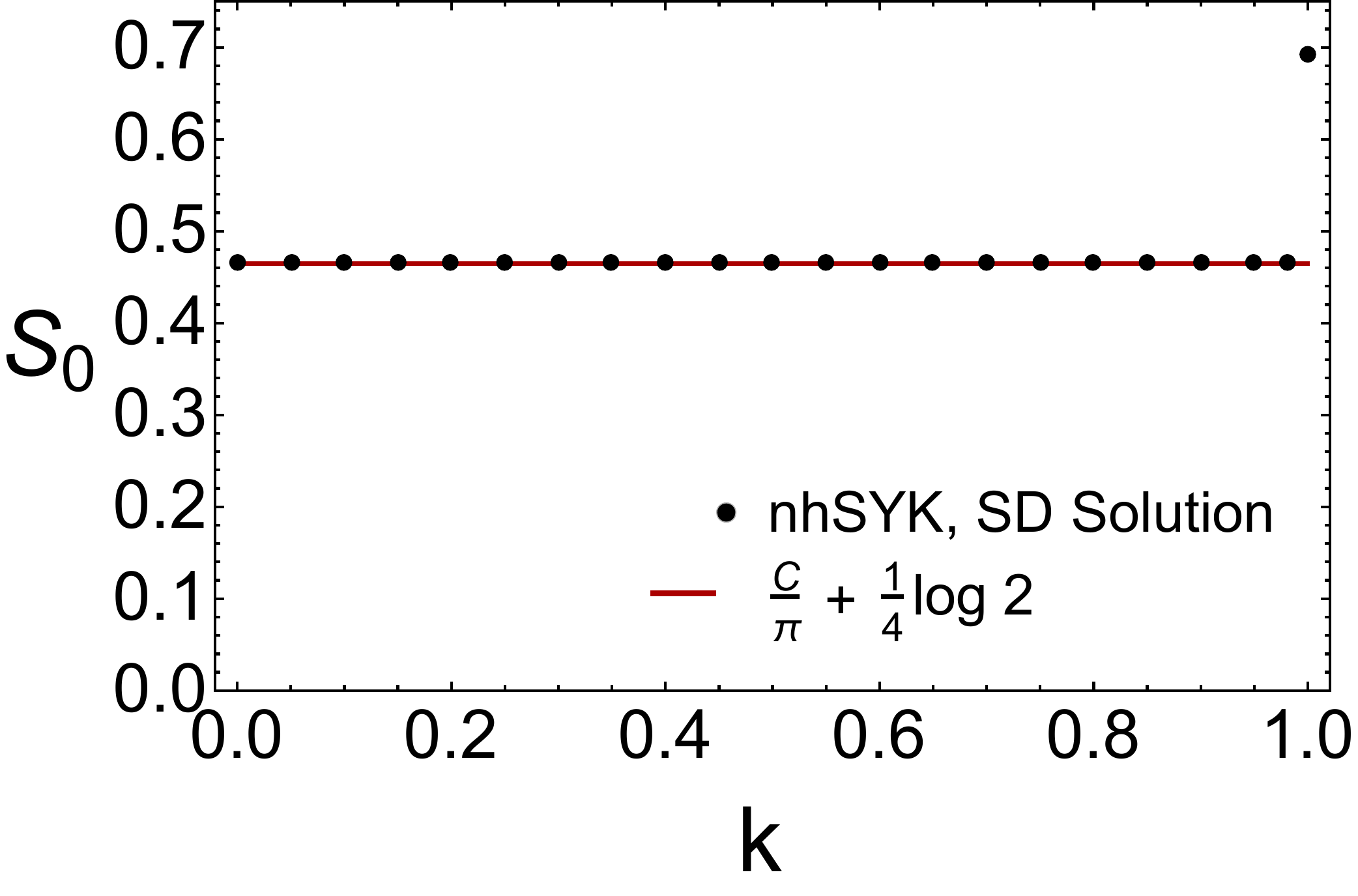}
   \includegraphics[width=8cm]{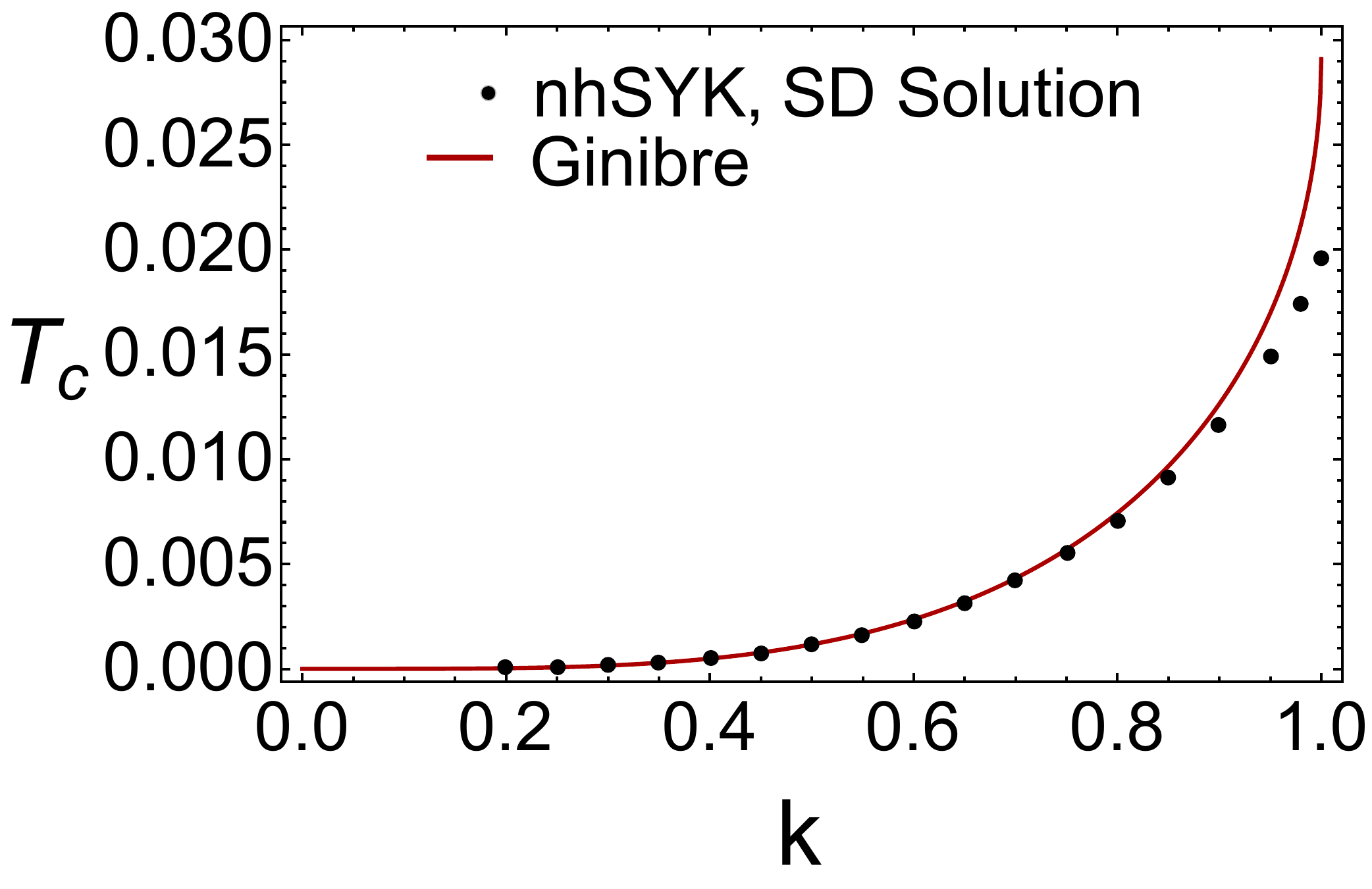}}
       \caption{The  zero-temperature entropy per particle $S_0(k)$   versus $k$ for the black hole phase calculated
         from the slope of the free energy (left), compared to $C/\pi +\frac 14 \log 2$
         with $C$ the Catalan constant (0.915966). The
         right figure compares the critical temperature obtained from
         the solutions of the Schwinger-Dyson equations to the formula from
         the elliptic Ginibre ensemble with the high-temperature entropy substituted by the entropy of the curve shown in the left figure. If we would have
         used the actual values of the entropy for $k=1$ (which is $\log 2$)
         the result of the SD equations would have agreed with the result from
         the Ginibre ensemble.
\label{fig:s-tc}}
 \end{figure}

   Next we consider the free energy of the high-temperature phase. Its intercept
   with the $T=0$ axis $\gamma(k)$ is compared to the focal point of the
   ellipse containing the eigenvalues obtained for the elliptic Ginibre model
   \be
   \gamma(k) = - \frac 2N \sqrt{E_0^2(k) - y_0^2(k)}.
   \ee
   Again the agreement is excellent without any fitting (see Figure \ref{fig:geometry} (right)). The free energy
   of the high-temperature phase is approximately linear in $T$,  but the entropy per particle (see Figure~\ref{fig:s-tc} (left))
   is
   only equal to $\log 2$ for $k=1$  contrary to
   the expectation from the Ginibre model.
   For $k<1$ the zero-temperature entropy of the high-temperature phase is  a constant equal to
   the $k=0$ value of $C/\pi + \frac 14 \log 2$
   (red curve in Figure~\ref{fig:s-tc}, left) where $C$ is the Catalan constant\footnote{This comes from the zero-temperature entropy density formula  \cite{kitaev2015,maldacena2016}
   \begin{equation*}
     {S_0} = \frac{1}{2}\log2 -\int_0^{1/q} \pi(\frac{1}{2}-x)\tan(\pi x) dx
   \end{equation*}
for the one-site Hermitian SYK model. }, but jumps to
   $\log 2$ close to $k=1$. Even for $k=0.98$ the zero-temperature entropy  is very close
   to the $k=0$ value.

   The critical temperature is determined by the intersection of the
free energy of the    low-
and high-temperature phases. The results are given in
Figure~\ref{fig:s-tc}. We also show the result for the Ginibre ensemble
(blue curve) that can be obtained from $\epsilon(k)$ and $\gamma(k)$. However,
it is clear that this cannot work because the slope
of the free energy of the high-temperature phase is always $\log 2$ for
the elliptic Ginibre
model. If we substitute $\log 2$ by the
actual slope for $k <1$, $S_0(k) = 0.464848$, we obtain
\be
T_c(k) = \frac{e_0}{S_0(k)} \left (1 - \frac {\sqrt{1-k^4}}{\sqrt{1+k^2}}\right ),
  \ee
  which is depicted by the red curve in Figure \ref{fig:s-tc}.
  If we use the actual value of $S_0(k)$ for $k=1$ we also
  find agreement with the result for the Ginibre ensemble.
 The agreement with the Ginibre ensemble provides
 strong support to the physical picture of RSB configurations dominating
 the free energy in the low-temperature limit and inducing a first-order phase transition.

\begin{figure}[t!]
       \centerline{
         \includegraphics[width=8cm]{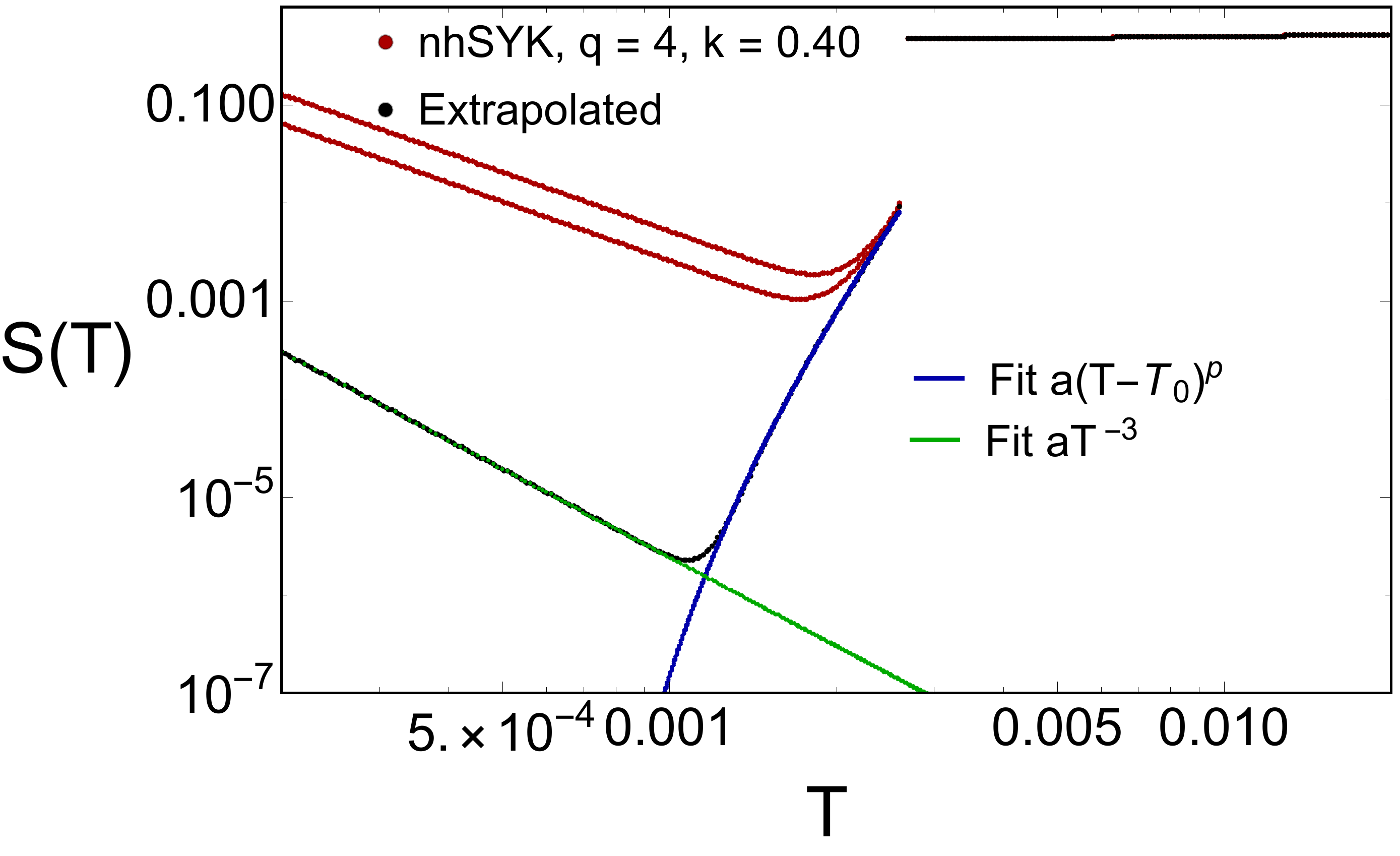}
         \includegraphics[width=8cm]{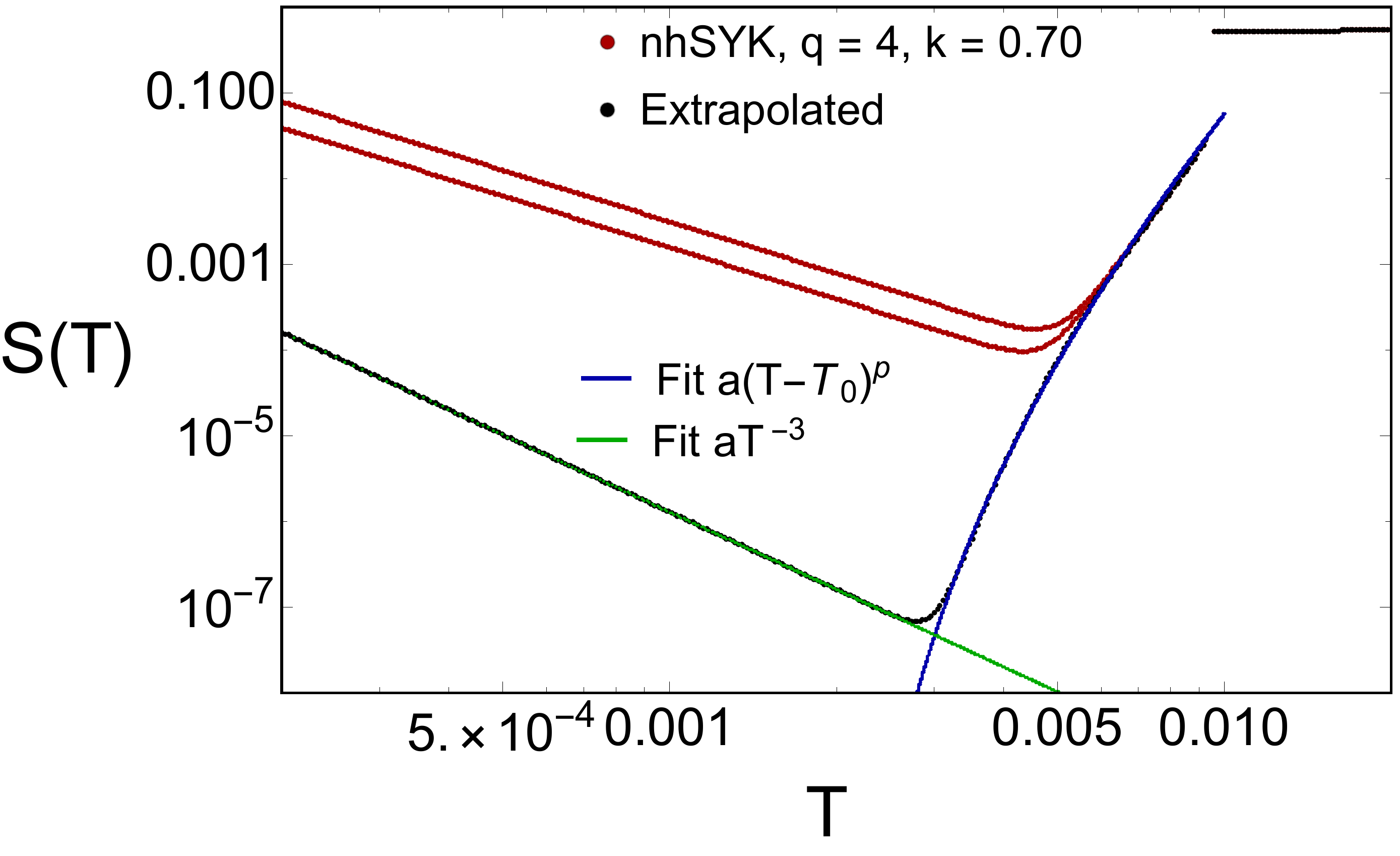}}
        \centerline{\includegraphics[width=8cm]{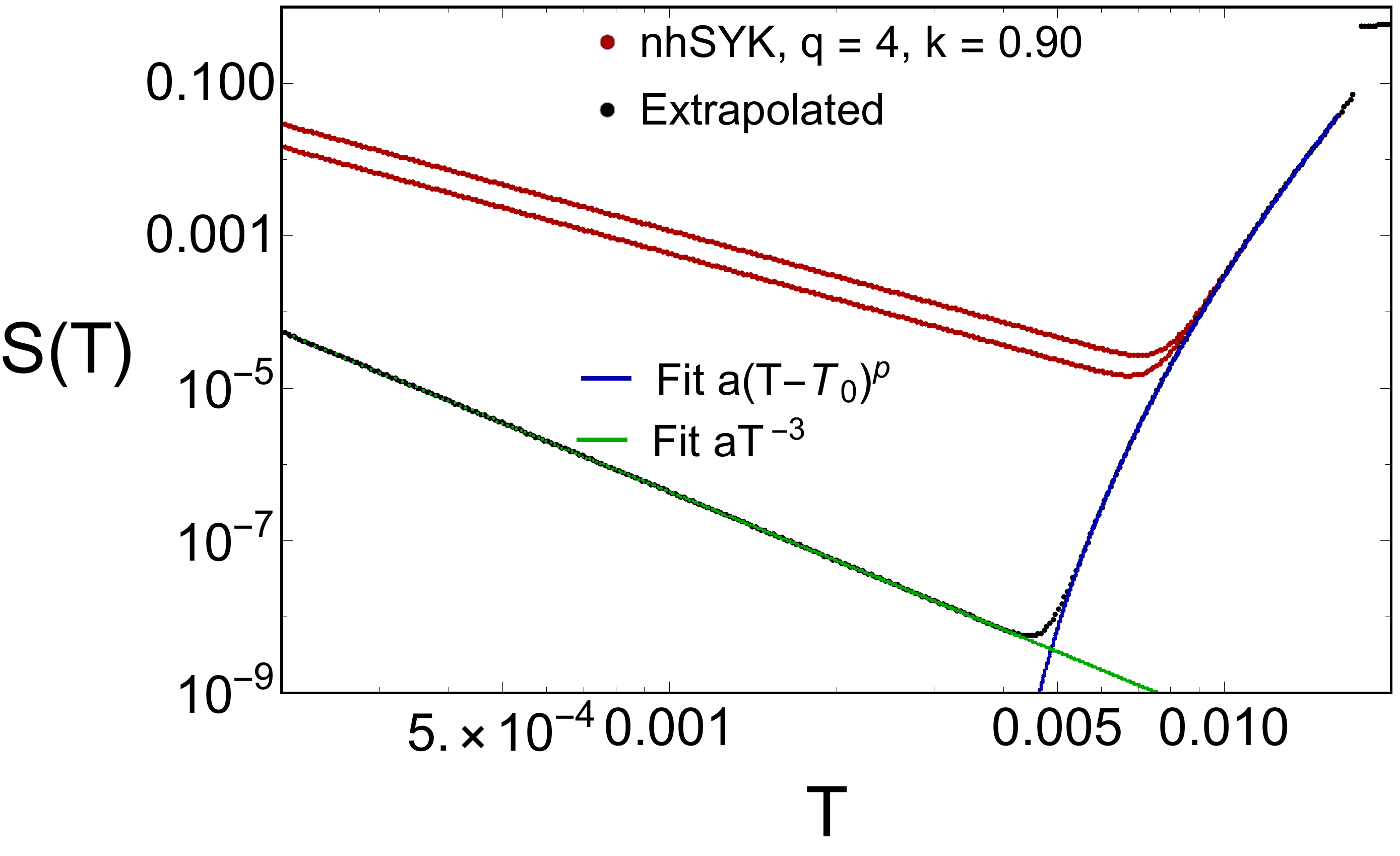}
         \includegraphics[width=8cm]{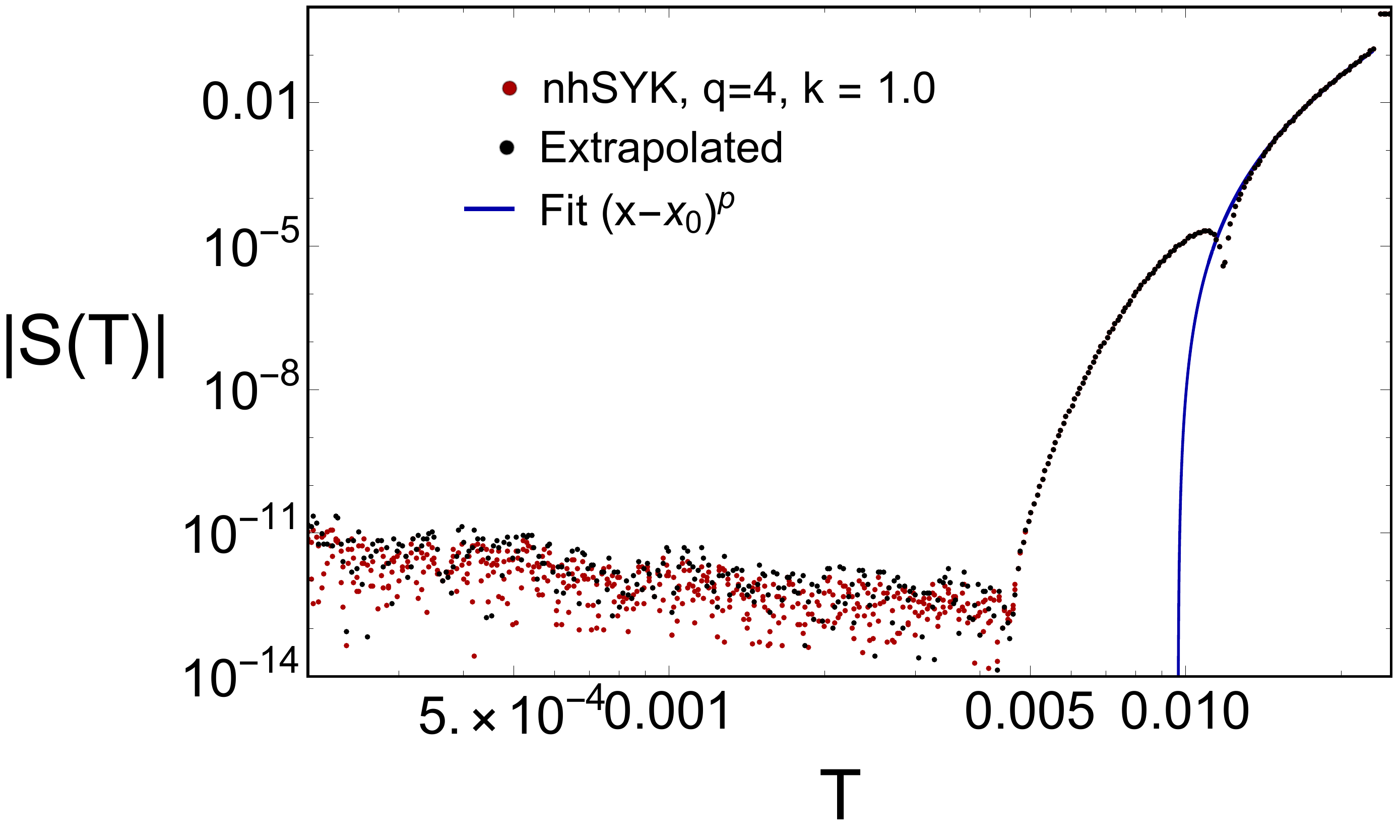}}
         
\caption{Log-log plot of the entropy of the two-site non-Hermitian $q=4$ SYK model as a function
  of the temperature for $k=0.4$ (upper left), $ k=0.7$ (upper right),
   $ k=0.98$ (lower left)  and $k=1$
       (lower right).
       The red dots show results obtained from a numerical
       solution of the Schwinger-Dyson equations for
       $10^5$ discretization points on $[0,\beta]$, and the black dots are
       a Richardson extrapolation \cite{Richardson:1911}  from the solutions with $5\times 10^4$ and $10^5$ discretization points. At the first-order phase transition point the entropy jumps from 0.010 to 0.474, from  0.032 to 0.506, from 0.070 to 0.569, and from
       0.132 to 0.693 ($\log 2$) 
       for $k = 0.4$, $k=0.7$, $ k=0.9$ and $k=1$, respectively.
       For $k=1$ the entropy becomes negative on the interval [0.0046,0.0118] and vanishes
       up to the accuracy of the calculation for $T<0.0046$. For $k=1$
       the results do not depend on the discretization step until it is increased
       by a factor of about 1000. The blue curve for $k=1$ is a fit for
       $T> 0.0118$ until the first-order transition point.      
       \label{fig:slow} }
     \end{figure}

 We already have seen that the entropy of the two-black-hole phase is not given
 by the Ginibre ensemble. Also for the low-temperature phase we observe
 deviations from the Ginibre ensemble which gives a vanishing entropy.
 Indeed, if we plot the entropy per particle
 \be
 S(T) = -\frac 2N \frac {dF}{dT}
 \ee
 on a log-log scale (see Figure~\ref{fig:slow}) we find a clear temperature
 dependence. For each of the four $k$ values, $k =0.4$, $k=0.7$, $k=0.9$
 and $k=1$,
 we observe a strong first-order phase transition at $T_c$, discussed earlier
 in this section. At this point
 the entropy per particle jumps from a small positive value to a value in the
range $[0.5, \log 2]$ 
  (see caption of Figure~\ref{fig:slow}).
   For $k= 1$ we observe a second critical temperature, $T_0$,  below which the
   entropy vanishes to the accuracy of the calculation.
   Between this temperature and $T_c$ the entropy becomes
   a small nonzero positive number  after first becoming negative.
    Changing the discretization steps by a factor 2,
   or even a factor 1000,  does not change this picture for $k=1$. Note that the
   apparent jump between $T_0$  and $T_c$ is due to plotting $|S(T)|$ on a log-log scale. 
   For $k=0.4$, $k=0.7$ and $k=0.9$ (in fact for $k<0.95$)
   the entropy remains positive and we
   plot $S(T)$ rather than $|S(T)|$. 

       The discretization error is expected to be of second order in the
       discretization step 
       $\Delta t =\beta/M$.  Indeed, at low temperatures the entropy behaves
       as $1/T^2$ (see Figure~\ref{fig:slow}). After Richardson extrapolation
       (see black points in Figure~\ref{fig:slow})
\be
       S_{\rm extrapolated}(T) = 2 S(M,T)- S(M/2, T),
       \ee
       the leading order dependence on the stepsize is canceled and the discretization
       error is of order $(\Delta t)^3$. Indeed the extrapolated result (black
       dots in Figure~\ref{fig:slow} behave as $1/T^3$ for temperatures
       below the kink (see green lines). We conclude that
for temperatures below the kink the nonzero value of 
the entropy is due to finite size effects. We expect that in the continuum
limit the entropy will also vanish in this region for $k<1$.
  Because $G_{LR}$ is
  continuous at $t=0$ and $t=\beta$ the finite size effects for
  contributions involving $G_{LR}$ are very small. On the other hand $G_{LL}$ is discontinuous
       at $t = 0$ and $t =\beta$ which results in large finite size effects
       of contributions of $G_{LL}$ to the free energy. This explains why
       the entropy for $k=1$, which depends only on $G_{LR}$, does not
       depend on the discretization step for a large range of $M$ values.
       We also notice that by choosing half-integer discretization points
       the discretization errors are reduced by an order of magnitude
       with respect
       to choosing integer discretization points. 

   The entropy is also given by \cite{maldacena2016} (i.e. using $F= U - TS$) 
   \be
   \frac {dF}{dT} = \frac FT +{\cal J}^2 \beta ^2 \frac 1M \sum_{n=1}^M G_{LL}(n-\frac 12)+\widetilde {\cal J}^2 \beta ^2 \frac 1M \sum_{n=1}^M G_{LR}(n-\frac 12).
       \ee
       Each of the terms can be calculated separately from the free energy and
       the Green's functions. This identity (which is valid at finite $M$)
       is satisfied numerically to 3 or 4 significant digits
       (except when the entropy
       is very small and large cancellations occur in the right-hand side). We could
       not fully explain why this identity is not satisfied with greater
       accuracy, but it could be due large finite size effects in $G_{LL}$
       for $t $ close to 0 or $\beta$. For $k > 0.95 $ the entropy becomes
       negative. However, its magnitude is very small -- the monotony of the
       free energy is only violated by about $10^{-6}$ of its value. Within a
       wide range of the parameters, it also
       does not depend on the size of the discretization step and
       the convergence criterion.
       However, we cannot exclude that the negativity
       of the entropy may be  an artifact of the algorithm. 

\begin{figure}[t!]
       \centerline{
         \includegraphics[width=8cm]{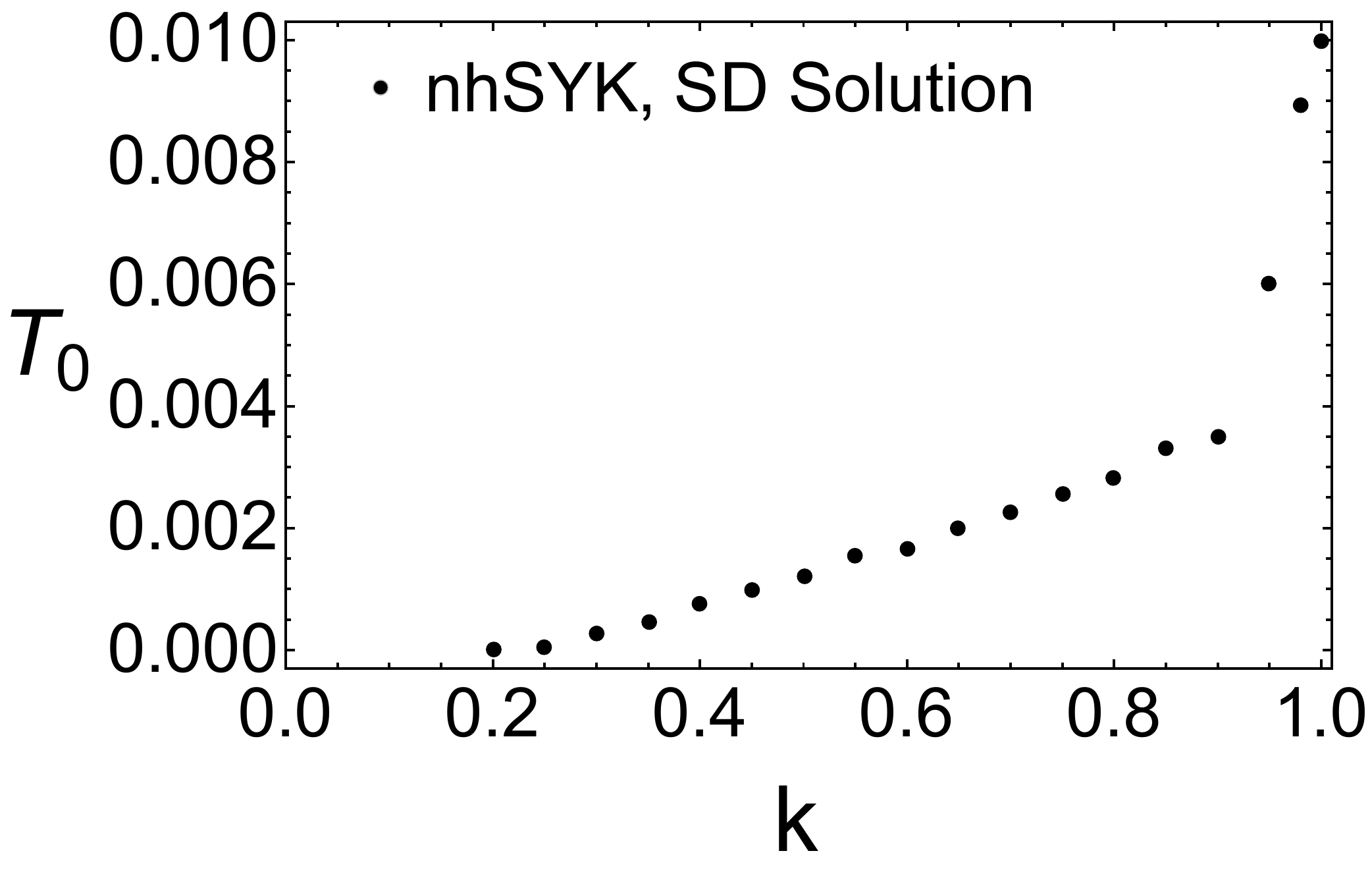}
                  \includegraphics[width=7.3cm]{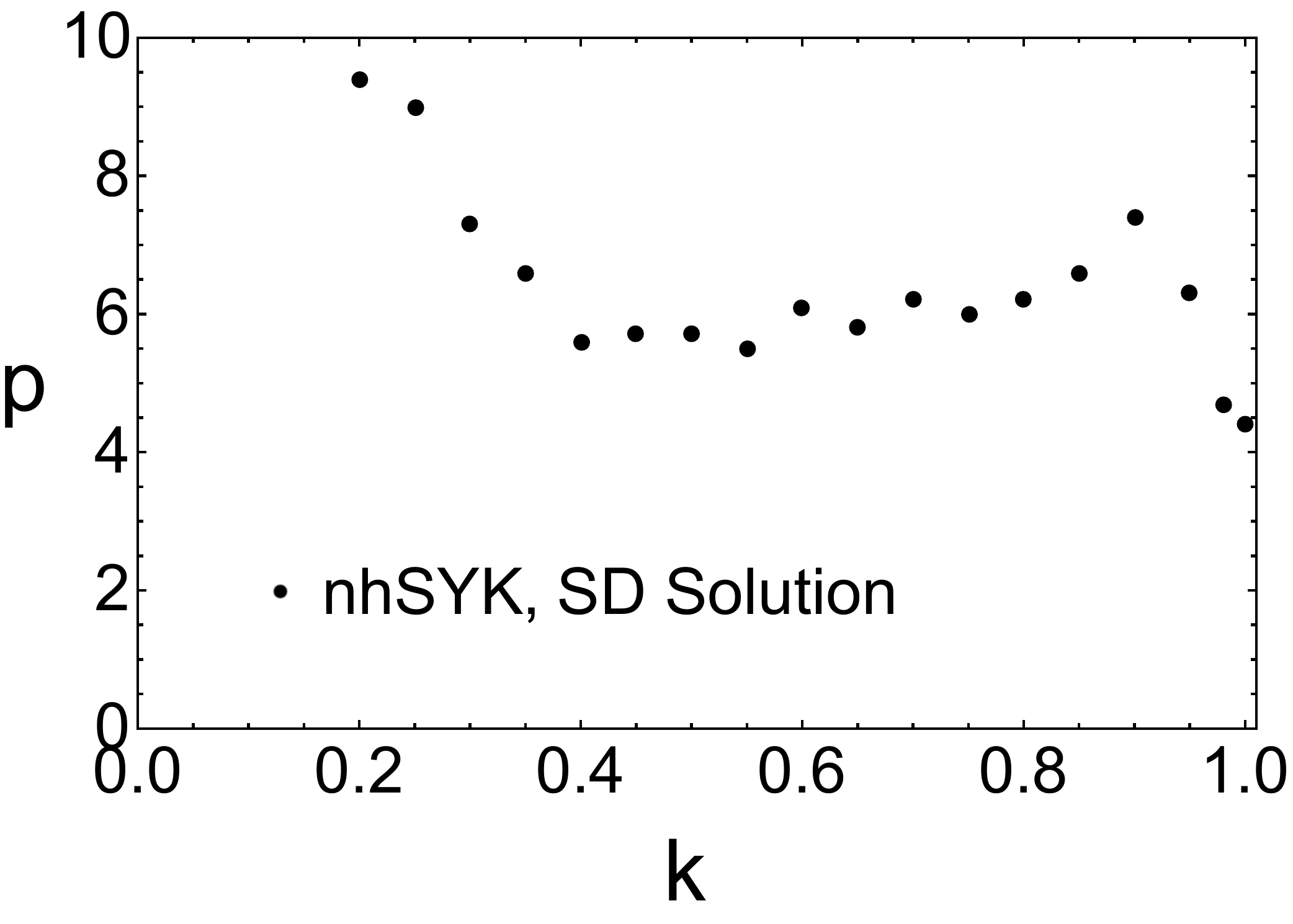}
  }
         
\caption{       The fitting parameters $T_{0}$ and $p$ as a function of $k$,
\label{fig:tc2}
}
\end{figure}

       To identify the value of the second critical point we fit the logarithm
       of $a(T-T_{0})^p$ to the logarithm of the extrapolated entropy
       between $T_{0}$ and $T_c$
well away from the  end points to reduce
       finite size effects. The results are shown in Figure
       \ref{fig:tc2}.
     For $k=0.98$ and $k=1$, we only fit in the region where the entropy is positive. Examples of the fitted curves are shown in Figure \ref{fig:slow} (blue curves). Taking these fits at face value would indicate a high order continuous
     phase transition. 
     Because of the smallness of the entropy and the magnitude of the
     finite size effects for $k<1$ we cannot exclude that such conclusion
     is an artifact of the algorithm we are using.

 \subsection{Decay of $G_{LR}(\tau)$ and the gap}
 In order to understand the nature of the RSB configurations,
 we investigate the behavior of $G_{LR}(\tau)$ in more detail.

 We are particularly interested in its exponential decay rate with $\tau$ which, for traversable wormholes, or weakly coupled two-site SYK models with real coupling (Maldacena-Qi model \cite{maldacena2018}),  is directly related to a gap $E_g$ in the spectrum.
 Contrary to the Maldacena-Qi model, where the gap $E_g$ is equal to the energy difference
 of the first excited state and the ground state, the non-Hermitian SYK
 Hamiltonian does not have a genuine gap, but as we will see below,  $G_{LR}(\tau)$
 still decreases exponentially for large $\tau$.
 
 The gap (decay rate) is computed by fitting the long-time behavior of the propagator $G_{LR}(\tau)$,
 or $G_{LL}(\tau)$, for sufficiently low temperature with an exponential Ansatz.
 More specifically, taking into account the symmetries of the solution (see section~\ref{symmetries} and appendix~\ref{subsec:symmetry_fourier_coefficients}), we employ the Ansatz
 \be
 \label{eq:mass_gap_ansatz}
 G_{LR} (\tau) &\sim& \sinh \left( E_g (\beta/2 - \tau)\right) ,\nn\\
 G_{LR} (\tau) &\sim& \cosh \left( E_g (\beta/2 - \tau)\right) ,
 \ee
 where the gap $E_{g}$ is a fitting parameter.
\begin{figure}
	\centering
	\includegraphics[width=8cm]{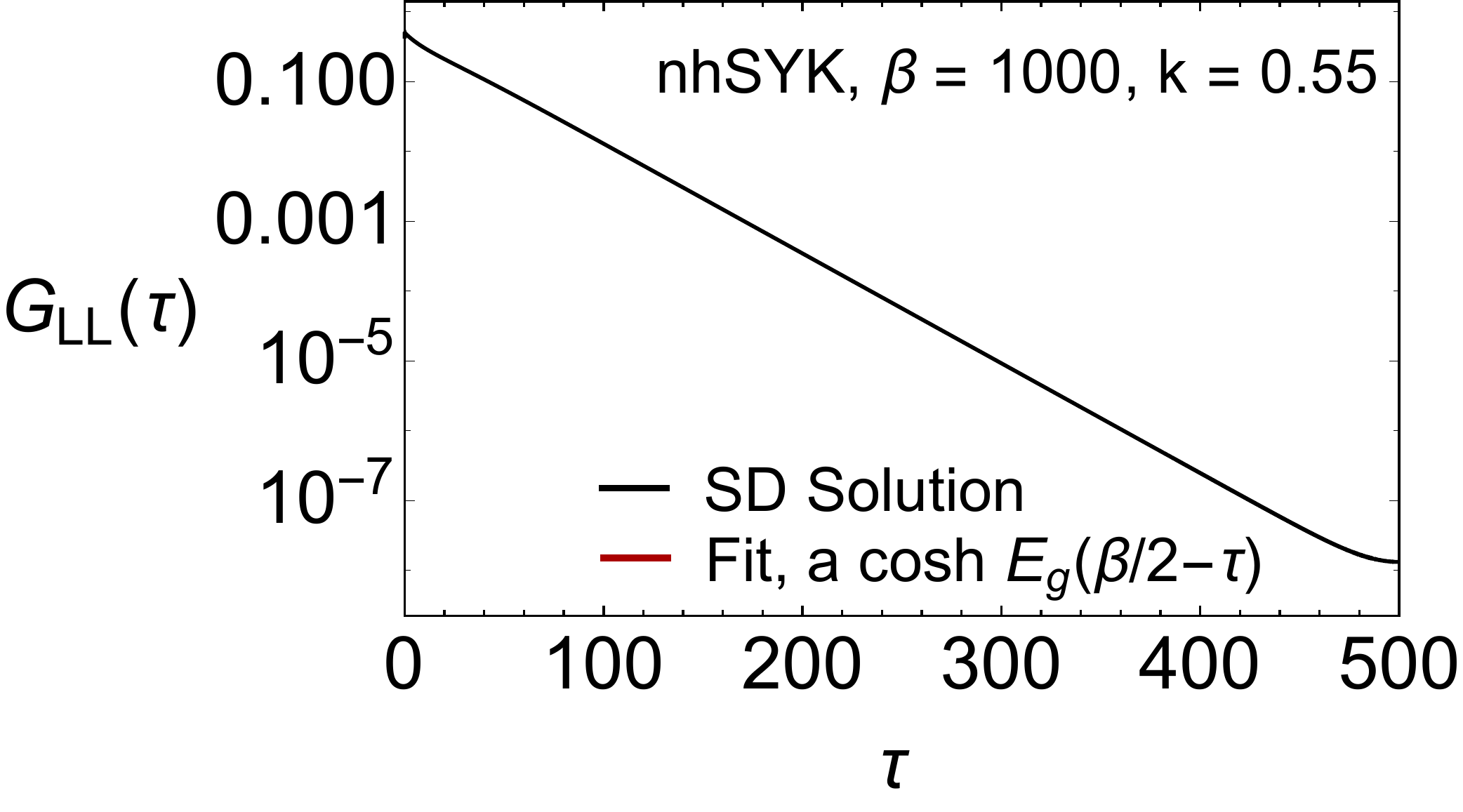}   
	\includegraphics[width=8cm]{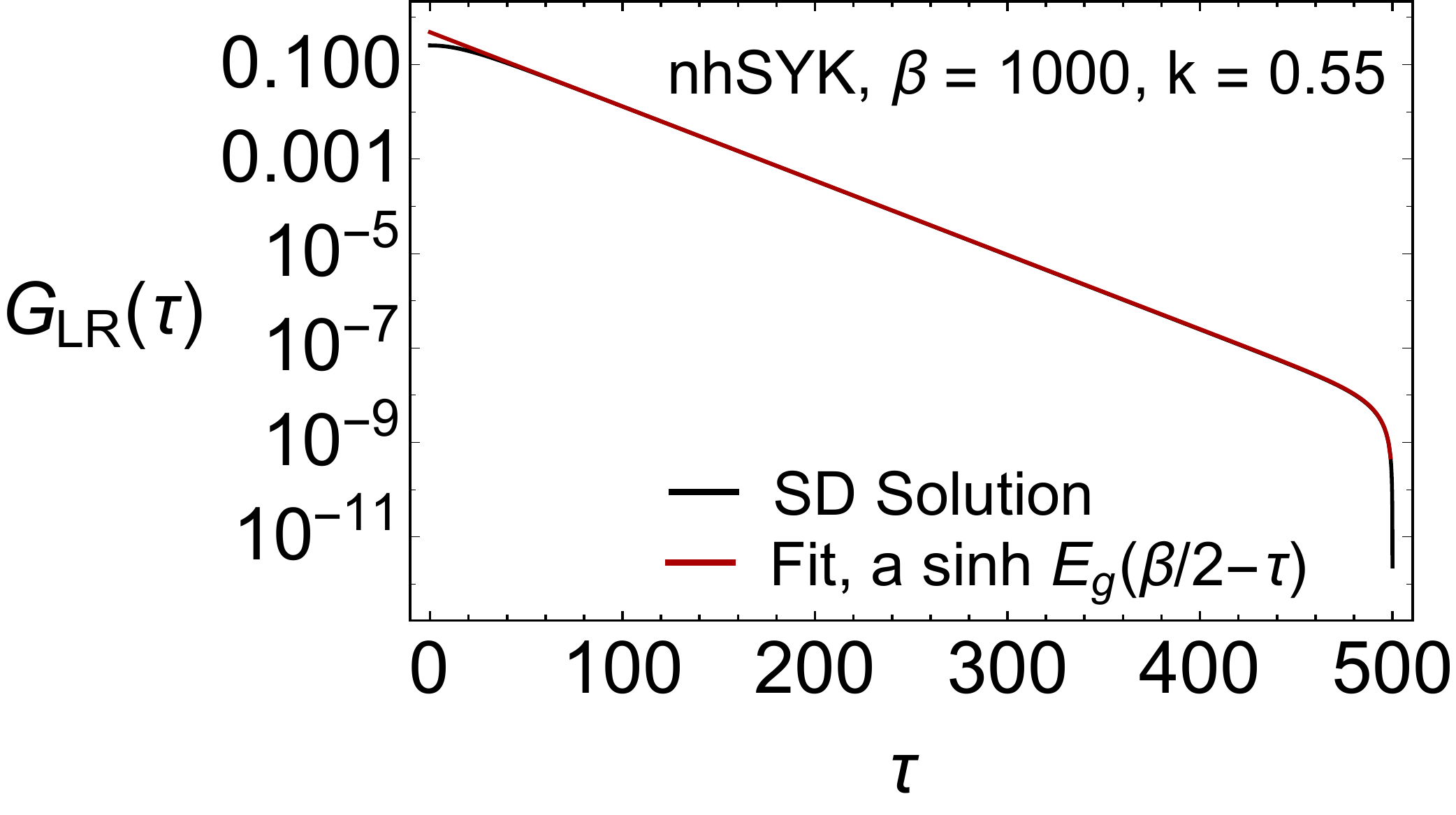}   
	\caption{The non-trivial solutions of the SD equations
          for $G_{LL}$ (left)  and $G_{LR}$ (right), for
          $T = 0.001$ and $k = 0.55$. The SD solutions (black curve) are compared to the free Green's functions  (red curves)
          with a fitting parameter  that sets the scale (the gap $E_g$) and an overall
          constant. The exponents of $G_{LL}$ and $G_{LR}$ agree to six
          significant digits.
          \label{Fig:GLRandfitting}}
\end{figure}
A feature to note is that these non-trivial solutions for $G_{LR}$
continue to exist for a range of temperatures
when they no longer minimize the  free energy
(see Figure \ref{fig:freek}). The two-black-hole solutions exist for all
temperatures.
We have checked that  at small
 temperatures this Ansatz reproduces well the 
 behavior of the propagator for $0\ll |\tau - \beta/2| \ll \beta/2$ (for
 small $k$ it also agrees well for  $\tau$ around $\beta/2$
 see Figure~\ref{Fig:GLRandfitting}. Close to the end points
 we see significant deviations from the free propagator in particular for
 $G_{LR}(\tau)$.
 
 \begin{figure}[t!]
 	\centering
 	\includegraphics[width=7.5cm]{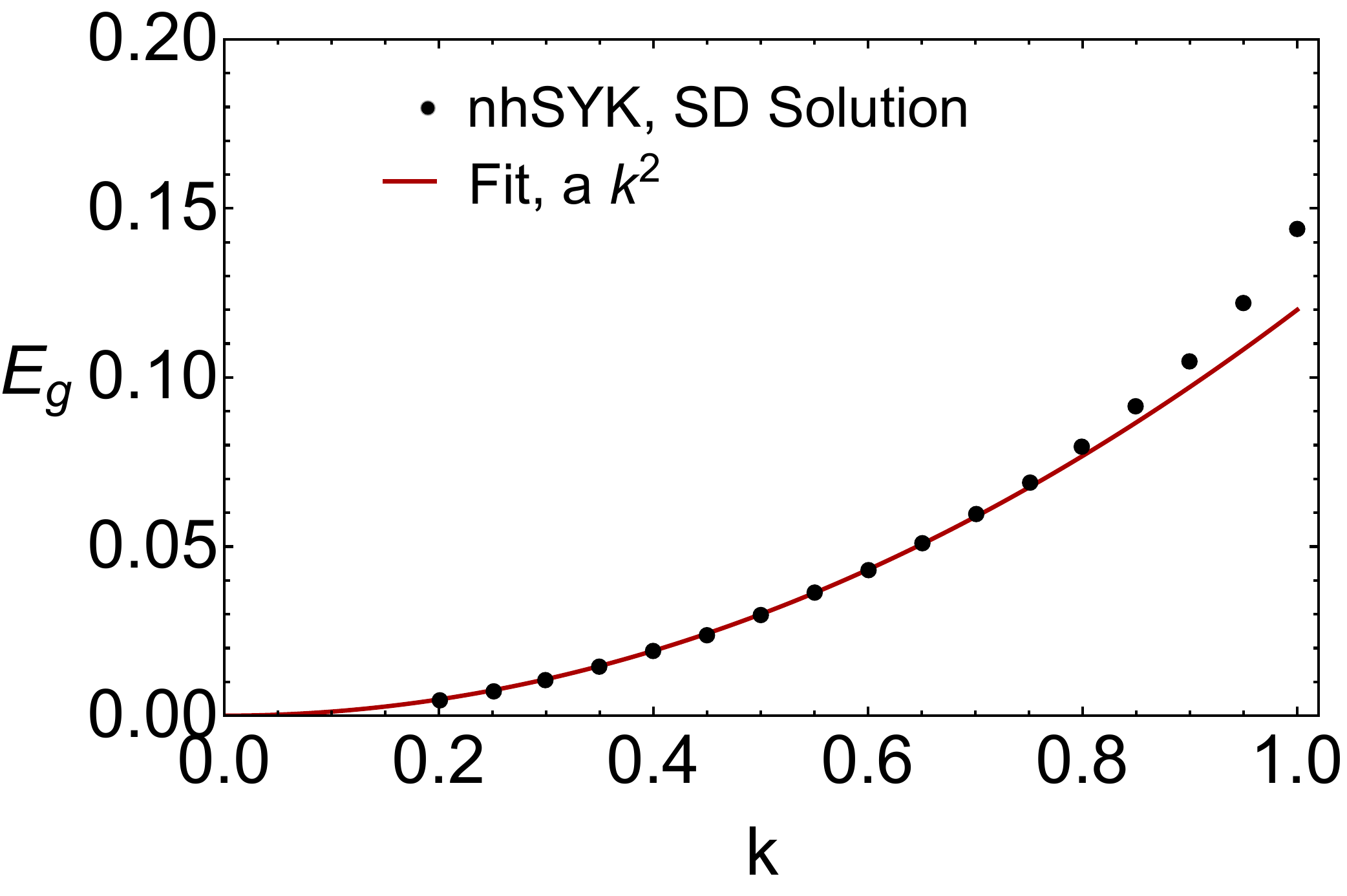}  
 	\includegraphics[width=8.3cm]{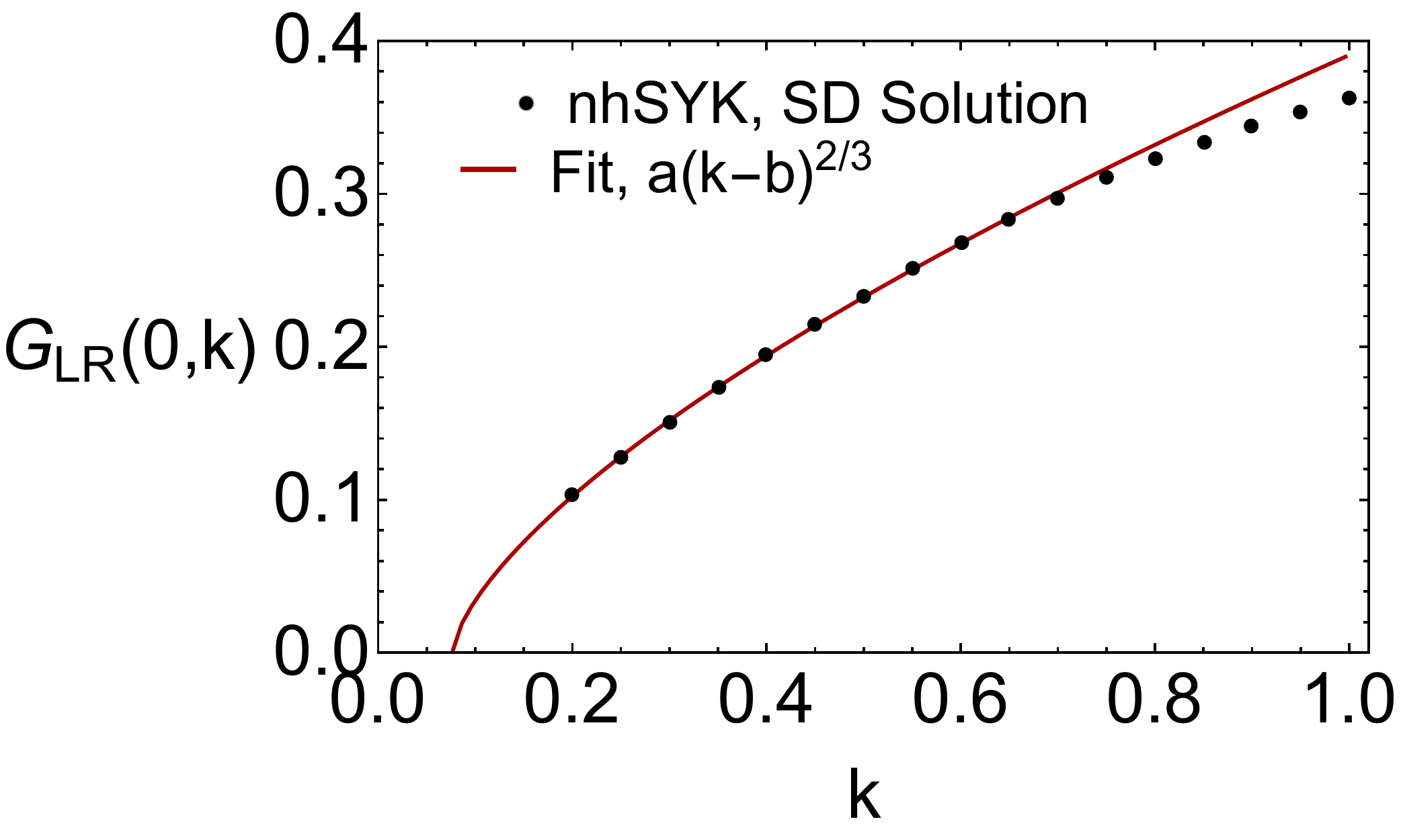} 
 	\caption{Top: The energy gap $E_g$ , namely, the rate of exponential decay of $G_{LR}(\tau)$ (left), and $G_{LR}(0)$ (right) as a function of the $k$
          calculated from the propagator for $T = 0.0005$. For small $k$
          we observe a quadratic dependence of $E_g$   on $k$.
          The exponent of the $k$-dependence of $G_{LR}(0)$ is not well determined, but an $a(k-b)^{2/3}$ gives a reasonable fit. 
                    	\label{Fig:energy-gap}
}
 \end{figure}
 
 The value of the decay rate $E_g$ as  function of the coupling $k$ in
 the low-temperature limit is shown in Figure ~\ref{Fig:energy-gap} (left).
 For small $k$ it depends quadratically on  $k$, but there are significant
 deviations for $k > 0.8$.

\begin{figure}
 	\centering
 	\includegraphics[width=8cm]{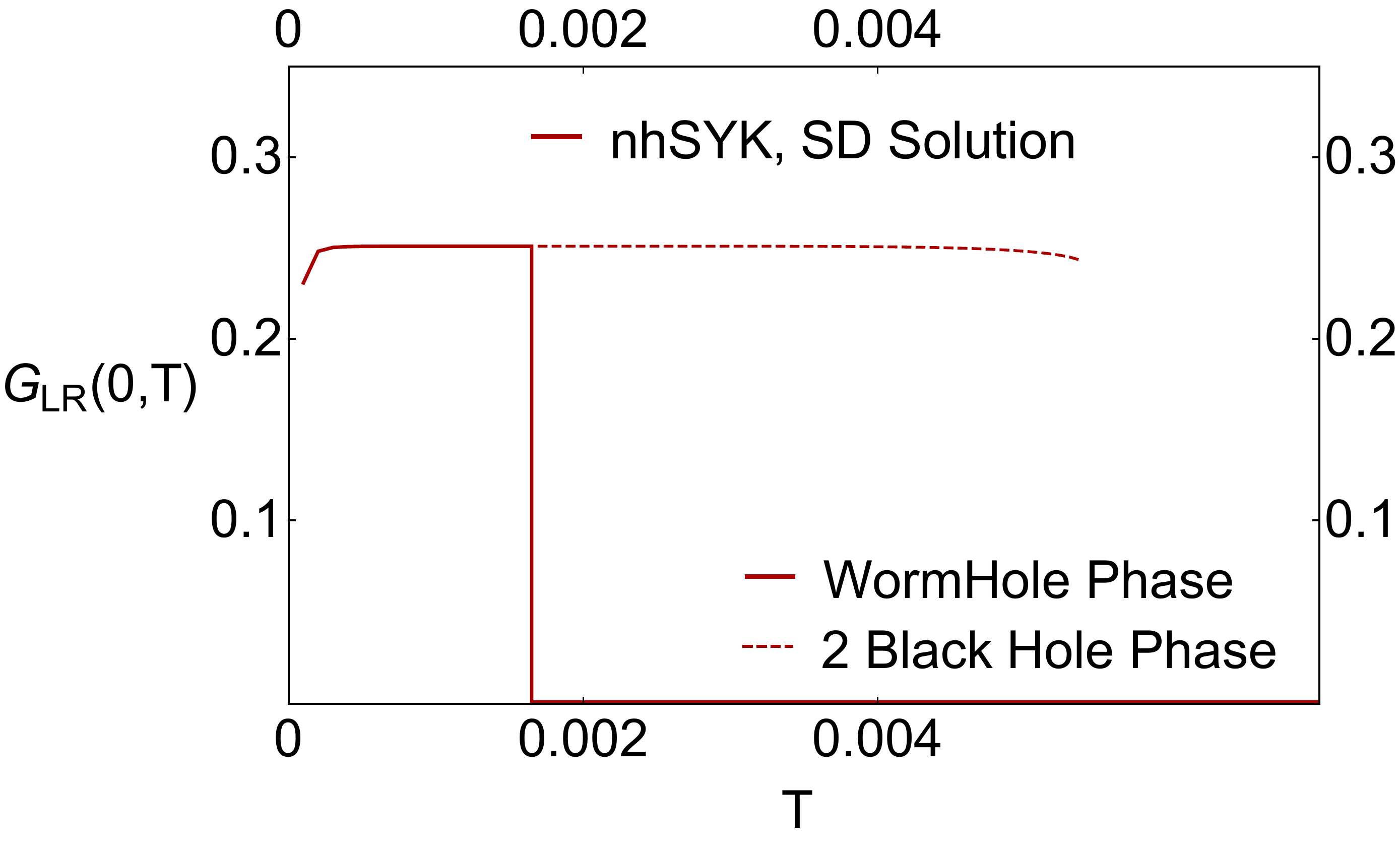}
        \caption{
          The behavior of the $G_{LR}(0)$, as a function of the temperature for $k = 0.55$. As expected in a first-order transition, it drops to zero abruptly at the transition which suggests that it can be considered an order parameter of the transition. The dashed red curve indicates the value of $G_{LR}(0,T) $ for
          $T> T_c$.
\cite{maldacena2018}.
 	\label{fig:glrt}}
 \end{figure}

Except for very low temperatures,  the value of $G_{LR}(0)$  is almost
constant as a function of the temperature in the RSB phase, and  vanishes beyond the critical temperature,  see  Figure~\ref{fig:glrt}.
 Therefore, $G_{LR}(0)$ can be considered as the order parameter of a first-order
 phase transition. The thermodynamic limit of this order parameter will be analyzed in detail  in the next subsection.
The wormhole solution continues to exist for $T>T_c$ until $T\approx 0.006$.
 The $k$-dependence of  $G_{LR}(0)$
 is shown in the right panel of Figure~\ref{Fig:energy-gap}. The small $k$
 behavior can be fitted by $\sim (k-b)^{2/3}$. The value of $b=0.076$ is consistent with the value of $k$ below which we can no longer obtain a wormhole solution
 from the SD equations.
  
 Qualitatively, the observation of an exponential decay, see Figure \ref{fig:glrt}
 (left), with a decay rate $E_g$ gives further support
to the physical picture of RSB configurations as tunneling events
 connecting different replicas. This is also the interpretation of
 wormholes in the gravity partition function. We note that the use of the term ``gap'' for the decay rate $E_g$ is more by analogy with the
 traversable wormhole case where it can be demonstrated rigorously that $E_g$ is the difference between the ground and the first excited state.
 In our case, the Hamiltonian is non-Hermitian and its spectrum does
 not have a gap.
 Therefore this interpretation can only be applicable to the associated replica field theory.

\subsection{Order Parameter for the Phase Transition}

In this subsection we study the  behavior of the mixed propagator at the origin,
$G_{LR}(0)$. We will show that it is an order parameter of the first-order
phase transition discussed before. Since the eigenstates of the Hamiltonian
are degenerate for ${\rm mod}(N/2,8) \ne 0$  the value of $G_{LR}(0)$ is basis
dependent. For example, if the eigenstates are chosen to be also eigenstates
of the chirality operators $\gamma_5^L = i^{N/4(N/2-1)} \prod_j^{N/2}\psi_L^j$ and $\gamma_5^R= i^{N/4(N/2-1)} \prod_j^{N/2}\psi_R^j$, the Green's function $G_{LR}(0)$ vanishes
identically.

At finite temperature, $G_{LR}(0)$ is given by
\be
G_{LR}(0) = \frac 1Z \frac 1{N/2}\left \langle \Tr\left [e^{-\beta H} \sum_k \psi_k^L \psi_k^R \right] \right \rangle.
\ee
As discussed before, this quantity vanishes for $\epsilon = 0$ in the action \eref{eq:MQ_model_action}
and is
purely imaginary for nonzero values of $\epsilon$. The zero-temperature limit
is given by the ground-state expectation value
\be\label{eq:spi}
\langle 0 |S |0\rangle \equiv \langle 0 |  \frac 1 {N/2}\sum_k \psi_k^L \psi_k^R |0\rangle.
\ee
For ${\rm mod}(N/2,4) =2$
  (Gaussian Unitary Ensemble universality class), we have four degenerate
  ground states which can be characterized by the chirality of the
  L and R SYK models (see appendix \ref{app:degeneracy}). Eigenstates with these quantum numbers are obtained
 by adding an infinitesimal term $\sim \gamma_{5,L} \gamma_{5,R}$ 
  to the Hamiltonian.
  In this basis, the spin operator with basis states $|++\rangle,\; |--\rangle,\;|+-\rangle,\;|-+\rangle$
  is given by
  \be
  \langle \chi_L \chi_R |iS | \chi_L \chi_R \rangle
  =
    \left (
  \begin{array}{cccc}
    0 & i\alpha & 0 & 0 \\
    -i\alpha & 0&0&0 \\
    0 &0 &0 & i \beta \\
    0 & 0 &  -i \beta & 0
     \end{array} \right ),
  \ee
where $ \chi_L ,  \chi_R = \pm$ are the possible chiralities of the left and right ground states.  The anti-symmetry follows from the Hermiticity and the anti-commutation properties of
  the gamma matrices.
Using the representation 
 \be
 \psi^k_L  &=& \gamma_k \otimes 1,\nn\\
\psi^k_R &=& \gamma_5 \otimes \gamma_k,
\label{gam:rep}
\ee
 we can write the constants $\alpha $ and $\beta$ as 
  \be
  \alpha = \sum_k \langle + | \gamma_k |-\rangle^2 =0\nn\\
  \beta = -\sum_k |\langle - | \gamma_k |+\rangle|^2\ne  0,
\label{albe}
  \ee
  and the minus sign is due to the chirality. For the $\gamma_k$ in \eref{gam:rep} we use the representation
  \begin{equation}\label{eqn:gammaTildeExplicit}
\begin{split}
{\gamma}_{2k-1} &=2^{-1/2} \overbrace{\sigma_3 \otimes\cdots\otimes\sigma_3}^{k}  \otimes\sigma_{1}\otimes\overbrace{\sigma_0 \otimes\cdots\otimes\sigma_0}^{\frac N2-k-1},\\
{\gamma}_{2k} &=2^{-1/2} \overbrace{\sigma_3 \otimes\cdots\otimes\sigma_3}^{k}  \otimes\sigma_{2}\otimes\overbrace{\sigma_0 \otimes\cdots\otimes\sigma_0}^{\frac N2-k-1},
\end{split}
\end{equation}
For each term contributing to the sum in \eref{albe} containing
a $\sigma_1$ there is a corresponding
gamma matrix with a $\sigma_2$ at the same position in the tensor product. The non-vanishing
matrix elements differ by $\pm i$ so the sum over the squares of the matrix elements, which
gives $\alpha$, vanishes. Since $\beta$ is equal to the sum of the absolute value of the matrix elements, it does not vanish.
  The ground state is thus given
  by
  \be
  |G\rangle = \frac 1{\sqrt2} (|+-\rangle \pm i|  -+\rangle).
  \ee
  We thus find that for $\epsilon \ne 0$, the ground state expectation value
  of $G_{LR}(0)=\mp i \beta$ is non-vanishing. Note that the sign of $G_{LR}(0)$
  is determined by the sign of $\epsilon$.

  Next we consider the case ${\rm mod}(N/2,8) =0$. Then the ground state
  of the single-site SYK  is unique  and can be in each of the two chirality sectors.
  Therefore, the ground state of the two-site SYK model has either $|++\rangle$ or
  $|--\rangle$ as ground state in terms of the chiralities of each SYK.
  This means that the expectation value of $S$ vanishes. For a nonzero value
  of $\epsilon$, a perturbative calculation  yields
  \be
  \label{glr0eps}
  G_{LR}(0) = \frac\epsilon{E_{++} - E_{--}}
  \ee
  with $E_{++}$ and $E_{--}$ the lowest energy with both chiralities positive or negative,
  respectively.
  In the thermodynamic limit, the spacing of $E_{++} - E_{--} \sim \exp[-\frac12 N S_0]$
  with $S_0$ the zero-temperature entropy density,
  and  a finite value is possible if the thermodynamic limit is taken before
    the limit $\epsilon \to 0$.

 \begin{figure}[t!]
\centerline{\includegraphics[width=8cm]{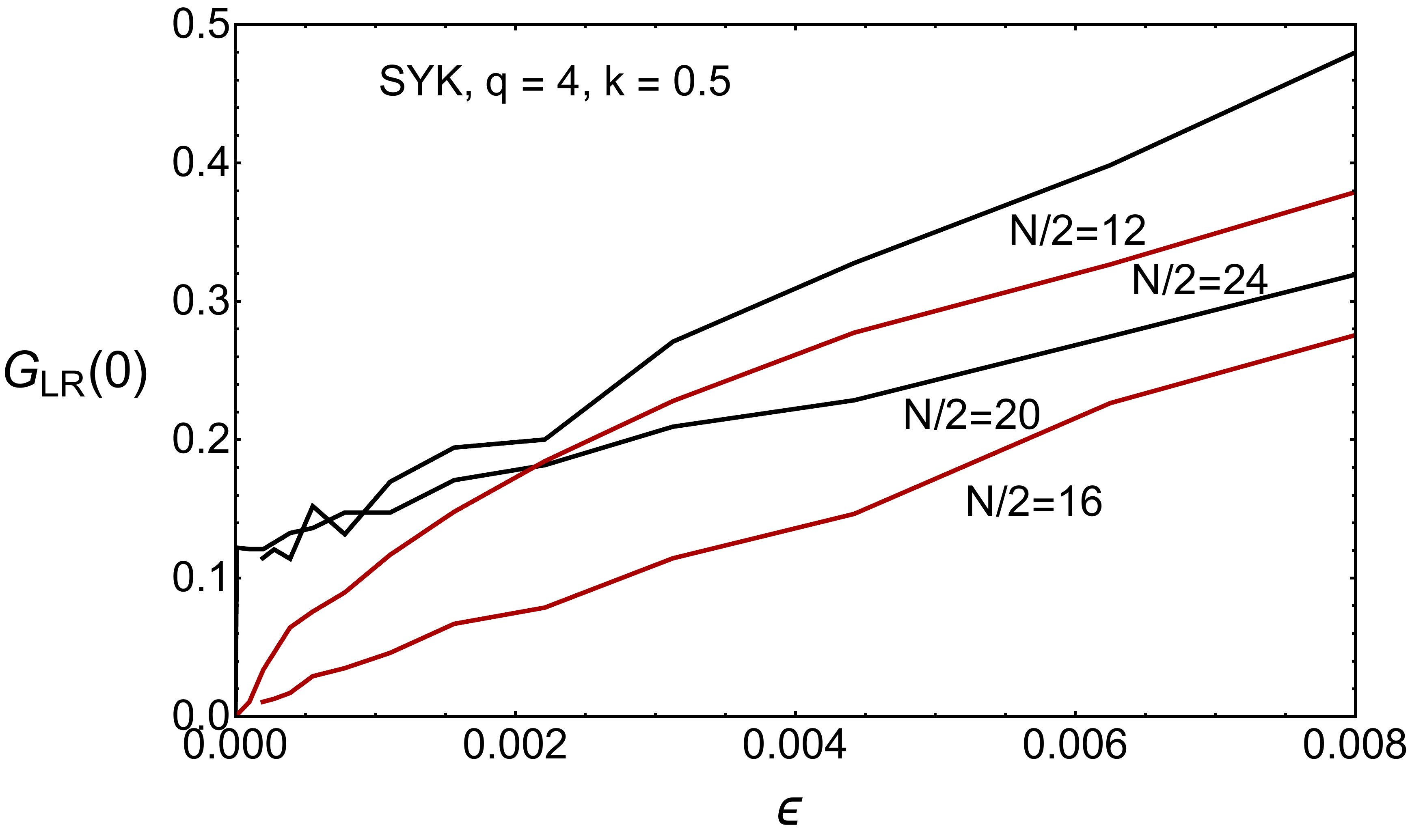}}
  \caption{The ground state expectation value of $S$ at zero temperature, namely, $G_{LR}(0)$, as a function of
    $\epsilon$ for $k=0.5$. 
 } \label{fig:glr0mu}
  \end{figure}
 
    Finally, we consider the case $  {\rm mod}(N/2,8) =4$ (Gaussian Symplectic Ensemble universality
    class). In this case the levels of each SYK are doubly degenerate (see Appendix \ref{app:degeneracy}).
    Therefore, we have four degenerate ground states. Since the charge conjugation
    matrix is the product of an even number of gamma matrices 
    both states of a Kramer's degenerate pair have the same chirality. We
    conclude that all four ground states have either the chirality $|+\;+\rangle $
    or $|-\;- \rangle$ so that the ground state expectation value
    of $S$ vanishes. For small $\epsilon$, the component of the wave function
    with the opposite chiralities is again given by first-order perturbation theory, and $G_{LR}(0)$ is again of the form \eref{glr0eps}. We expect to obtain a finite value if the thermodynamic limit is taken before the $\epsilon \to 0$. These arguments also apply to the original MQ model with real couplings.
  
    In Figure \ref{fig:glr0mu} (left), we show the $\epsilon$ dependence of
    $G_{LR}(0)$ for different values of $N$. As was expected, only for $N/2 = 12$ or $N/2= 24$
    we observe spontaneous symmetry breaking.
    
\begin{figure}[t!]
   \centerline{\includegraphics[width=7cm]{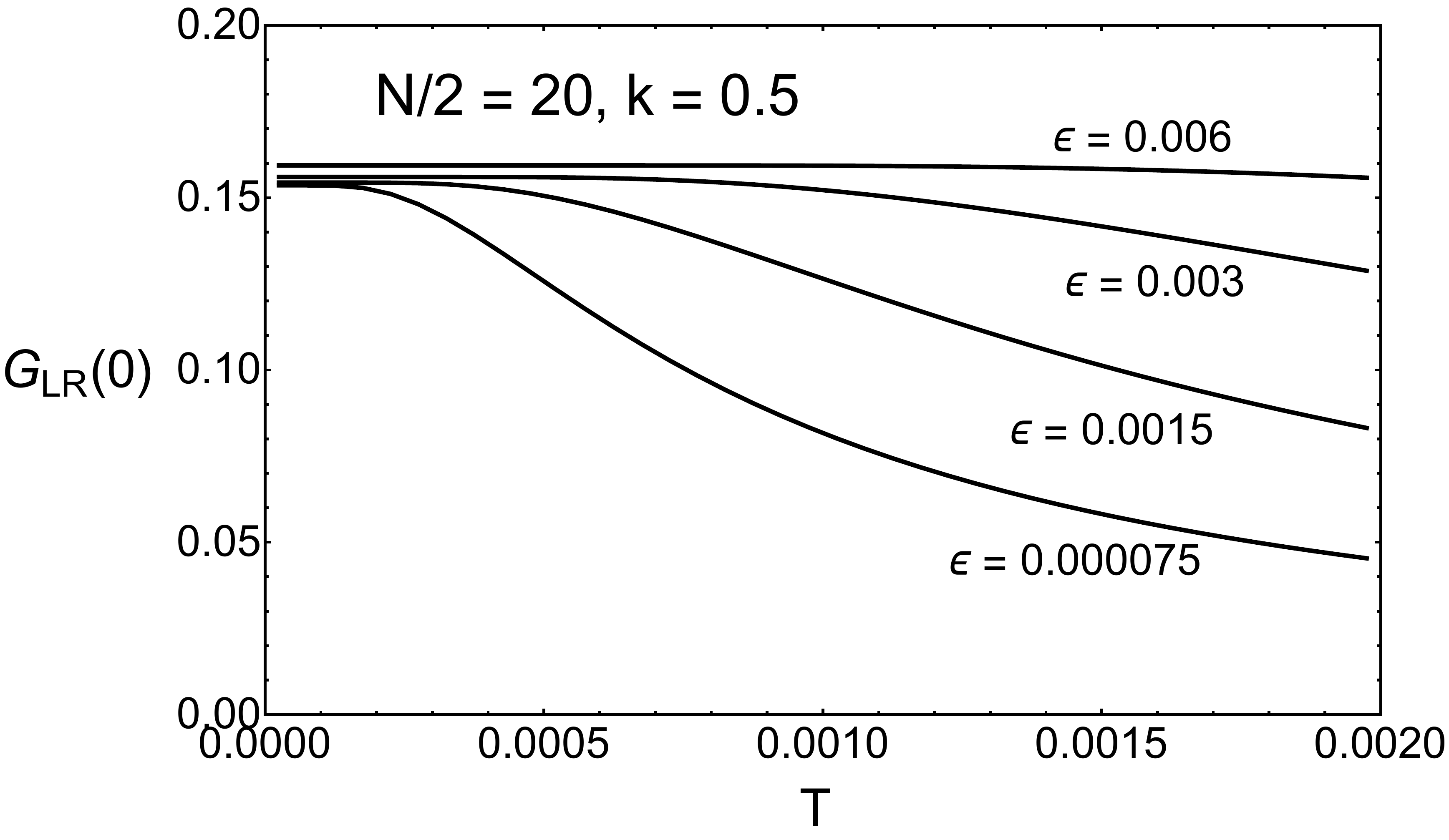}
   \includegraphics[width=7cm]{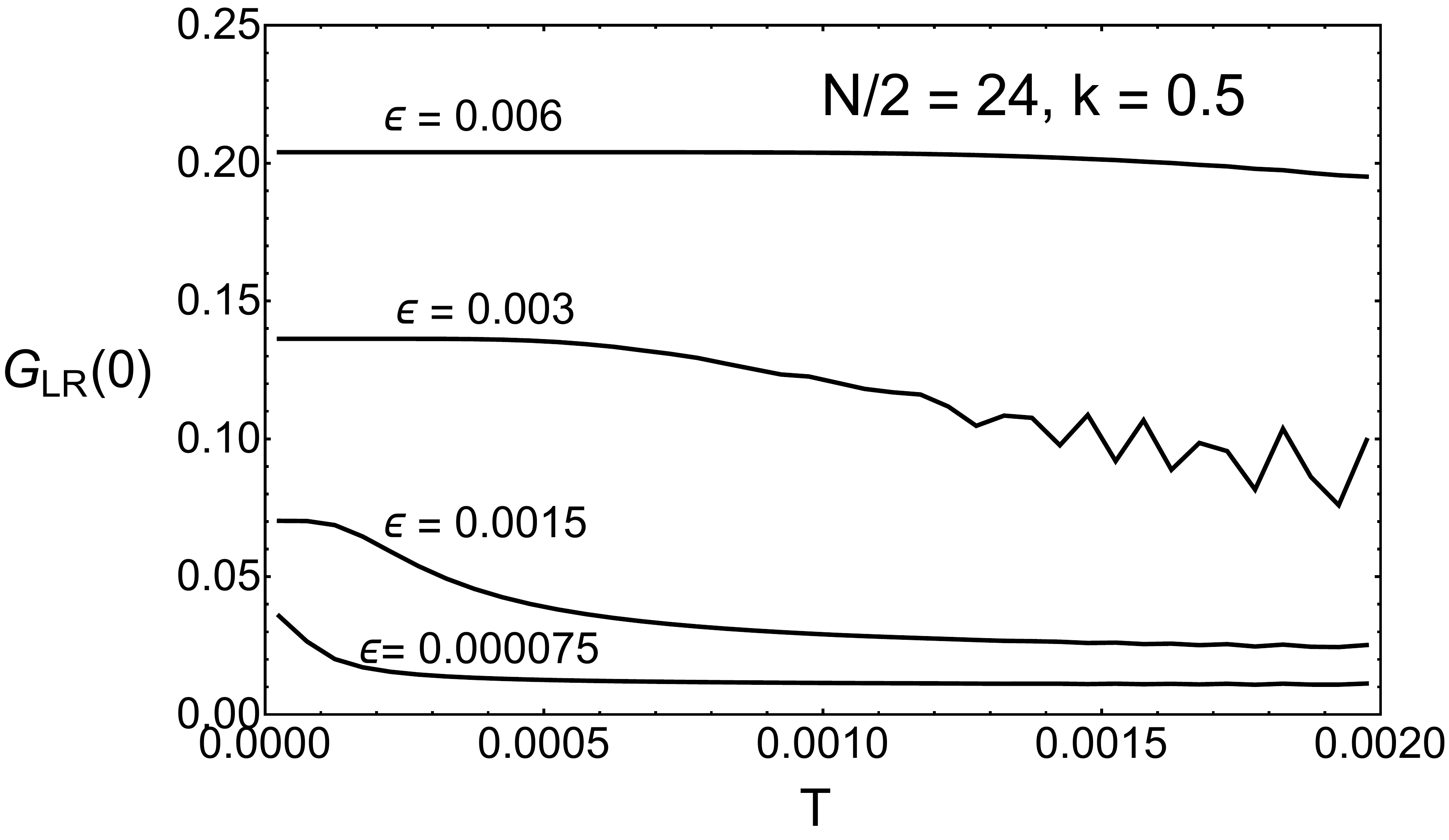}}
   \caption{The temperature dependence of $G_{LR}(0)$
     as a function of the temperature
     $T$ for $k=0.5$ and $N=20$ (left) or $N=24$ (right). The values of
     $\epsilon$ are indicated in the figures.
     %figures produced by steep.nb
  \label{fig:glr0t} }
  \end{figure}

The theoretical arguments given in this section are consistent with the
numerical calculation of $G_{LR} (0)$ from the solution of the SD equations (see Figure \ref{fig:glr0t}) which identifies $G_{LR} (0)$  as the order parameter of a first-order
phase transition.

   We now show that an exact finite $N$ calculation of the order parameter $G_{LR}(0)$ based on the exact diagonalization of the Hamiltonian yields similar results. 
 Since a non-Hermitian matrix can be diagonalized by a similarity
 transformation, $ e = V^{-1}H V$, we have that
 \be
 G_{LR}(0) = \frac 1Z \frac 2N \Tr \left [ e^{-\beta e}V^{-1} S V \right  ].
 \ee
 In Figure \ref{fig:glr0mu}, we show the zero-temperature limit of
 this quantity as a function of $\epsilon$ 
 for $N/2 = 12$, $N/2=16$, $N/2=20$ and
 $N/2=24$, all for $k =0.5$. The RMT universality class of a single SYK is
 GUE, GOE, GUE and GSE, in this order.
In agreement with the above arguments,
 in the GUE class the symmetry is
 spontaneously broken, but in the GOE and the GSE class it is not clear
 whether a finite result can be obtained for large $N$ and small $\epsilon$.
 The temperature dependence of $G_{LR}(0)$ for  $N=20$  and $N=24$ is
 shown in Figure~\ref{fig:glr0t}. Again we observe that the GUE universality
 class (for $N=20$) and the GSE universality class (for $N=24$)
 behave qualitatively different. In the
 GUE universality class (left), the low-temperature limit of $G_{LR}(0)$ saturates to
 a finite value, while the pseudo-critical temperature seems to be
 proportional to $\epsilon$. For the GSE universality class, the zero-temperature value is proportional
 to $\epsilon$ while the critical temperature seems to scale as $\epsilon^2$.
 Also the numerical value of the zero-temperature limit of $G_{LR}(0)$ is
 below the result obtained from the SD equations which may be due to the slow
  convergence of the large $N$ limit.

 \section{Outlook and conclusions}\label{sec:outlook}
 By downgrading the condition of Hermiticity to only $PT$ symmetry in random quantum systems,
 the saddle point equations have RSB solutions connecting different replicas with lower energy than the replica-symmetric ones.
 The nature of these solutions is strikingly similar to that of wormholes in JT gravity. With the free energy as observable, we have identified a first-order transition in two examples where the dynamics is quantum chaotic, the two-site Ginibre model and the two-site non-Hermitian SYK model.

 The free energy of the two-site non-Hermitian SYK model was calculated in two ways: by
 explicit diagonalization of the Hamiltonian at finite $N$, and by solving the Schwinger-Dyson
 equations in the thermodynamic limit. 
 A strong first-order transition at finite $T$, separates a low-temperature phase where the free energy is dominated by RSB configurations (the wormhole phase)
 from the high-temperature phase controlled by replica symmetric configurations (the two-black-hole phase). Although we did not present explicit results relating the infrared limit of this SYK model with a gravity theory, this transition is reminiscent of an Euclidean wormhole-to-black-hole transition \cite{Garcia-Garcia:2020ttf}.  The solutions
 of the Schwinger-Dyson equations also indicate that there is a second phase transition
 at a lower temperature, which is continuous,
 below which the free energy becomes strictly constant. In between these two phase transitions,
 the free energy increases rapidly until it jumps to the value of the two-black-hole phase
 at the first-order phase transition temperature. Because of substantial finite size effects,  
 only for $k=1$, the existence of a phase with a constant free
 energy is established unambiguously. For $k < 1$ we cannot exclude that the derivative of the free energy remains non-vanishing all the way to zero temperature. Another remarkable observation is that the zero-temperature
 entropy of the black hole phase does not depend on the degree of non-Hermiticity as long as
 $k < 1$ but jumps to the high-temperature value at $k = 1$.
 We have no good explanation what causes the discontinuity of the $k$-dependence of the
 entropy.

  Although we have restricted our analysis to $q = 4$, we expect that this transition is universal provided that $q > 2$ when the dynamics is
 quantum chaotic and therefore its spectral correlations are expected 
 to be  those of the Ginibre ensemble. This expectation is based on the
 following facts: i) we have found excellent agreement between the two-site elliptic
 Ginibre ensemble and the $q = 4$ two-site non-Hermitian SYK model -- in some sense the Ginibre model is an SYK model with $q\sim N$  and ii) for real couplings, spectral correlation do not depend qualitatively on the value of $q > 2$. 
 
 A natural question arises: is it a requirement for this first-order phase 
 transition to happen that the dynamics is quantum chaotic?  Indeed for the
 two-site $q = 2$ SYK model which is integrable  \cite{maldacena2016,Jia:2022reh}
 we only have second-order phase transitions  \cite{Jia:2022reh}.
However, as is shown in
 Appendix \ref{rdisk}, for uniform uncorrelated random eigenvalues
 inside the complex unit disk we do find a first-order phase transition, but
 in the low-temperature phase the free energy depends linearly on the temperature.

 The results of this paper open several promising research avenues. First of all, it remains to establish unambiguously the existence of the second continuous phase transition mentioned above.
 Because of the smallness of the entropy in the wormhole phase, this requires a
 new  algorithm for solving the SD equations that greatly reduces the finite size effects,
 allowing us to study the details of this phase transition. An improved algorithm will
 also help us to analyze the discontinuity of the $k$-dependence of the entropy.

 It would also be interesting to explore whether a generalized Schwarzian is still the effective description of the infrared limit of both, a perturbed JT gravity theory (related to Euclidean wormholes) and the two-site non-Hermitian SYK investigated in this paper. If this is the case, like for  traversable wormholes \cite{maldacena2018}, it would be a strong indication that RSB configurations are the field theory equivalent of
 Euclidean wormholes which may be relevant in the solution of the factorization problem in holography.  It would also be interesting to include in our model an explicit Maldacena-Qi coupling in order to study transition/crossover from Euclidean to traversable
wormholes  \cite{Garcia-Godet-2022}.
 
In light of our results, and recent developments in the resolution of the information paradox \cite{almheiri2020,penington2020}, an interesting research direction 
is to investigate the growth of entanglement entropy in a setting based on the non-Hermitian SYK. Of special interest is the contribution of multi-replica wormholes in the late stages of the time evolution. 
 Another problem that deserves further attention is a more exact delimitation of the
conditions to observe RSB configurations even within $PT$-symmetric systems.  Is  quantum chaos always a necessary and sufficient condition, beyond the $q = 2$ example  discussed? If not so, is it possible to characterize the existence of RSB configurations as a function of the range of interactions?
 Are many-body correlations important or can similar results be obtained in single-particle, non-interacting two-site disordered systems? 
 Is the extension of these results to higher spatial dimensions straightforward? If so, does the existence of RSB configurations depend on the strength of disorder or the hopping range in real space? Can they occur in the presence of Anderson or many-body localization? We plan to address some of these questions in the near future.

 \acknowledgments{
 	JJV and YJ  acknowledge partial support from U.S. DOE Grant
 	No. DE-FAG-88FR40388.  YJ is also partly funded by an Israel Science Foundation center for excellence grant (grant number 2289/18), by grant no. 2018068 from the United States-Israel Binational Science Foundation (BSF), by the Minerva foundation with funding from the Federal German Ministry for Education and Research, by the German Research Foundation through a German-Israeli Project Cooperation (DIP) grant "Holography and the Swampland", by Koshland fellowship and by a research grant from Martin Eisenstein.
 	AMG was partially supported by the National Natural Science Foundation of China (NSFC) (Grant number 11874259), by the National
 	Key R$\&$D Program of China (Project ID: 2019YFA0308603) and also acknowledges financial support from a Shanghai talent program. 
 	DR acknowledges the support by the Institute for Basic Science in Korea (IBS-R024-D1).
 	AMG acknowledges illuminating correspondence with Victor Godet, Zhenbin Yang, Juan Diego Urbina and Klaus Richter.}

\appendix

\section{Calculation of the free energy for the two-site Ginibre ensemble}\label{a:ginibre}
In this appendix we evaluate the partition function of the two-site Ginibre model. For a Ginibre ensemble of $D\times D$ matrices, the eigenvalue kernel is given by \cite{ginibre1965,mehta2004}
\be
K(z_1,z_2) = \frac{e^{-z_1 z_2^*}}\pi \sum_{k=0}^{D-1} \frac{(z_1 z_2^*)^k}{k!}.
\ee
The eigenvalue density reads
\be
\rho(z) = K(z,z),
\ee
and the connected two-point correlation function is equal to
\be
\rho_{2,c}(z_1,z_2) = -K(z_1,z_2)K(z_2,z_1) + \delta(z_1-z_2) K(z_1,z_1),
\ee
where the second term is due to the self-correlations. The spectral density
is normalized to $D$ and the eigenvalues are located in a circle of radius
$\sqrt D$. To be able to adjust the overall scale of the eigenvalues, we
include a factor $\sigma$ in the definition of the partition function
\be
Z_2(\beta) = \int d^2 z_1 d^2 z_2 \rho_{2,c}(z_1,z_2) e^{-\beta (z_1+z_2^*)/\sigma}+
|Z_1(\beta)|^2.
\ee
The second contribution is due to the disconnected part of the two-point function with $Z_1$ given by
\be
Z_1(\beta)=\int d^2z \rho(z) e^{-\beta z/\sigma}.
\ee

We first calculate the disconnected contribution.
The one-site partition function requires the integral
 \be
 Z_1(\beta) = \int \frac{d^2z}\pi e^{-\beta z/\sigma} e^{-|z|^2}
 \sum_{k=0}^{D-1}\frac{ (z z^*)^k}{k!}.
 \ee
 Only the first term of the Taylor expansion of $\exp(-\beta z)$  gives
 a nonvanishing result, and after changing to polar coordinates we obtain
 \be
Z_1(\beta) = \int_0^\infty ds e^{-s} \sum_{k=0}^{D-1} \frac {s^k}{k!}
 =D.
 \ee
  Next we calculate the contribution due to self-correlations. It is given by
 \be
Z_{\rm self}(\beta)  &=&\frac 1\pi \int d^2z e^{-\beta (z+z^*)/\sigma} e^{-|z|^2}
 \sum_{m=0}^{D-1}\frac{ (z z^*)^m}{m!}\nn\\
 &&
 = \sum_{m=0}^{D-1} \sum_{l=0}^\infty
 \left (\frac {\beta}{\sigma}\right )^{2l}\frac{(m+l)!}{m!l!l!}\nn\\
 &&= \sum_{m=0}^{D-1} e^{\beta^2/\sigma^2}L_m^0(-\beta^2/\sigma^2)\nn\\
 &&=e^{\beta^2/\sigma^2}L^1_{D-1}(-\beta^2/\sigma^2),
 \label{self-gin}
 \ee
 where we have used a summation formula for associated Laguerre
 polynomials:
 \be
 \sum_{m=0}^n L_m^a(x)=L_n^{a+1}(x).
 \label{sum-lag}
 \ee
 The asymptotic behavior of the Laguerre polynomials is given by
 \be
 L_n^\alpha(-x) \sim n^{\alpha/2} (-x)^{-\alpha/2}e^{-x/2} J_\alpha(2i\sqrt{nx}).
 \ee
 For $\alpha =1 $ and $x=\beta^2/N$ this gives
 \be
 L_D^1(-\beta^2/\sigma^2) \sim\frac \sigma\beta \sqrt D  e^{-\beta^2/2\sigma^2}
 I_1(2\beta \sqrt D/\sigma).
\label{L-asym}
 \ee
 It is instructive to calculate the large $D$ asymptotics by expressing the sum in the first line
 of \eref{self-gin} as an incomplete $\Gamma$-function:
\newcommand{\erfc}{{\rm erfc}}
 \be
 Z_{\rm self}(\beta) &=&2 \int_0^\infty rdr \frac {\Gamma(D,r^2)}{\Gamma(D)} I_0(2\beta r/\sigma).
   \label{zsap}
    \ee
    For large $D$ the incomplete $\Gamma$-function can be approximated by
    \be
    \frac {\Gamma(D,r^2)}{\Gamma(D)} \approx \frac 12 \erfc((r-\sqrt D)\sqrt 2).
      \ee
      Inserting this in equation \eref{zsap} we obtain after a partial integration
      \be
      Z_{\rm self}(\beta) &=&\frac{\sigma}{\beta} \sqrt{\frac2\pi} \int_0^\infty r dr
      e^{-2(r-\sqrt D)^2} I_1(2\beta r/\sigma)\nn \\
        &\approx& \frac{\sigma}{\beta} \sqrt{\frac D 2}  
        e^{{\beta^2}/2\sigma^2} I_1(2\beta \sqrt D/\sigma)
\label{z-self}
        \ee
        For large $D$ we have that $\frac 12 \erfc((r-\sqrt D) \sqrt 2) \to \theta(\sqrt D- r)$,
        but with this asymptotic, we would have missed the $\exp(\beta^2/2 \sigma^2)$
        factor.

 The contribution of the genuine two-point correlations can be worked out
 in the same way. We obtain
 \be
 &&-\frac {1}{\pi^2} \int d^2z_1d^2z_2 e^{-(\beta/\sigma) (z_1+z^*_2)} e^{-|z_1|^2-|z_2|^2}\sum_{m=0}^{D-1}\frac{ (z_1 z^*_2)^m}{m!}\sum_{n=0}^{D-1}\frac{ (z_2 z^*_1)^n}{n!}\nn\\
 &&=  -\sum_{n=0}^{D-1} \sum_{m=0}^n {n\choose m}\frac 1{m!} \left ( \frac{ \beta^2}{\sigma^2}\right )^m\nn\\
 &&=- \sum_{n=0}^{D-1} L_n^0(-\beta^2/\sigma^2)\nn\\
 &&= -L_{D-1}^1(-\beta^2/\sigma^2),
 \label{two-gin}
 \ee
 where we have again used the summation formula \eref{sum-lag}.
 We are interested in a scaling limit where $\beta/\sigma \ll 1$. Then
 we can expand the exponent in \eref{self-gin}. The first term in the
 expansion is  canceled by the two-point correlations \eref{two-gin}.
 For the partition function we then  obtain the result
 \be
 Z_2(\beta) = D^2 + \frac{\beta^2}{\sigma^2} L^1_{D-1}(-\beta^2/\sigma^2).
 \ee
 To obtain a free energy density $\log Z/N$ that is stable in the large $N$ limit,
 we choose $D=2^{N/2}$. Inserting on the right-hand side the large $D$ limit of the Laguerre polynomial
 given in    \eref{L-asym} we find
\be
 Z_2(\beta) = D^2 + \sqrt D\frac{\beta}{\sigma} I_1                                                           (2\beta \sqrt D/\sigma).
 \ee 
   This contribution scales in the same way with $N$ as the disconnected part if
   we choose $\sigma \sim \sqrt D/N$.
Ignoring logarithmic corrections, we obtain the free energy
   \be
   \frac {F(T)}N& =& -\frac {T}N \log\left (D^2 + 
   I_1(2\beta\sqrt D/ \sigma)\right )\nn\\
   &\approx&  -\frac {T}N
   \log\left (D^2 +  e^{2\beta\sqrt D/ \sigma}\right )
   .   \ee

   The two exponents are equal at
   \be
   T_c =\frac {2\sqrt D / \sigma}{2\log D}.
   \ee
   Choosing $\sigma = \sqrt D /N $ we obtain $ T_c =2/\log 2 $, and  the
   free energy is equal to
   \be
   \frac {F(T)}N& =& -\frac {T}{\log 2}\theta(T-T_c) -2\theta(T_c-T),
   \ee
in agreement with the $k\to 1$ limit of \eref{eq:hamiltonian_antonio}.

The integrals for the contributions of the self-correlations \eref{self-gin} and genuine two-point correlations \eref{two-gin} generally cannot be calculated
exactly, and we must rely on an approximate evaluation. To do that, we
assume that the eigenvalue correlations are in the universality class of
the Ginibre ensemble. Noting that the integral over $z_1$ and $z_2$ can
be written as an integral over the center of mass and the difference
of $z_1$ and $z_2$, we assume that the integral of $z_1-z_2$ can be extended
to entire complex plane, while the integral over $(z_1+z_2)/2$ is replaced
by the large $D$ limit of the spectral density.
Since the eigenvalue correlations are short ranged, we
expect that the extension of the integral over $z_1-z_2$ to the entire complex
plane and using the large $D$ limit of the two-point function
is a good approximation. However, we have seen earlier in this section that this approximation
does not determine the exponent $\alpha$ of the prefactor $\exp[\alpha \beta^2/\sigma^2]$.
In particular, contributions from the boundary of the eigenvalue disk  may change the value of
$\alpha$ but this does not affect the free energy in the thermodynamic limit.

In the last part of this section we compare the exact and approximate expressions for
the replica breaking part of the partition function.
For the approximate calculation of the contribution of the self-correlations to the partition function
we replace the spectral density
by its large $D$ limit
\be
\frac 1\pi e^{-|z|^2} \sum_{k=0}^{D-1} \frac{(zz^*)^k} {k!}\to
\frac 1\pi \theta(\sqrt D -|z|),
  \ee
  which after changing to polar coordinates results in
  \be
  Z_{\rm self, \;app} &=& \frac 1\pi \int_0^{\sqrt D} rdr
  \int_0^{2\pi} d\phi e^{-\frac{2r\beta}\sigma \cos\phi}\nn\\
&=&  2\int_0^{\sqrt{D}} r dr I_0(2r \beta/\sigma)
  = \frac {\sigma \sqrt D}{\beta} I_1(2\beta\sqrt D /\sigma).
  \ee
  This differs by  factor $\exp(\beta^2/2\sigma^2)$ from the exact result
  \eref{z-self}, and we have seen earlier in this section that this is due to contributions
  from the boundary of the eigenvalue region.

  The approximate result for the contribution of the genuine two-point correlations  is given by
  \be
  Z_{c,\rm app}
  &&=-\frac {1}{\pi^2} \int_{|z|<\sqrt D}  d^2\bar z
  \int  d^2 \eta e^{-(\beta/\sigma) (\bar z+\bar z^* +i {\rm Im}(z_1-z_2))}
  e^{-|z_1-z_2|^2},
  \ee
  where $\bar z = (z_1+z_2)/2$ and $\eta = z_1-z_2$, and the two-point correlation function
  is replaced by its
  large $D$ limit.   The integral over the center of mass is replaced by
  an integral over   a disk with constant density which is justified in the
  large $D$ limit. The integral over ${\rm Im}(z_1-z_2)$ can be performed by
  completing squares while the integral over  ${\rm Re}(z_1-z_2)$ is a simple
  Gaussian. This results in
  \be
  Z_{c, \rm app}
  =-\frac {e^{-\beta^2/4\sigma^2}}{\pi} \int_{|z|<\sqrt D}  d^2\bar z
   e^{-(\beta/\sigma) (\bar z+\bar z^*)}.
   \ee
   The integral over the center of mass is the same as the integral that enter in the calculation of the contribution of the self-correlations.
   We thus find
     \be
  Z_{\rm c,app}
  =-e^{-\beta^2/4\sigma^2}
\frac {\sigma \sqrt D}{\beta} I_1(2\beta\sqrt D /\sigma),
  \ee
  which also does not reproduce the prefactor $\exp[-\beta^2/\sigma^2]$
  of the large $D$ limit of the exact calculation \eref{two-gin}.
The sum of the approximate result for the contribution of the self-correlation
and the contribution of the genuine two-point correlations is
given by
\be
 Z_{\rm self, app} +Z_{\rm c,app}
  =\frac{\beta^2}{4\sigma^2}
\frac {\sigma  \sqrt D}{\beta} I_1(2\beta\sqrt D /\sigma),
  \ee
  which differs by a factor four from the exact result. However, this prefactor
  does not contribute to the large $D$ limit of the free energy.

  In the main text we  used the same approximation to calculate the
  partition function for the elliptic Ginibre ensemble, and we expect that also in
  that case, only constant prefactor is affected by our approximations.

\section{Numerical solution of the  SD equations for the non-Hermitian SYK}\label{a:nSD}

We proceed iteratively as follows:
\begin{itemize}
\item Start from an appropriate initial Ansatz for $G_{LL}(\omega_n)$ and $G_{LR}(\omega_n)$. To be explicit, we found that the disconnected solutions, dominant at large temperatures, could be easily reached by starting with two copies of the free Ansatz, used to solve the standard SYK model. On the other hand, RSB solutions can be found starting from the solution for the two-site model in presence of an explicit coupling and sending the coupling to zero.
  
	\item Using a fast Fourier algorithm, one computes the corresponding correlator in the time domain.
	\item The self-energies are calculated according to  \eqref{eq:self_energies_general}, 
          and then $\Sigma_{LL}(\omega_n)$ and $\Sigma_{LR}(\omega_n)$ are obtained by an inverse
          Fourier transform.
	\item The propagators in frequency space are updated using the weighted rule
	\begin{align}
	\label{eq:updated_rule}
	& G_{LL}^{\mathrm{new}}(\omega_n) = (1 - x) \, G_{LL}^{\mathrm{old}}(\omega_n) -x \,  \frac{i \omega_n + \Sigma_{LL}(\omega_n)}{(i \omega_n + \Sigma_{LL}(\omega_n))(i \omega_n + \Sigma_{LR}(\omega_n)) + \Sigma_{LR}^2(\omega_n)} \ , \nn \\
	& G_{LR}^{\mathrm{new}}(\omega_n) = (1 - y) \, G_{LR}^{\mathrm{old}}(\omega_n) + y \,  \frac{\Sigma_{LR}(\omega_n)}{(i \omega_n + \Sigma_{LL}(\omega_n))(i \omega_n + \Sigma_{LR}(\omega_n)) + \Sigma_{LR}^2(\omega_n)} \ ,
	\end{align}
	where $x$ and $y$ are real parameters between $0$ and $1$ introduced to prevent
        over-relaxation.
	In practice we initially fix them at $0.5$ and further reduce them if the updated propagators
        start running away.
      \item The procedure is repeated until we reach convergence when difference of
        the absolute value of the Fourier coefficients becomes less than $\epsilon$.
        In most calculations we take $\epsilon = 10^{-10}$. We have checked the
        convergence by taking $\epsilon$ as small as $10^{-14}$.
        
\end{itemize}

The time domain $[0,\beta]$ is discretized as $\tau_k = \beta(k-1/2)/M$, $k= 1, \cdots, M$,
where $M$ is chosen to be $2\times 10^5$ and $10^5$ which allows us to extrapolate to the continuum limit. The corresponding Matsubara frequencies are equal to
\be
\omega_n = \frac {2\pi(n+\frac 12)}{\beta}
\ee
with $ n = -M/2, -M/2+1, \cdots M/2-1$.

\vspace*{2cm}
\noindent
{\it Symmetry properties of the correlators}\\
\label{subsec:symmetry_fourier_coefficients}
As we have shown in the main text, the propagators $G_{LL}(\tau)$ and $G_{LR}(\tau)$  satisfy the following symmetry properties:
\begin{itemize}
	\item $G_{LL}(\tau)$ is real and is symmetric about  $\beta/2$.
	\item $G_{LR}(\tau)$ is purely imaginary and is anti-symmetric about $\beta/2$.
\end{itemize}

Let us analyze the implications of the above conditions on the Fourier components of
$G_{LR}(\tau)$ since they are  relevant
for the calculation of the order parameter.
The Fourier decomposition of the propagator given by
\begin{align}
\label{eq:G_LR_series_expansion}
G_{LR} (\tau) = \frac 1\beta \sum_{n = - M/2}^{M/2 - 1} \exp \left(- i \, \frac{2 \pi}{\beta} \left(n + \frac 12\right) \tau\right) G_{LR}^{(n)} 
\end{align}
is purely imaginary
if
\begin{align}
\sum_{n = - M/2}^{M/2 - 1} \exp \left(- i \, \frac{2 \pi}{\beta} \left(n + \frac 12\right) \tau\right) G_{LR}^{(n)} = - \sum_{m = - M/2}^{M/2 - 1} \exp \left( i \, \frac{2 \pi}{\beta} \left(m + \frac 12\right) \tau\right) \overline{G_{LR}^{(m)}} \ .
\end{align}
This requires that the Fourier coefficients satisfy
\begin{equation}
\label{eq:G_LR_imaginary_condition}
G_{LR}^{(m)} = - \overline{G_{LR}^{(-m - 1)}} \ . 
\end{equation}
Similarly, we can impose the condition that $G_{LR}$ be anti-symmetric about $\beta/2$. 
This results in the requirement that the coefficients $G_{LR}^{(m)}$ are \textit{purely imaginary}
\begin{equation}
\label{eq:G_LR_periodic_condition}
G_{LR}^{(m)} = - \overline{G_{LR}^{(m)}} \ .
\end{equation}

\section{Spectral density of the non-Hermitian SYK model and Q-Hermite polynomials}\label{a:denSYKq}
\begin{figure}[t!]
	\centerline{\includegraphics[width=7cm]{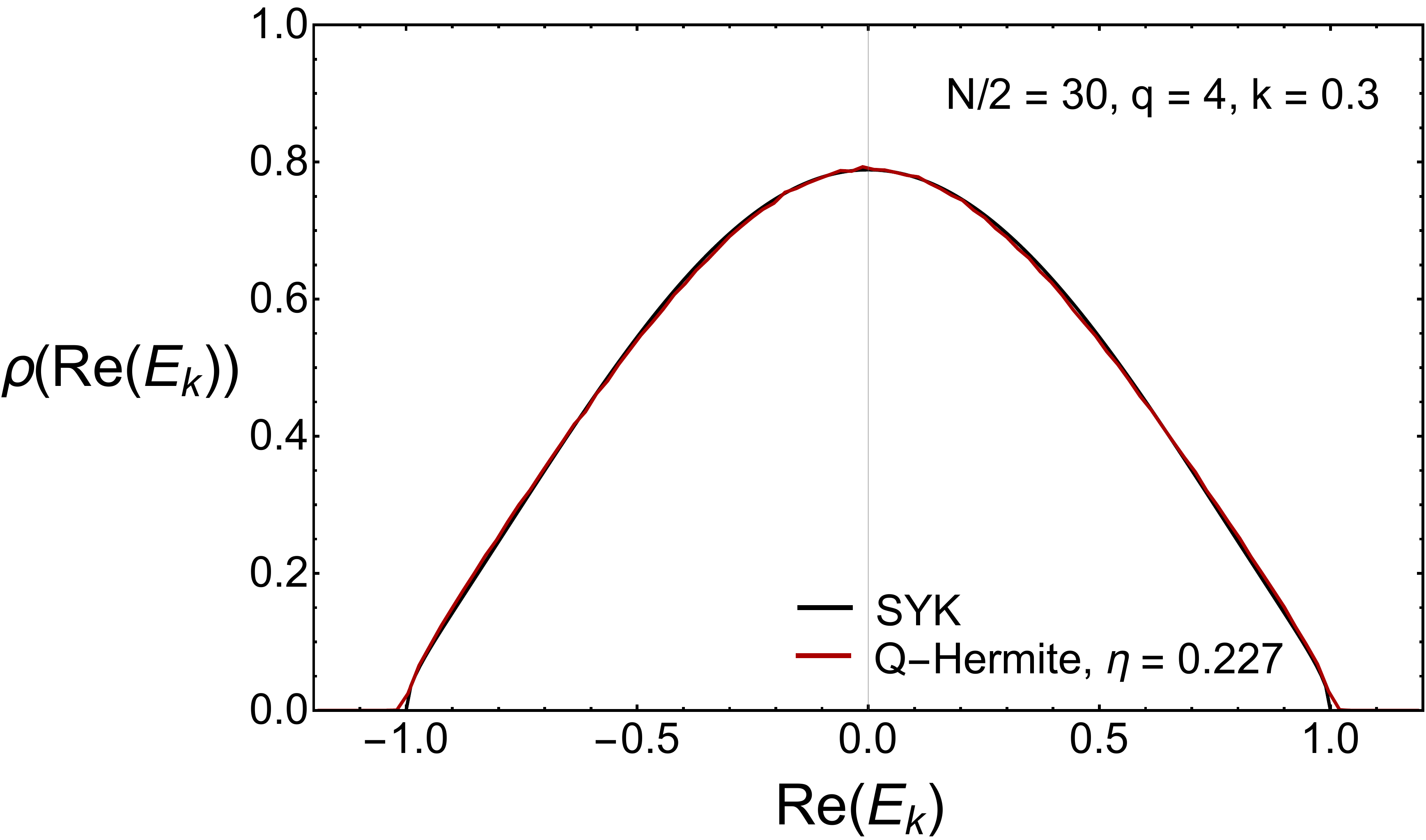}
		\includegraphics[width=7cm]{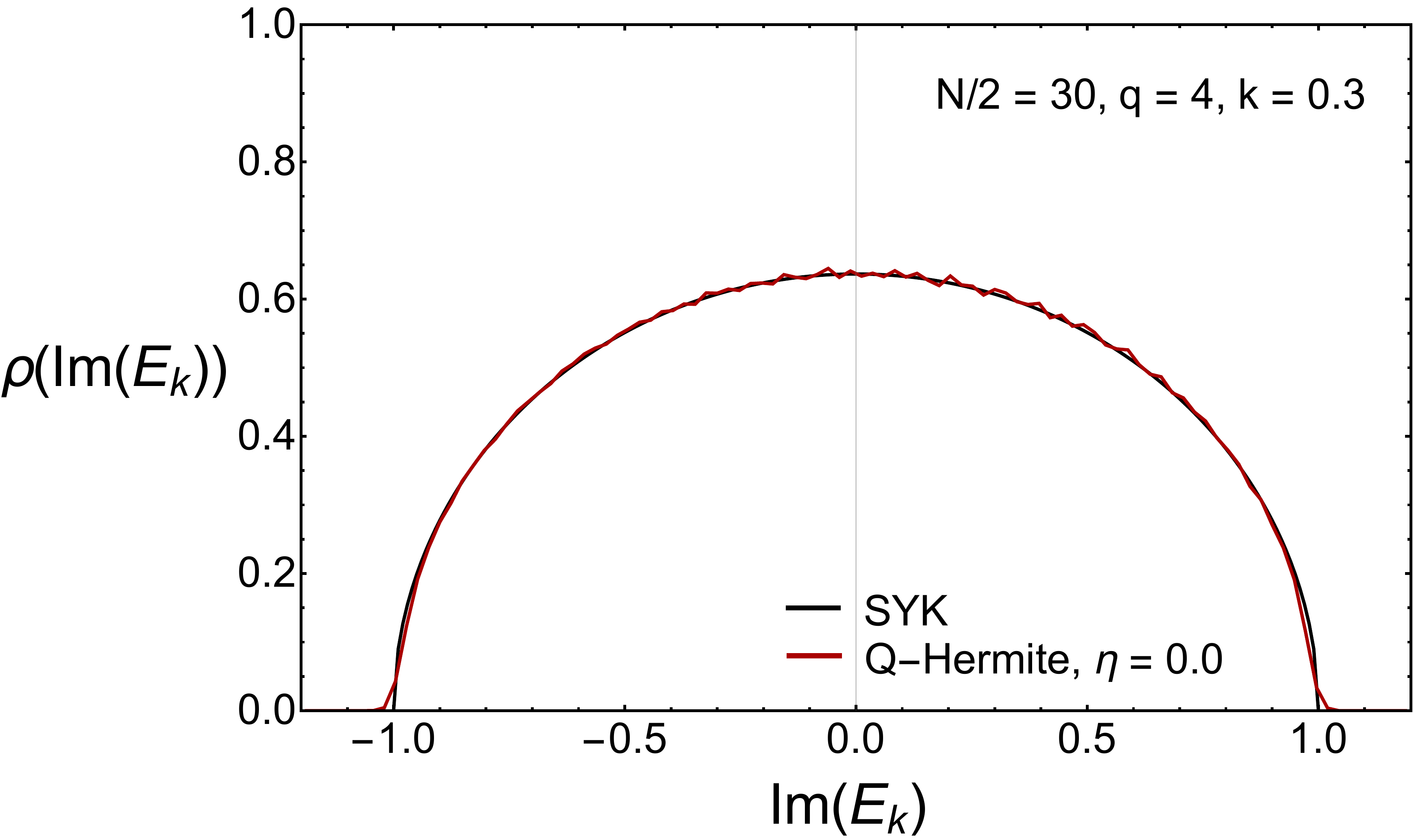}}
	        \centerline{\includegraphics[width=7cm]{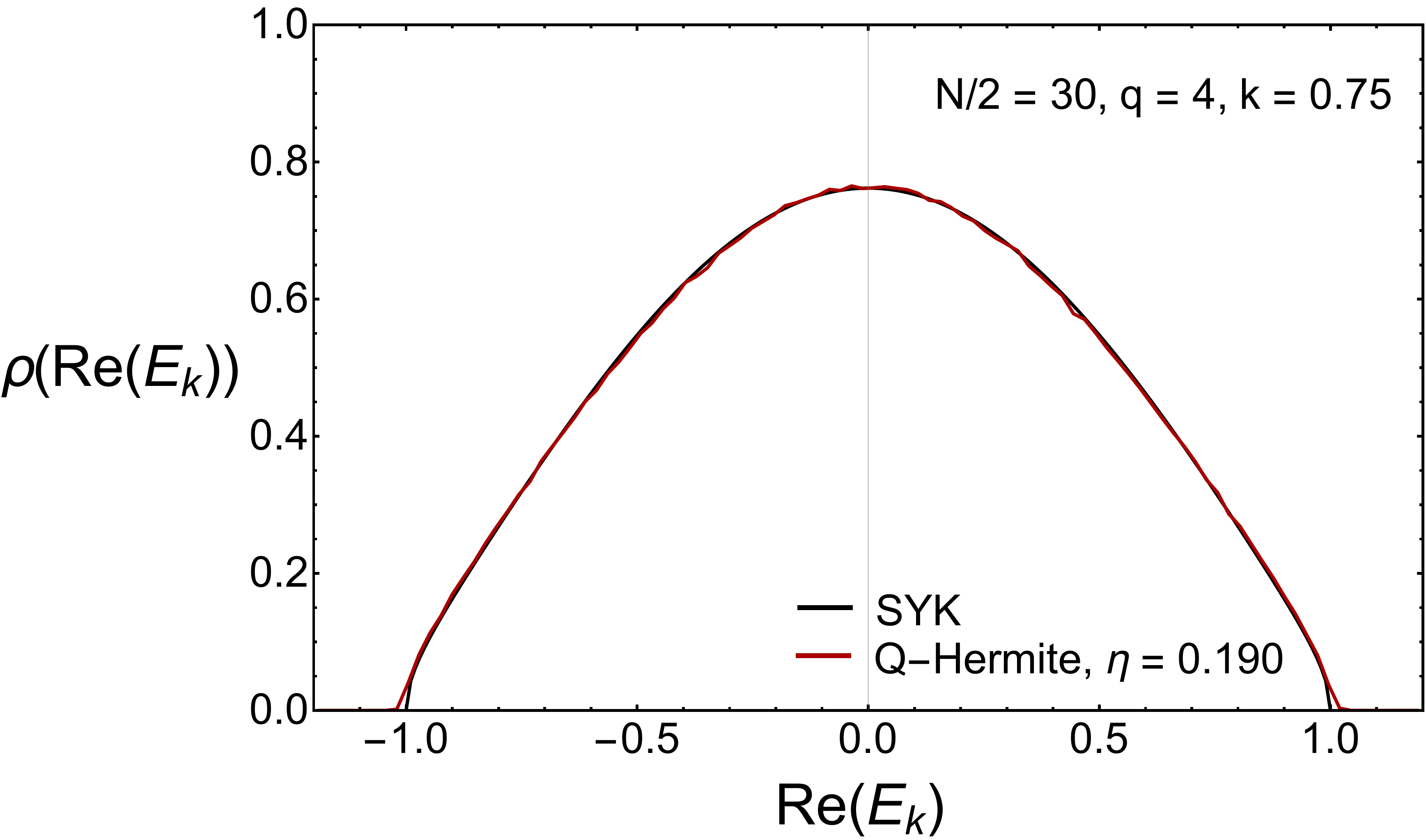}
		\includegraphics[width=7cm]{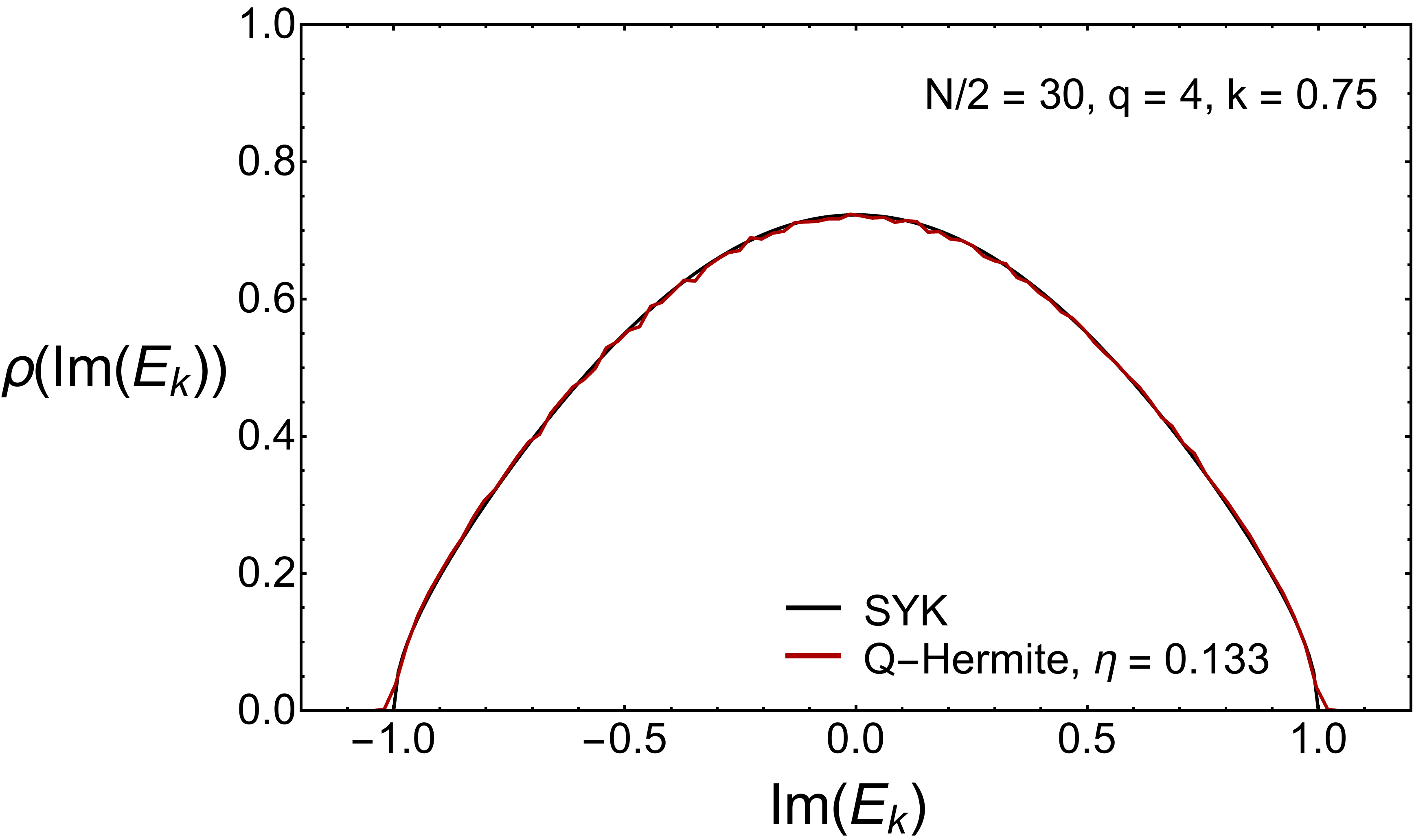}}
	\centerline{\includegraphics[width=7cm]{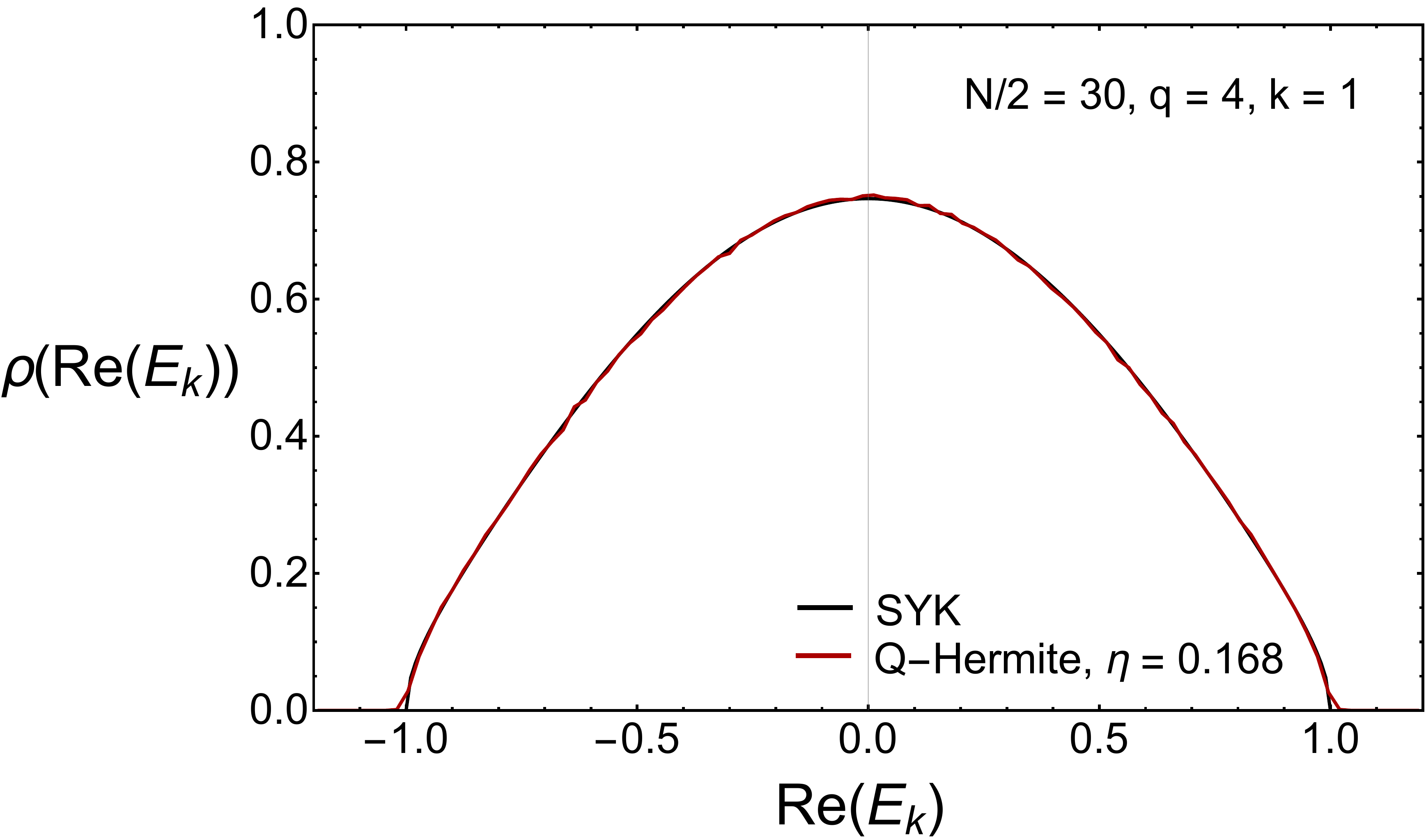}
		\includegraphics[width=7cm]{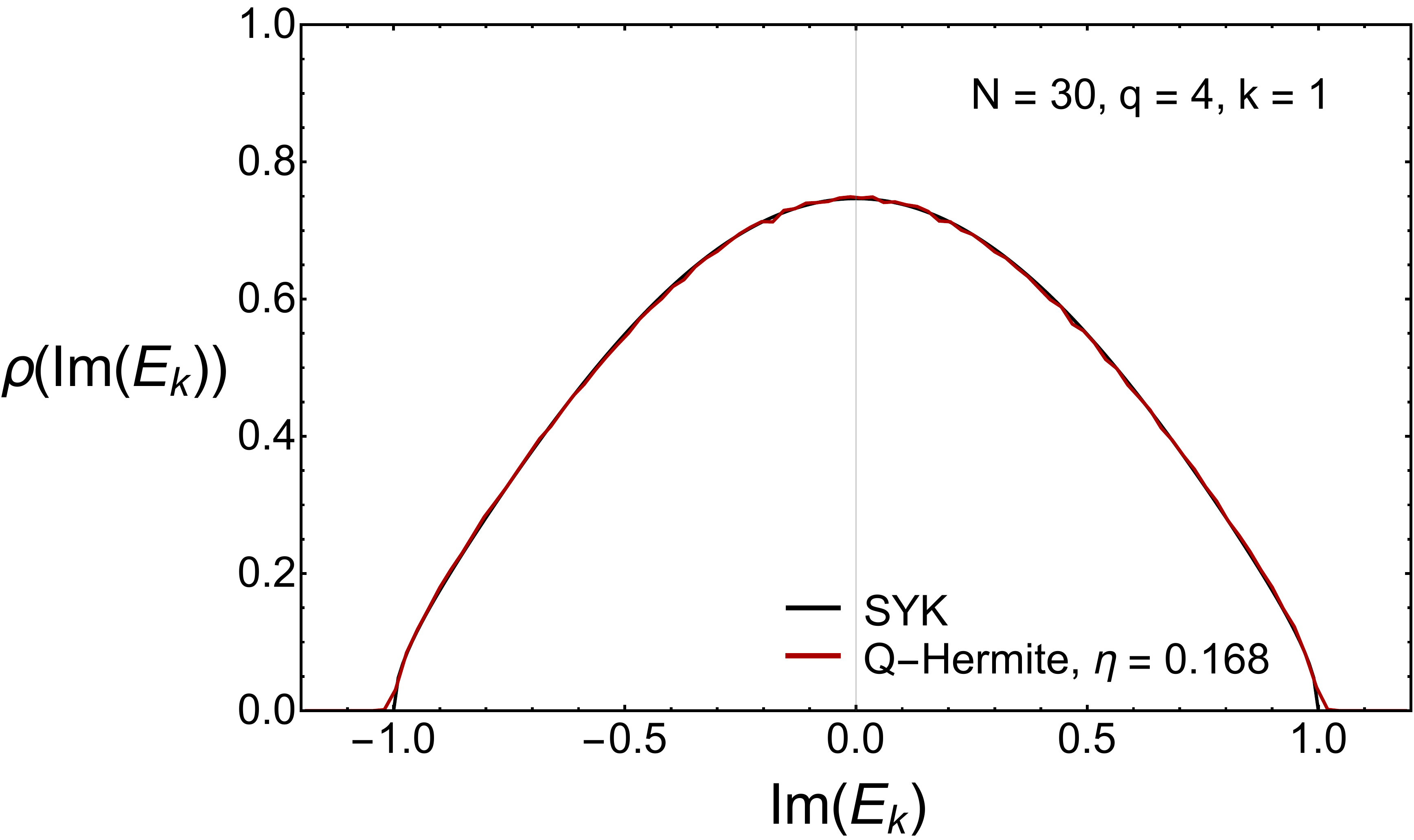}}
	\caption{The distribution of the real (left) and imaginary
		parts of the eigenvalues of the non-Hermitian one-site
		SYK model with $N/2=30$, $q=4$ and $k$ as given in the legend
		of the figure (black curve). The red curves represent a fit of the Q-Hermite density function with a value of $\eta$ given in the legend of the figure.
	\label{fig:reim}}
\end{figure}

 In this appendix, we discuss the distribution of the real and
imaginary parts of the eigenvalues. In Figure~\ref{fig:reim} we show
results for $k=0.3$, $k=0.75$ and $k=1$ (see caption). For $k=1$ the distribution of the real part (left) is the same as the distribution of the imaginary part (right). The real and imaginary parts of the  eigenvalues have been normalized by the length of the long and short axis of the  ellipses containing the
eigenvalues.
The eigenvalue distribution is fitted to the Q-Hermite spectral density
using the $Q$-parameter $\eta$ as a fitting parameter. For $k=0$ this parameter is given by
\be
\eta = {N/2\choose 4}^{-1}\sum_{m=0}^4 {4\choose m} {N/2-4 \choose 4-m},
\ee
which is equal to  $\eta =0.233$ for $N/2=30$. The values we obtain by fitting
are lower, in particular for the distribution of the imaginary part, see
legend of Figure~\ref{fig:reim} . We also point out that for $k=0.3$ the distribution of the imaginary parts
of the eigenvalues is very close to semi-circular. We expect that this also will be the case for
$k < 0.3$.

\section{Partition Function for the SYK model at $k>1$}\label{a:pfg1}
In this appendix, we relate  the free energy  for $k >1$ to the free energy for $k<1$.
The single-site Hamiltonian for $k>1$ can be written as
\be
H_L(k)=H_1 + ik H_2 = ik(H_2 - \frac ik H_1).
\ee
Since the probability distribution of  the Hamiltonian is invariant for
$H_1 \to-H_2$ and $H_2 \to H_1$, as far as ensemble averaged observables are concerned, the Hamiltonian of this ensemble can be written as
\be
ik (H_1 +\frac ik H_2) = ik H_L(1/k).
\ee
Note however the above change of variables produces an overall minus sign for  $H_R$, therefore the relation for the two-site Hamiltonian reads
\begin{equation}\label{eqn:two-siteBigk}
H(k) = ik [H_L(1/k)-H_R(1/k)].
\end{equation}
 The one-site partition function is equal to
\be
Z_L(\beta) =\vev{ \Tr e^{-\beta H_L(k)}} =\vev{\Tr e^{-i\beta k H_L(1/k)}},
\label{zk-sc}
\ee
so it is equivalent to the partition function for non-Hermiticity parameter
$1/k$ at an imaginary inverse temperature $ik\beta $.
For $k \to \infty$,  we see from equation \eqref{eqn:two-siteBigk} that the two-site partition function becomes the spectral form factor
of a Hermitian SYK model. For $\beta \to \infty$, the form factor (or the partition function)
will be dominated by the self correlations and is thus given by
\be
Z(\beta) =Z_L Z_R = \sum_{mn} e^{i\beta k (E_m -E_n)} \approx D = 2^{N/4},
\ee
 \begin{figure}[t!]
	\centerline{\includegraphics[width=9cm]{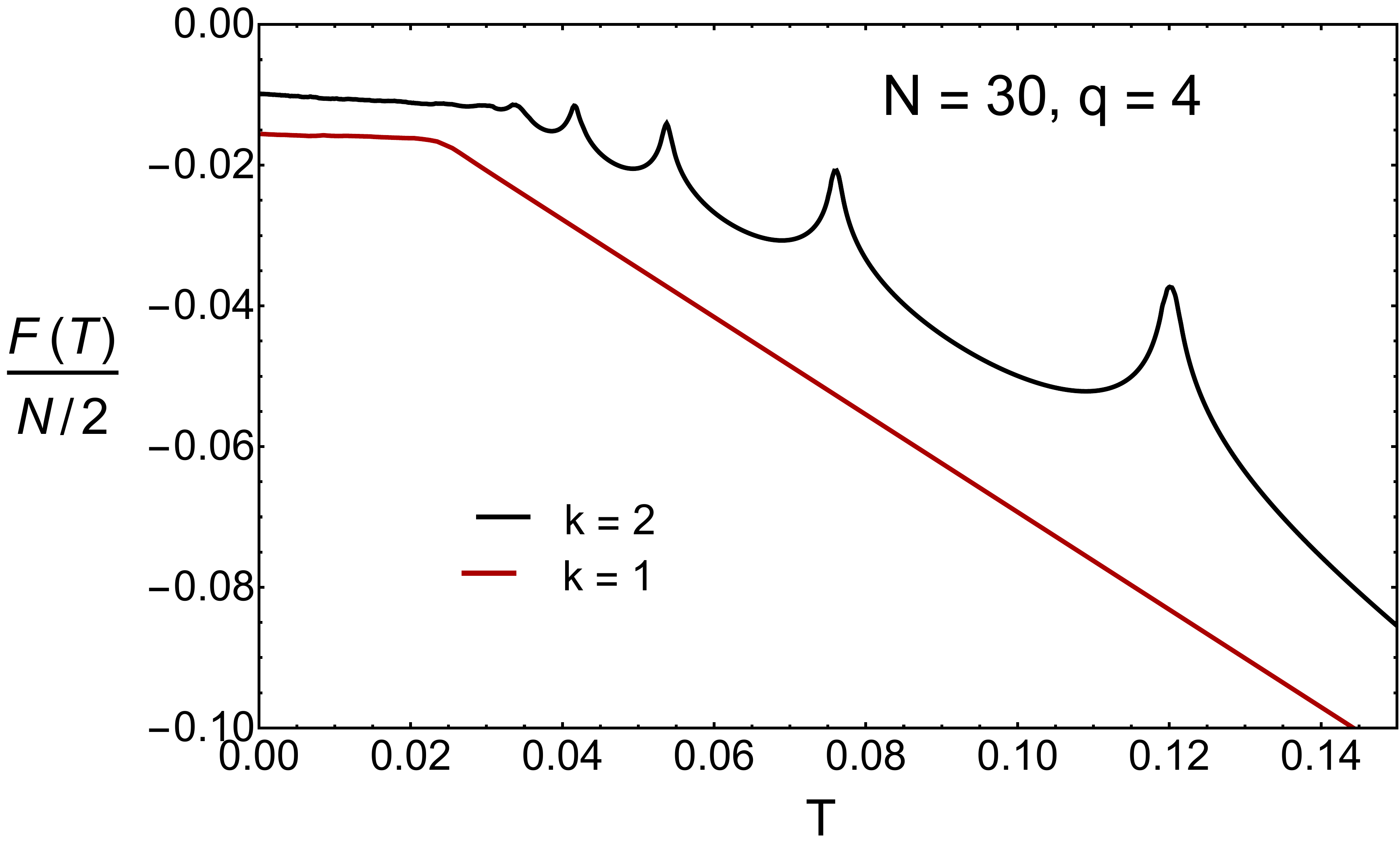}}
	\caption{The quenched free energy per particle as a function of the temperature
          for the $q=4$ non-Hermitian SYK model with  $N=30$  and the non-Hermiticity parameters
          $k=1$ (red) and $k=2$ (black). The position of the peaks are related to the zeros
          of $J_1(x)$ with the left most peak corresponding to the smallest nontrivial zero
        of $J_1(x)$.
	\label{fig:frk2}}
 \end{figure}
 resulting in a free energy of $F(T)/(N/2) = - T/2 \log 2 $ as compared to the high-temperature
limit of the free energy given by $F(T)/(N/2) = - T \log 2 $.
In Figure~\ref {fig:frk2} we show the free energy as a function of the temperature for
$k=1$ red and $k=2$ (black). The ratio of the intercepts with the $y$-axis is 1.578
while from the Ginibre ensemble we get $\sqrt{5/2}=1.58114$. If we assume that for $k>1$
the zero-temperature slope of the free energy does not depend on $k$ as is the case for $k < 1$,
an estimate for the critical
temperature can be obtained by equating $-T/2\log 2 = E_0 =-0.009849$. This gives $T_c=0.0284$,
which is in good agreement with Figure \ref{fig:frk2}.
The position of the peaks
is approximately inversely proportional the position of the zeros of $J_1(x)$ with the left most
peak corresponding to the smallest nontrivial zero, which will be explained
in the next paragraph.

We can work out the $k>1$ partition function in more detail for the elliptic
Ginibre ensemble (which provides a good approximation of the spectra of the non-Hermitian
SYK model). Using the scaling \eref{zk-sc}
the eigenvalues of this ensemble are distributed
homogeneously inside an ellipse given by
\be
x= E_0 \cos \phi, \qquad y = y_0 \sin \phi
\ee
with $E_0$ and $ y_0$ given by equation \eref{Ey} with $k \to 1/k$, namely 
   \be
        E_0 = \frac{k \sigma_0}{\sqrt{1+k^2}\sigma(1/k)},
        \qquad y_0 = \frac{ \sigma_0/k}{\sqrt{1+k^2}\sigma(1/k)}.
        \ee
The one-site  partition function is then given by
\be
Z_L(\beta) &=&
\frac {2D}{ik\beta \sqrt{E_0^2-y_0^2} } I_1\left(ik\beta \sqrt {E_0^2-y_0^2}\right)=
\frac {2D}{k\beta \sqrt{E_0^2-y_0^2} } J_1\left(k\beta \sqrt {E_0^2-y_0^2}\right),
\label{zlg1}
\ee
which vanishes at the zeros of the Bessel function. If
the zeros of $J_1$ are given by $z_n$, the partition function vanishes when
\be
\beta\sqrt{E_0^2-y_0^2} =z_n.
\ee
The free energy is singular at the critical temperatures
\be
T_n= \frac{\sqrt{E_0^2-y_0^2}}{z_n},
\ee
which are of the order of the system size in the normalization that $E_0$ scales linearly
with the number of particles. 

Since the exponent of the disconnected  part  is purely imaginary 
it does not contribute to the free energy in the thermodynamic
limit. The disconnected part of the two-site partition function is thus given by
\be
Z_{\rm disconnected}^{k>1} \sim D^2.
\ee
Using the arguments leading to \eref{eqn:z2cCircular} the connected part of the
partition function is equal to
\be
Z_{\rm connected}^{k>1} \sim e^{2\beta |E_0|}.
\ee
Equating the two gives the critical temperature
\be
T_c(k>1) = \frac{2|E_0(k)|}{\log D^2} = \frac {|E_0(k)|}{N/2 \log 2} .
\ee
This is the result for the Ginibre ensemble. The temperature dependence of the free energy
of the high-temperature phase of the SYK model is not linear in $T$ for $k \ne 1$ (apart
from the peaks due to the zeros of the Bessel function). As is the case for $k<1$ we expect
that the zero-temperature slope is smaller than $\log 2$. Earlier in this appendix we have
argued that it is equal to $\frac 12 \log 2$ resulting in a critical temperature of
\be
T_c(k>1) = \frac {2|E_0(k)|}{N/2 \log 2},
\ee
which is in good agreement with Figure~\ref{fig:frk2}.
\section{Degeneracies of the non-Hermitian SYK models}
\label{app:degeneracy}

Numerical diagonalization reveals that the non-Hermitian one-site SYK model has exactly the same energy level degeneracies as the Hermitian model:
\begin{itemize}
\item $N/2 \text{ mod } 8 = 0$,  $T^2 =1$, no degeneracy.
\item $N/2 \text{ mod } 8 = 2, 6$,  $T^2 =\pm1$, two-fold  degeneracy.
\item $N/2 \text{ mod } 8 = 4$,  $T^2 =-1$, two-fold  degeneracy.
\end{itemize}
Here, $T$ is the time reversal operator with
\be
T = C K
\ee
with $C$ the product of the odd or even gamma matrices, and $K$ the complex conjugation operator.
In the Hermitian case, a proof can be found in the appendix A of
\cite{garcia2016}.
In the first case the $T$ operator commutes with the chirality matrix $\gamma_c$ and with
the product of four gamma matrices in the Hamiltonian. This does not impose any conditions on the
eigenvalues of the two blocks also in the non-Hermitian case so that the eigenvalues are
non-degenerate.
In the latter two cases the degeneracy proof  of \cite{garcia2016} needs to be modified for
the non-Hermitian case.

\subsubsection{$N/2 \text{ mod } 8 = 2, 6$}
In this case the degeneracy can be shown by a light modification of the
proof of \cite{garcia2016}.  The time reversal operator anti-commutes with the chirality matrix $\gamma_c$ and hence, in the chiral basis, it has the form
\begin{equation}\label{eqn:GUEtimeRev}
T = \begin{pmatrix}
0 & c K \\ c^* K &0
\end{pmatrix}
\end{equation}
where $K$ is the complex conjugation and $cc^* = \pm 1$.  Since $\gamma_c$ commutes with the Hamiltonian we have
\begin{equation}\label{eqn:GUEblockHami}
H =  \begin{pmatrix}
A & 0 \\ 0 &B
\end{pmatrix}.
\end{equation}
In the non-Hermitian model time reversal transforms the Hamiltonian as 
\begin{equation}
T^{-1} H T = H^ \dagger.
\end{equation} 
Then eqns. \eqref{eqn:GUEtimeRev} and \eqref{eqn:GUEblockHami} imply
\begin{equation}
c^{-1}A c = B^T
\end{equation}
and hence we conclude $A$ and $B$ have the same eigenvalues, so $H$ is generically two-fold degenerate.  Notice that the degeneracy comes form two different chirality sectors, so $\vev{ \Omega| \psi_L \psi_R| \Omega}$ in the corresponding two-site model is not necessarily zero if we take $|\Omega \rangle$ to be a linear combination of two different chiralities. 

\subsubsection{$N/2 \text{ mod } 8 = 4$}
This would be the GSE case for the Hermitian SYK model. The two-fold degeneracy is a Kramers degeneracy and the proof is simple for the Hermitian case which we now briefly recap.  Since $T$ is a symmetry in the Hermitian SYK,
 if $ |v \rangle$ is an eigenstate then $|T v \rangle$  is also an eigenstate. To prove $|T v \rangle$  is linearly independent from $ |v \rangle$ we use the fact that $T$ is anti-unitary and $T^2=-1$:
\begin{equation}
\vev{ v| Tv} = \vev{ v| T^{-1}T^2v} =   \vev{T v| T^2v}^* = - \vev{T v| v}^* = - \vev{ v| Tv}. 
\end{equation}
Hence $|T v \rangle$ is orthogonal to $|v \rangle$ and we conclude two-fold degeneracy.

In the non-Hermitian case,  $T$ is no longer a symmetry and we only have
\begin{equation}
T^{-1} H T = H^ \dagger,
\end{equation} 
so  $|T v \rangle$ is not an eigenstate of $H$ even if  $|v \rangle$ is.  However the spectrum is still two-fold degenerate, and the proof for Kramers degeneracy needs modification.  The gist is a proof by contradiction: if there exists an eigenvalue of $H$ that is not degenerate, then $T$ cannot be invertible and hence contradicts its anti-unitarity. 

Suppose  $H$ is a $D \times D$ matrix and the complete set of eigenvectors of $H$ is $\{|v_1\rangle,  |v_2\rangle, \ldots, |v_D\rangle\}$.  Suppose   $|v_1\rangle$ is a vector with a nondegenerate eigenvalue $\lambda_1$:
\begin{align}
H |v_1\rangle  &= \lambda_1 |v_1\rangle ,\\
H^\dagger | T^{-1}v_1\rangle & = \lambda_1^* | T^{-1}v_1\rangle ,
\end{align}
where the second equality follows from $T^{-1} H T =H ^\dagger$. We still have $\vev{ v_1| Tv_1}=0$ for the same reason as in the Hermitian case.  For any other eigenvector    $|v_i\rangle$ ($i \neq 1$),  we have that
\begin{equation}
H |v_i\rangle  = \lambda_i |v_i\rangle \implies \langle v_i | H^\dagger = \langle v_i |  \lambda_i ^*.
\end{equation}
Combining the above equations we can deduce that for $i\neq 1$
\begin{equation}
  \lambda_i ^*  \langle v_i |  T^{-1}v_1\rangle =   \langle v_i | H^\dagger  T^{-1}v_1\rangle =   \lambda_1 ^*  \langle v_i |  T^{-1}v_1\rangle, 
\end{equation}
since $\lambda_1\neq \lambda_i$ by assumption , we conclude $\langle T v_i |v_1\rangle= \langle v_i |  T^{-1}v_1\rangle^*=0$. 

We have now  proven that  if $|v_1\rangle $ has a nondegenerate eigenvalue then $|v_1\rangle $ is orthogonal to all $T|v_i\rangle $ ($i =1,2,\ldots, D$), namely  $T$ brings the full Hilbert space into the orthogonal complement of $|v_1\rangle $, this contradicts the fact that $T$ is an invertible operator.  Hence by contradiction we have proven that every eigenvalue is at least two-fold degenerate.  Given that there is no symmetry mechanism to enforce an even higher degeneracy,  we will see two-fold degeneracy for a generic realization of the ensemble.   

It is worth noting that the above proof applies to each chiral block Hamiltonian because both $H$ and $T$ commute with $\gamma_c$ so in chiral basis  they are simultaneously block-diagonal.   So degenerate eigenstates always have the same chirality and $\vev{ 0| \psi_L \psi_R|0}$ in the corresponding two-site model (without the $i\epsilon$ symmetry breaking term) must vanish regardless which linear combination one takes.

\section{Random Eigenvalues in a Disk}\label{rdisk}

In this appendix we discuss the partition function of uniform
uncorrelated random eigenvalues in a disk with level density
given by
\be
\rho(z) =\frac D{\pi R^2} \theta(R -|z|).
\ee
The radius $R$ will be adjusted in order to get a stable large $D$ limit.

The one-site annealed
partition function is the same as the one for the Ginibre model:
\be
Z_1(\beta)= \int \rho(z) e^{-\beta z} d^2z = D.
\ee

\begin{figure}[t!]
	\centerline{   \includegraphics[width=8cm]{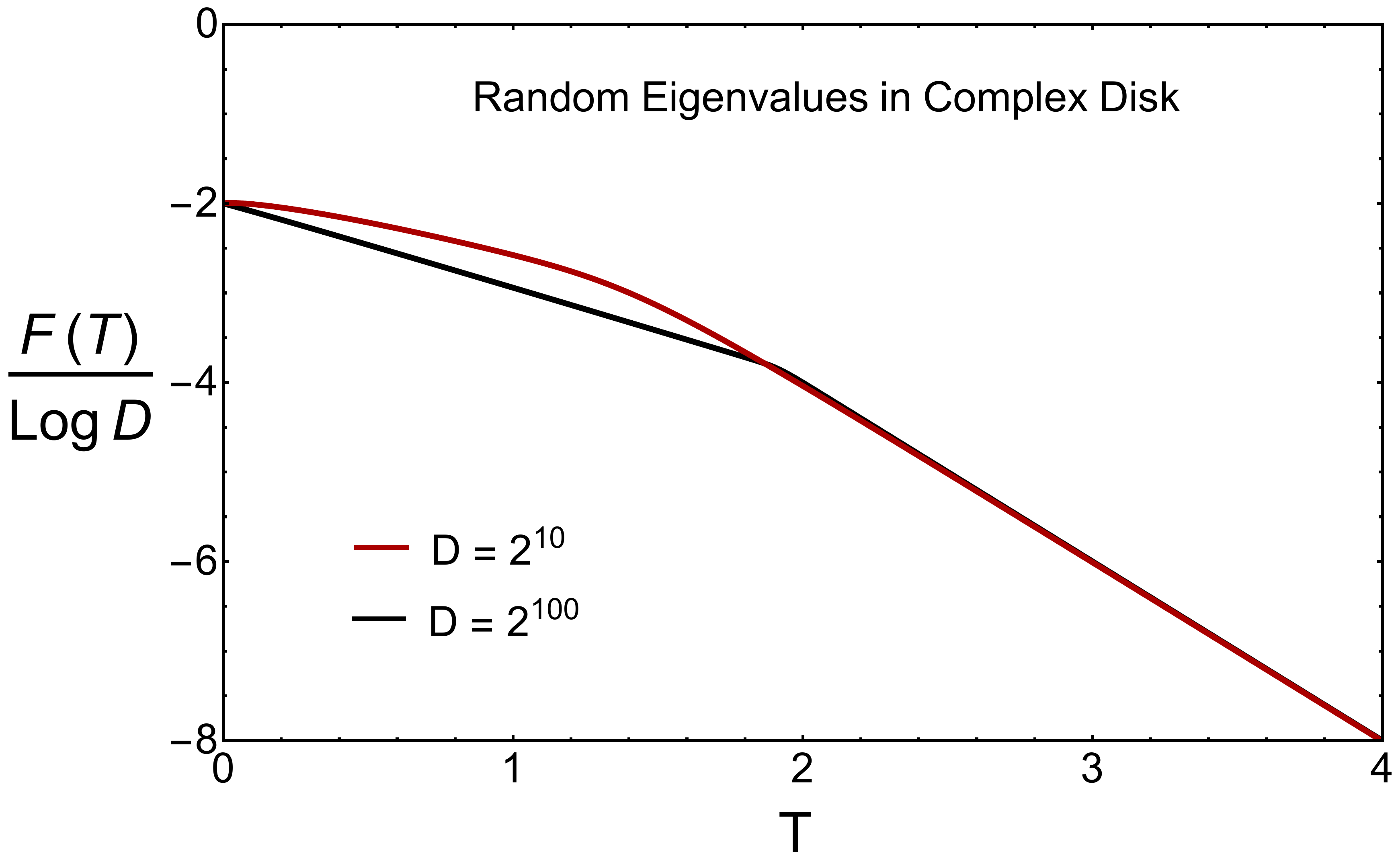}}
	\caption{Annealed two-site free energy of uniformly distributed random eigenvalues in a complex disk
		of radius $\log D$.
		Results are plotted for $D=2^{10}$ (red) and $2^{100}$ (black).
	\label{fig:disk}}
\end{figure}
The two-site partition function corresponding to one replica
and one conjugate replica is given by
\be
Z_2(\beta) &=& \left \langle \sum_{kl} e^{-\beta (E_k + E_l^*)}\right \rangle\nn\\
&=& \left \langle \sum_{k\neq l} e^{-\beta (E_k + E_l^*)}\right \rangle
+ \left \langle \sum_{k} e^{-\beta (E_k + E_k^*)}\right \rangle
\ee
Since different eigenvalues are uncorrelated, and all eigenvalues have the same
distribution.  We obtain in the large $D$ limit
\be
Z_2(\beta) &=& \frac{D(D-1)}{\pi^2 R^4} \int_{\cal D} d^2z  e^{-\beta z}\int_{\cal D} d^2 z  e^{-\beta z^*}
+ \frac{D}{\pi R^2}  \int_{\cal D} d^2z e^{-\beta (z+z^*)}
\nn \\
&=& D(D-1) + \frac D{R \beta } I_1(2 R \beta ).
\ee
The  large $D$ limit of the free energy density is given by
\be
\frac{F(T)}{\log D} = -  \lim_{D\to \infty} \frac{T}{\log D}\log( Z_1^2(\beta) +Z_2(\beta)) = 
-\lim_{D\to \infty}   \frac{T}{\log D}\log \left (e^{2\log D} +  \frac D{R \beta } I_1(2 R \beta )\right ).
\label{free-disk}
\ee
A nontrivial large $D$ limit is obtained if we scale $R$ as $\log D$, and we will
choose $R=\log D$. At the critical temperature $T_c$ the two leading exponents
are equal so that
\be
2\log D = \log D + 2 \beta_c \log D
\ee
resulting in
\be
T_c =2,
\ee
and the free energy is given by
\be
\frac{F(T)}{\log D} =(-2-T) \theta(T_c-T )  - 2T \theta(T-T_c).
\label{disk-asym}
\ee
In Figure \ref{fig:disk} we show the free energy \eref{free-disk}
versus the
temperature. The convergence to the asymptotic result \eref{disk-asym} is slow, but
for $D=2^{100}$ (black curve) a kink becomes visible.

\bibliography{librarynh}
 
\end{document}